\documentclass[iop,twocolappendix,numberedappendix]{emulateapj}
\usepackage{natbib}
\usepackage{graphicx, amssymb, amsmath}
\usepackage{comment}
\usepackage{xspace}
\usepackage{defs}
\usepackage{cleveref}
\usepackage{hyperref}
\usepackage{listings}
\usepackage{adjustbox}
\crefname{section}{§}{§§}
\begin{document}

\newcommand{\Dn}{$D_{n}(4000)$\xspace}
\newcommand{\Hd}{$H\delta$\xspace}
\newcommand{\OII}{$[OII]$\xspace}
\newcommand{\ewHd}{EW($H\delta$)\xspace}
\newcommand{\ewOII}{EW($[OII]$)\xspace}
\newcommand{\DoNothingA}[1]{#1}
\newcommand{\tbf}{\DoNothingA}

\title{Star Formation In The Cluster Merger DLSCL J0916.2+2953}
\author{A. S. Mansheim\altaffilmark{1}, B. C. Lemaux\altaffilmark{1}, W. A. Dawson\altaffilmark{2}, L. M. Lubin\altaffilmark{1}, D. Wittman\altaffilmark{1,3}, S. Schmidt\altaffilmark{1}}
\affil{$^{1}$Physics Department, University of California, Davis, One Shields Avenue, Davis, CA 95616, USA, asmansheim@ucdavis.edu}
%\email{asmansheim@ucdavis.edu}
\affil{$^{2}$Lawrence Livermore National Laboratory, P.O. Box 808 L-210, Livermore, CA 94551, USA}
\affil{$^{3}$Instituto de Astrof\'{i}sica e Ci\^{e}ncias do Espa\c{c}o, Faculdade de Ci\^{e}ncias, Universidade de Lisboa, Lisbon P-1649-004, Portugal}

\begin{abstract}

We investigate star formation in DLSCL J0916.2+2953, a dissociative merger of two clusters at z=0.53 that has progressed $1.1^{+1.3}_{-0.4}$ Gyr since first pass-through. 
We attempt to reveal the effects a collision may have had on the evolution of the cluster galaxies by tracing their star formation history. We probe current and recent activity to identify a possible star formation event at the time of the merger using \ewHd, \ewOII, and \Dn measured from the composite spectra of 64 cluster and 153 coeval field galaxies.
We supplement $Keck$ DEIMOS spectra with DLS and \textit{HST} imaging to determine the color, stellar mass, and morphology of each galaxy and conduct a comprehensive study of the populations in this complex structure.
Spectral results indicate the average cluster and cluster red sequence galaxies experienced no enhanced star formation relative to the surrounding field during the merger, ruling out a predominantly merger-quenched population.
We find that the average blue galaxy in the North cluster is currently active and in the South cluster is currently post-starburst having undergone a recent star formation event.
While the North activity could be latent or long-term merger effects, a young blue stellar population and irregular geometry suggest the cluster was still forming prior the collision. While the South activity coincides with the time of the merger, the blue early-type population could be a result of secular cluster processes. 
The evidence suggests that the dearth or surfeit of activity is indiscernible from normal cluster galaxy evolution.
%The dearth or surfeit of recent activity can only be tied to the merger environment with knowledge of the prior states of both clusters.

\end{abstract}

\section{Introduction}
\label{sec:intro}

\tbf{It is well} established that environment plays a key role in the evolution of galaxies \citep{Hogg2004}. \citet{Dressler1980} found a local morphology-density relation implying high density regions of the universe such as clusters have larger fractions of early-type galaxies.
Similarly, \citet{ButcherOemler78} observed a significantly higher fraction of blue cloud galaxies in clusters at intermediate redshifts \hbox{(0.3 $\lesssim$ z $\lesssim$ 0.5)} than in present-day clusters. More recently, \cite{P99} found a higher fraction of passive red sequence galaxies in clusters when compared to field galaxies at the same redshift. 
These observations reveal that the unique conditions within clusters of galaxies can catalyze processes that occur more slowly in field, and possibly group, environments. In this paper, we go a step beyond clusters by looking at the most significant event in the lifetime of a cluster: the cluster-cluster merger. These collisions are the most energetic events in the universe, second only to the Big Bang. As a result, the evolutionary path of a cluster galaxy during a merger may be subject to drastic change. 

Simulations suggest cluster mergers were more common in the past, making them important to understanding the results of hierarchical structure formation \citep{CohnWhite05}. \cite{Edge1990} related the frequency of cluster mergers in the past to the richness and evolution of clusters we see today. 
%Moreover, \citet{Beers1990} asserted that most clusters are not in dynamical equilibrium, making the effect of an unrelaxed environment on constituent galaxies of great relevance. 
The extreme merger environment provides a laboratory for the study of induced or quenched star formation and the physical mechanisms that govern galaxy evolution.
Astronomers have yet to establish how cluster mergers affect star formation. Conflicting results indicate slow (e.g. galaxy harassment or strangulation) and fast (e.g. ram pressure stripping) processes could either trigger \citep{MillerOwen2003, Owen2005, Ferrari2003, HwangLee2009, Ma2010}, quench \citep{P2004, Chung2010}, or have no effect on star formation \citep{Chung2009}. These mechanisms can alter not only the star formation activity, but the morphology of a galaxy.

Triggering mechanisms in cluster galaxies can occur via several types of encounters, at different radii and over different timescales. 
Galaxy mergers and tidal interactions can cause a burst in star formation on short timescales, even though mergers are less frequent in typical cluster environments due to high relative velocities \citep{Toomre1972,D99}.
As a galaxy experiences a time varying potential, simulations suggest tidal interactions can transfer gas from the disk to the nucleus, triggering star formation \citep{Bekki1999}. On longer timescales, torquing of galaxy groups or pairs can trigger star formation. %Source??
The high density of galaxies in a cluster environment is conducive to harassment, which is the cumulative effect of many high-speed close encounters as a galaxy moves towards the cluster core. Over long timescales ($\tau\geq$1 Gyr) this process can change a galaxy's morphology and gas content. On short timescales, each passage can transfer gas to the nucleus, triggering star formation \citep{Moore1996,Moore1998,Fujita1998,Haynes2007}.
As clusters pass through each other during a merger, the intracluster medium (ICM), member galaxies and dark matter potential are all disrupted. The trajectory of galaxies can be altered, facilitating major and minor galaxy mergers and tidal interactions, as well as warping as they move through changing potentials.

During a cluster merger, the ICM can be both compressed and rarefied, with bulk flows of gas increasing around galaxies as they move through the changing environment. In a dissociative merger like DLSCL J0916.2+2953, the gas distribution is fundamentally altered, detaching completely from the both cluster potentials.
As cluster galaxies infall into the dense ICM, the pressure can induce bursts of star formation in gas rich galaxies as molecular clouds collapse and form stars \citep{DG1983,Evrard1991}. This short-timescale effect becomes a quenching mechanism on slightly longer timescales, known as ram pressure stripping \citep[$\tau\sim$100 Myr,][]{Fujita1999}.
Strangulation (starvation) can occur over long timescales as a galaxy's hot halo is slowly removed by the ICM, before the gas can cool to fuel star formation \citep[$\tau>>$1 Gyr,][]{Treu2003,2000ApJ...540..113B,Larson1980}. These mechanisms that drive galaxy evolution in clusters can be dramatically impacted during a cluster merger as the ICM is decoupled from the cluster galaxies.
The unique environment of a cluster merger introduces another triggering mechanism, tidal shocks \citep{Byrd1990}.

%The majority of galaxies are subject to both slow-working and fast-acting processes as they interact with the changing environment. 
%baryonic and non-baryonic cluster components. 
%These components are all impacted by a cluster-cluster merger. 

%As clusters pass through each other, the intracluster medium (ICM), member galaxies and dark matter potential are all disrupted. The trajectory of galaxies can be altered, facilitating major and minor mergers and tidal interactions, as well as warping as they move through changing potentials. 

%The ICM can be both compressed and rarefied, with bulk flows of gas increasing around galaxies as they move through the changing environment. In a dissociative merger like Musket Ball, the gas distribution is fundamentally altered, detaching completely from the both cluster potentials. 

%A concentrated and diverse population of galaxies exists within an individual cluster. Galaxies comprise only 3\% of the total mass of a cluster, however, and their properties are at the mercy of the cluster environment. 

DLSCL J0916.2+2953, nicknamed Musket Ball because it is older and slower than the iconic Bullet Cluster, is an ideal laboratory to investigate the role of star formation in a cluster merger \citep{Markevitch2002}. Its age allows for the signatures of a range of processes and its large projected separation allows for reliable deconvolution of members \citep{Dawson2012}. 
We can identify signs of elevated or diminished star formation by looking for a burst at the time of the merger, with a hindsight of $\sim$1 Gyr since the collision.

We probe the star formation history of Musket Ball by applying techniques developed to study galaxy evolution in field, cluster and group environments.	
There is a well-established history of the use of spectral signatures to trace galaxy evolution in cluster environments. %\citep{D99, P99})
\citet{D04} and \citet{P99} utilized the equivalent widths (EWs) of \Hd (4101 \AA) and \OII (3726 \AA$\ $ and 3729 \AA) to determine of how much of the light in a cluster galaxy is from active regions of star formation (instantaneous), how much is from A stars (recent). The measurement of \Dn (4000 \AA) supplements how much light is from stellar populations older than a few Gyr. 
We rely on this established framework to derive star formation histories measured from EW(\Hd), EW(\OII), and \Dn of various galaxy populations. The addition of high resolution imaging from \textit{The Hubble Space Telescope} (\textit{HST}) allows us to study morphology, and spectroscopic redshifts allow us to perform dynamical analysis to highlight substructure. 

In this paper, we introduce results from high quality spectral measurements of DLSCL J0916.2+2953 and their possible implications on the evolution of galaxies during a cluster merger. We analyze spectra, redshifts and imaging for a comprehensive study of the star formation histories of the two clusters, contrasted with a local sample at the same redshift. In Section \ref{sec:target} we discuss our target. In Section \ref{sec:obs} we discuss details of observations. In Section \ref{sec:structure} we discuss global structure. In Section \ref{sec:galpop} we discuss methods of data analysis. In Section \ref{sec:results} we discuss results. In Section \ref{sec:disc} we discuss global conclusions.

Throughout this paper, EW measurements are presented in rest frame, with the convention of negative values for emission and positive for absorption. We assume a flat Cold Dark Matter ($\Lambda$CDM) universe with $H_0 = 70$ km s$^{-1}$ Mpc$^{-1}$, $\Omega_M = 0.3$, and $\Omega_\Lambda = 0.7$. 
%Unless otherwise specified, magnitudes are in AB.
% where J, K, R, V, and B Vega offsets are +0.91, +1.85, +0.210, +0.006, and -0.105 respectively, and 
%${\rm mag}_{AB}$=${\rm mag}_{Vega}$+offset.
%$mag_{AB}$=$mag_{Vega}$+offset.
	
\section{Target} \label{sec:target}

The Musket Ball system is a dissociative merger in which two clusters have undergone an initial pass-through and the gas has consequently separated from the bulk of the dark matter and galaxies. It was first identified by its \tbf{Weak Lensing (WL)} shear signal, using the Deep Lens Survey (DLS) \citep{Wittman2002}. The system was confirmed to be at z=0.53 using \tbf{redshifts attained} with the Low Resolution Imaging Spectrometer (LRIS) on the Keck I (2007 January 16), and two clusters were identified to be at the same redshift using the DEep Imaging Multi-Object Spectrograph \tbf{\citep[DEIMOS;][]{2003SPIE.4841.1657F}} on Keck II (2011 March 2-3) \citep{Dawson2012}.

Sub-millimeter (The Sunyaev-Zel'dovich Array Survey) and X-ray ($Chandra$ ACIS-I) observations confirm that most of the gas is offset between the North and South subclusters \citep{Muchovej2011}. With high resolution mass maps from Subaru and \textit{HST}, \citet{Dawson2013} determine North and South cluster masses in the $2 􀀀-3\times10^{14}  M_{\odot}$ range, with a roughly equal mass ratio.

\begin{figure*}
	  \begin{center}
	  \includegraphics[height=69mm]{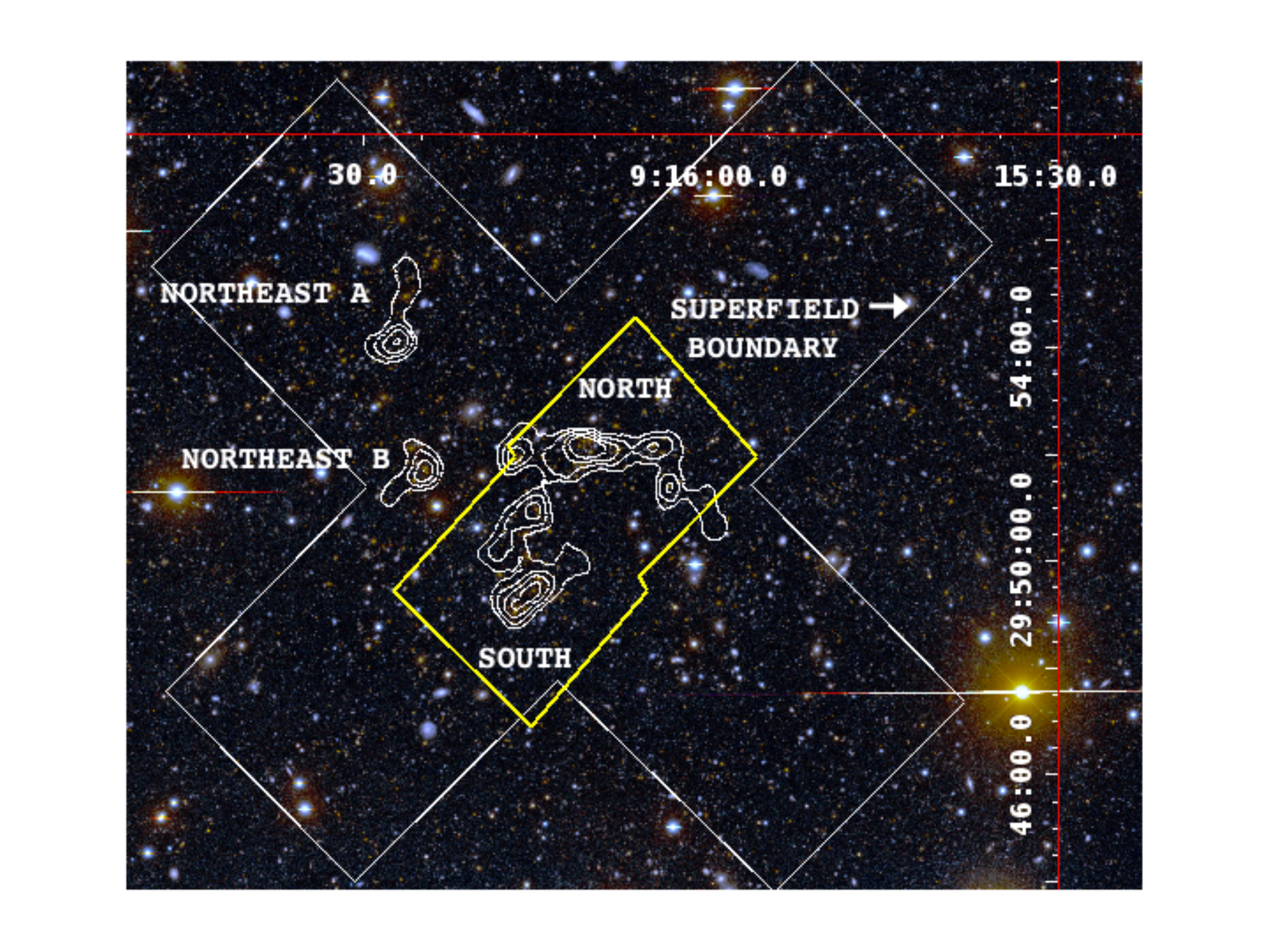}
	  \includegraphics[height=67mm]{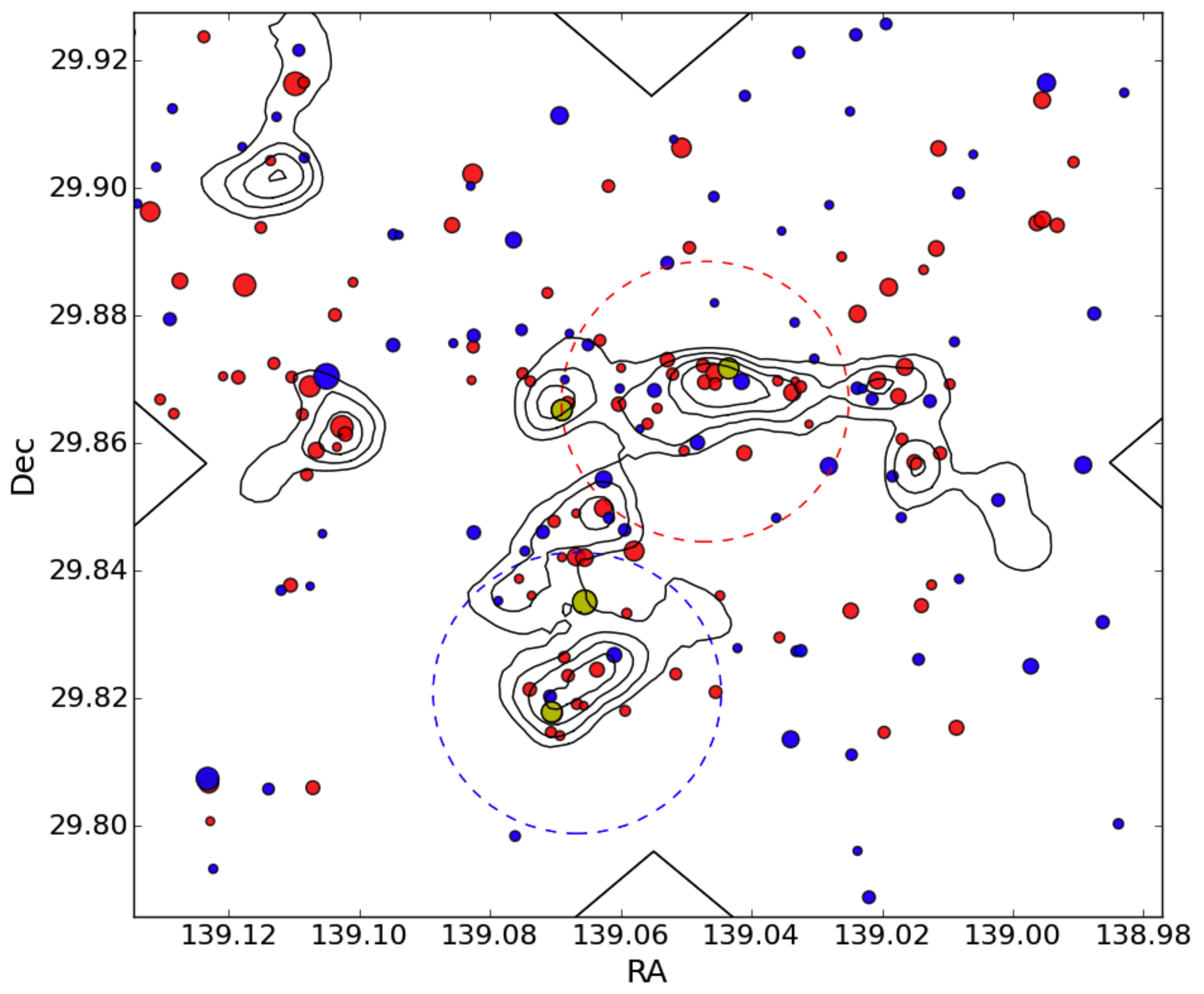}	  
	  \end{center}
	\caption{
	%$16'\times16'$ (7 Mpc x 7 Mpc) (15.77'x 5.9x5.9 see p63)
	\textit{Left:} Optical image of $16'\times16'$ ($\sim$ 7 Mpc x 7 Mpc) field of view with galaxy density \tbf{contours}. \tbf{The largest, white boundary follows the footprint made by two of our rectangular DEIMOS slitmasks which were overlayed making an X shape with one mask per arm at a time for eight masks. The central area where the masks intersect creates the \textit{area of maximum spectroscopic coverage} because each of the ten exposures includes this region. This area is shaped like a diamond (not outlined in the picture) and encompasses both clusters.} Isodensity contours are created using an empirically weighted galaxy number density, discussed in Section \ref{subsec:isodensity}. Contours delineate two clusters, \tbf{the} North and South, two smaller overdensities, referred to as \tbf{the} North East, and \tbf{the remaining} population once these overdensities are subtracted, referred to as Superfield. Outer cluster contours have a density of approximately 200 galaxies Mpc$^{-2}$, with increments of 50 galaxies Mpc$^{-2}$. The yellow boundary surrounding \tbf{the} North and South represents the \textit{HST} ACS footprint. 
	\textit{Right:} The North and South cluster contours overlayed with 0.5 Mpc radius circles centered on galaxy density centroids for scale. Red and blue filled circles are high quality, spectroscopically confirmed z$_{clust}$ objects used in our analysis, with radii scaled by observed-frame R-band luminosity. \tbf{The red or blue color assigned to each dot is} based on a fit to the red sequence, discussed in Section \ref{subsec:rsfit}. BCG/MMCG candidates are highlighted in yellow and discussed in Section \ref{subsec:BCG}. The North BCG candidate closest to the galaxy density peak is confirmed at our cluster redshift by the SHELS survey \citep{SHELS}.
	}
	\label{fig:map}	
\end{figure*}

The projected separation between the clusters is 1.1 Mpc, where the merger axis is defined as a line which connects the galaxy density peaks. Galaxy isodensity contours are discussed in Section \ref{subsec:isodensity}, and seen overlayed on the DLS composite image in Fig. \ref{fig:map}. The gas component is projected onto a local mass minimum and is elongated transversely to the merger axis. Simulations predict that this geometry is characteristic of a first pass-through with a small impact parameter \citep{2006MNRAS.373..881P,1993A&A...272..137S}.
The elongated gas is suggestive of a transverse orientation of the merger to our line of sight. The angle is constrained to $48^{+19}_{-20}$ deg using a biweight statistic with bias corrected confidence limits \citep{Dawson2013}. This orientation in the plane of the sky, along with the large projected separation of the cluster boundaries, allows for a straightforward disentanglement of cluster members, which is vital to determining the star formation histories of each cluster.

A Time Since Collision (TSC) of $1.1^{+1.3}_{-0.4}$ Gyr is calculated with 68\% confidence using a Monte Carlo dynamics code developed for dissociative mergers \citep{Dawson2013}. 
The TSC is comparable to the lifetime of an A star, which places Musket Ball in the temporal sweet spot for the detection of such populations, and thus benefits from their constraints on star formation history. 

\section{Observations and Reduction} \label{sec:obs}

\subsection{Photometry} \label{subsec:photoz}

Photometric imaging in B, V, R, $\textit{z}\arcmin$ (12, 12, 18, and 12ks) was done using the Mosaic cameras at Kitt Peak National Observatory (KPNO) \textit{Mayall}  4-m telescope as part of the Deep Lens Survey (DLS) \citep{Wittman2002}. Magnitude limits for 5$\sigma$ point sources in B, V, R, and $\textit{z\arcmin}$ are 26.0, 26.3, 26.5 and 23.8 mag, where median seeing in the vicinity of our targets was 1\farcs46, 1\farcs30, 1\farcs28, and 1\farcs12, respectively. Fig. \ref{fig:map} (\textit{left}) shows a $16'\times16'$ area of the BVR composite image centered on the merger.

Ground-based J and K band \tbf{imaging (both 3.6ks exposures)} was obtained with NOAO Extremely Wide-Field Infrared Imager (NEWFIRM) on the KPNO \textit{Mayall}  4-m telescope with 3$\sigma$ point sources to depths of 22.7 and 22.0, where median seeing was 1\farcs10 and 2\farcs39, respectively. 

\tbf{After standard instrumental signature removal, residual photometric variations within and between DLS fields were modeled and removed \citet{Wittman2012} with a global linear least squares method inspired by the \textit{Ubercal} method \citep{2008ApJ...674.1217P}.}
%Internal photometric calibrations were done using  in addition to DLS photometric calibration, detailed in  and \citet{Schmidt2013}}. 
Colors and magnitudes were measured using a version of Colorpro \tbf{\citep[][based on SExtractor by \citep{1996A&AS..117..393B}]{Coe2006}} modified to improve accuracy given a variable seeing and PSF between filters. 
Photometric redshifts were estimated using BPZ \citet{Benitez00}, 
and are discussed in detail for DLS in \citet{Schmidt2013}.

\subsubsection{Stellar Mass Measurements} \label{subsubsec:stellarmass}

Stellar masses, hereafter referred to as log(${M_{*}}$), are computed from K-band magnitudes using methods described in \citet{Lemaux2012}. K-band magnitudes are first converted to absolute magnitudes using spectroscopic redshifts, or photometric redshifts for untargeted objects. Rest-frame K-band luminosities are calculated with applied evolutionary and \textit{k}-corrections
%\citep{2003MNRAS.344.1000B}
\tbf{\citep[using a $\tau =$ 0.6 Gyr, $z_{f} =$ 3 Single Stellar Population model,][]{2003MNRAS.344.1000B,2009AJ....137.4867L}, then converted to stellar masses based on interpolated values of the mass-to-light ratio at our redshift \citep{Drory2004}. The scatter found in K-band derived log(${M_{*}}$) is found by \cite{Lemaux2012} to be minimal (0.26 dex) and to introduce no bias in stellar mass when compared to those derived using SED fitting with nine bands.}

\subsection{HST Observations} \label{subsection:HST}

\textit{HST} observations were done using \textit{ACS} / WFC with F606W and F814W filters (GO-12377, PI: W. Dawson) in a $2\times1$ pointing mosaic that covers the North and South clusters, seen as a yellow box in Fig. \ref{fig:map}. The exposure times for F606W and F814W are 2520 s and 4947 s per pointing in 6 orbits, with magnitude limits 
%with \tbf{5 $\sigma$ magnitude limits for points sources}  
of 26.5 and 27, respectively. Photometry was performed on each passband. SExtractor \citep{1996A&AS..117..393B} \tbf{was run in} dual image mode on an inverse variance weighted average of the two passband images, with CLEAN=Y with CLEAN PARAM=1.2, adopting MAG$\_$AUTO for magnitude measurements. Detections required at least 5 contiguous pixels above 1.5$\times$ sky rms values, similar to \citet{Jee09}. 

\subsection{Spectroscopy}

Observations in March 2011 were taken with DEIMOS using seven slitmasks over two half nights with three 1.2 ks exposures per mask. A grating of 1200 l mm$^{-1}$ was used with 1\arcsec$\ $ wide slits, tilted to a $\lambda_{c}$=6700 \AA$\ $with a Full Width Half Maximum (FWHM) of 1.7 \AA$\ $(68 km s$^{-1}$) and typical restframe wavelength coverage of 5500-7900 \AA. Follow-up observations \tbf{on DEIMOS} in January 2013 used one mask with three 1.8 ks exposures and a grating tilt of 6200 \AA.
Seeing on March 2$\&$ 3, 2011 was 0.6\arcsec-0.71\arcsec (5.04-5.96 pixels) and 0.59\arcsec-0.82\arcsec (4.95-6.88 pixels), respectively, and on January 16, 2013 was 0.75\arcsec-0.84\arcsec (6.27-7.06 pixels). 

Spectra were reduced using a modified version of the DEEP2 spec2d package \citep{Davis03}. The details of the reduction process are described in \citep{Newman2013}. Our quality ranking system is based on that used in the DEEP2 redshift survey. Only \tbf{objects with spectra of the highest quality rankings, q $= 3, 4, -1$,} were included in our analysis. The \tbf{galaxies} in this sample have secure redshifts, meaning one secure, high signal feature and one marginal or better quality feature, at the correct wavelengths. 
The DLS field F2 was also covered by the SHELS survey \cite{SHELS}. \tbf{We do not include the spectra for these objects in our analysis, but we utilize the} high quality spectroscopic redshifts, as discussed in Section \ref{subsec:CMD}.

\subsubsection{Spectroscopic Selection} \label{subsubsection:selection}

Spectroscopic targets were selected from within a $16'\times16'$ area centered on the merger. The DEIMOS slitmask footprint is outlined by the white cross in Fig. \ref{fig:map}. Targets were limited to R$<$ 23.29 based on completeness of DLS.

%DLS completeness falls off for objects fainter than R$<$ 23.79, so 
%(R$_{AB}<$23.29). w -0.21 correction

Target selection was weighted by probability of cluster membership using photometric redshifts, discussed in Section \ref{subsec:photoz}. When selecting targets for the 2011 observations, objects with best fit photometric redshifts within z$_{phot}$ = 0.53$\pm\sigma_{z_{phot}}$  \tbf{\citep[$\sigma_{z_{phot}}$=0.1][]{Schmidt2013}} were given twice the priority of galaxies outside this range. \tbf{The} follow-up observations \tbf{on DEIMOS} in January 2013 targeted only objects within this z$_{phot}$ range. 

March 2011 observations yielded 655 high quality spectra (0 $<$ z $<$ 1.2).
Follow-up observations in January 2013 yielded 93 high quality spectra. 
%targeted potential cluster members within the cluster z$_{phot}$ range. 
%0.525 $\leq$ z$_{clust}$ $\leq$ 0.54. 267 (222+45)
In total, 267 high quality spectra have been confirmed to be in our cluster redshift range, z$_{clust}$, where z$_{clust}$ = [0.525, 0.540] (see Section \ref{subsec:redshiftdist}).
%0.53$^{+0.010}_{-0.005}$
These galaxies populate the dominant redshift peak in our observed distribution, seen in the top histogram of Fig. \ref{fig:zhist}, which includes both clusters. The redshift distribution of z$_{clust}$ galaxies is shown in detail in the bottom of Fig. \ref{fig:zhist} and discussed in Section \ref{subsec:redshiftdist}.
%Spectra with null magnitudes for either V, R, J or K were excluded.

Our area of maximum and equal spectroscopic coverage is defined by the intersection of all slitmasks, forming a diamond in the center of Fig \ref{fig:map}. Within this area, we targeted and attained high quality redshifts for 69\% (164 out of 236) and 65\% (153) of available photometric objects with redshift in z$_{phot}$, respectively. Of these high quality redshifts, 78\% (119) were confirmed to be in z$_{clust}$. 

%Completeness and sampling are further evaluated in Section \ref{subsubsec:completeness}.
While we have excellent spectral coverage, it is important to account for uncertainties due to our selection method. Evaluating incompleteness allows us to compare the scientific properties of populations in areas with differing coverage, and gauge how well our spectral measurements represent the underlying galaxy populations. We simulate the maximum possible variance in our results that can be attributed to sampling, outlined in Appendix \ref{subsubsection:completeness}. 
Our data set is one of the largest collections of high quality spectra for an intermediate redshift, dissociative cluster merger.

%\begin{deluxetable*}{llllllllllll}
\begin{deluxetable*}{p{1.2cm}p{1.2cm}p{1.2cm}p{0.7cm}p{0.7cm}p{0.7cm}p{2cm}p{0.55cm}p{0.8cm}p{1cm}p{1cm}p{1cm}p{1cm}}
%\begin{adjustbox}{max width=\textwidth}
%\tabletypesize{\tiny}
%\tabletypesize{\footnotesize}
\tablecaption{\footnotesize{RED SEQUENCE LINE FITTING PARAMETERS}}
%\resizebox{\columnwidth}{!}{
\tablehead{ \colhead{\footnotesize{Region}}
 & \colhead{\footnotesize{RA}}
 & \colhead{\footnotesize{DEC}}
 & \colhead{\footnotesize{y0}}
 & \colhead{\footnotesize{m\tablenotemark{a}}}
 & \colhead{\footnotesize{$\sigma$}}
 & \colhead{\footnotesize{z\tablenotemark{b}}}
 & \colhead{\footnotesize{N$_{spec}$}}
 & \colhead{\footnotesize{$\sigma_{v}$}\tablenotemark{c}}  
 & \colhead{\footnotesize{$\sigma_{v}$ M$_{200}$}\tablenotemark{c}}
 & \colhead{\footnotesize{WL M$_{200}$}\tablenotemark{c}} 
 & \colhead{\footnotesize{SD$_{gal}$}\tablenotemark{d}} \\
 \colhead{{}}
 & \colhead{{J2000}}
 & \colhead{{J2000}}
 & \colhead{{}}
 & \colhead{{}}
 & \colhead{{}}
 & \colhead{{}}
 & \colhead{{}}
 & \colhead{{}}
% & \colhead{{(km s$^{-1}$)}}
% & \colhead{{(km s$^{-1}$)}}
 & \colhead{\footnotesize{(10$^{14}$ M$_{\odot}$)}}
 & \colhead{\footnotesize{(10$^{14}$ M$_{\odot}$)}}
 & \colhead{\footnotesize{(arcmin$^{-2}$)}}
}
\startdata
North & 09\fd16\fm11\fs & 29\fd51\fm59\fs & 0.619 & 0.047 & 0.033 &  0.53074$^{+0.00064}_{-0.00068}$ & \ 38 & 740$^{+130}_{-190}$ & 3.7$\pm2.3$ & 1.7$^{+2.0}_{-0.72}$ & 5.4$\pm1.6$ \\
South & 09\fd16\fm16\fs & 29\fd49\fm15\fs & 1.095 & 0.008 & 0.024 &  0.53414$^{+0.00064}_{-0.00065}$ & \ 27 & 770$^{+110}_{-92}$ &  4.1$\pm1.6$ & 3.1$^{+1.2}_{-0.79}$ & 6.6$\pm1.5$ \\ 
Superfield & ... & ... & 0.708 & 0.039 & 0.010 & ... & 153 & ... & ... & ... & 1.9$\pm1.0$ \\
NorthEast & ... & ... & 0.533 & 0.056 & 0.041 & ... & \ 13 & ... & ... & ... & 2.8$\pm0.0$ \\
\enddata
\label{tab:1}
%}
%\begin{adjustbox}{max width=\columnwidth}
\tablenotetext{a}{\footnotesize{Fit parameters used are based on CSMD (Section \ref{subsec:rsfit}).}}
\tablenotetext{b}{\footnotesize{Super\tbf{f}ield and NorthEast have two redshift peaks. NorthEast has too few objects to centroid.}}
\tablenotetext{c}{\footnotesize{From \citep{Dawson2013}, values are consistent with velocity dispersion measured using additional DEIMOS data (Section \ref{subsection:spectral}).}}
\tablenotetext{d}{\footnotesize{\tbf{Surface density of galaxies including only areas that received equal coverage. 
%This restriction reduces  which reduces N$_{spec}$ to 48 and 7 for the Superfield and North East samples, respectively. 
Errors are based on completeness and discussed in Appendix \ref{subsubsection:completeness}.}}}
%\end{adjustbox}
\end{deluxetable*}

\section{Global Structure Properties} \label{sec:structure}

%Convention suggests that the assembly of clusters is more likely to occur in dark-matter and galaxy-rich regions of space. 
%The discovery of our system reveals two merging clusters that are not isolated. We investigate clustering and subclustering with this in mind. 

\subsection{Redshift Distribution} \label{subsec:redshiftdist}

\begin{figure}
%	  \begin{left}
	  \begin{center}
	  \includegraphics[scale=.43]{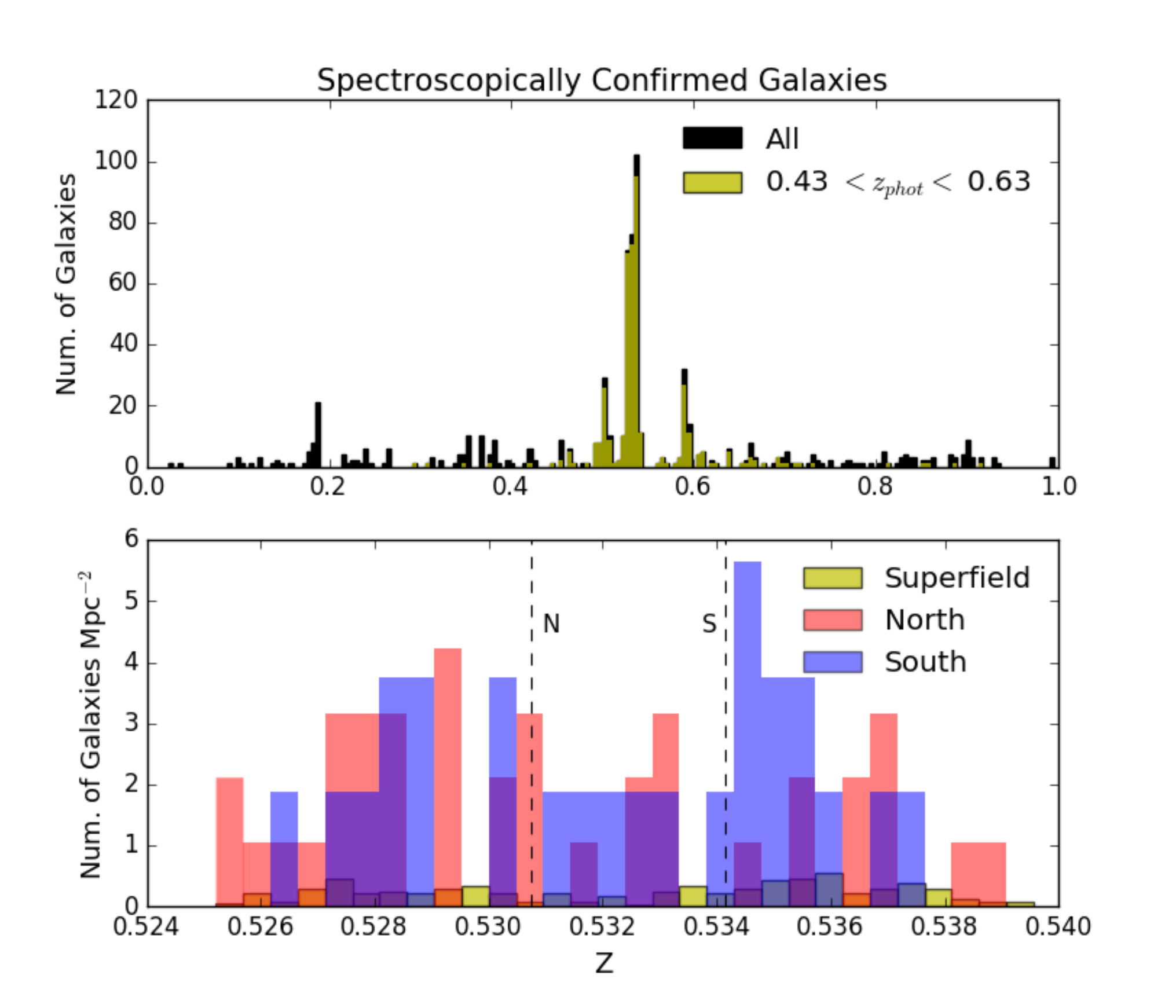}	
	  %\includegraphics[height=60mm]{zgeometry.png} if include with just repeat picture
	  %\includegraphics[scale=0.75]{zgeometry.png}
	  %\includegraphics[angle=45,width=52mm]{zgeometry.png}
	  %	  \resizebox{3.5in}{!}{\includegraphics{ndark15nfixedatone.png}}
	  \end{center}
	\caption{ 
	\textit{Top}: Redshift histogram for all spectroscopically confirmed Q=3, 4 objects in our DEIMOS footprint (Fig. \ref{fig:map}). Objects above z = 1.0 have been omitted from the plot. Small peaks at z$\sim$0.5 and z$\sim$0.65 show no coherent spatial structure when examined in spatial coordinates. 
%	\textit{Middle}: Redshift histogram of objects in the dominant redshift peak (seen above) color coded by cluster membership. Dotted lines are redshift centroids for North and South clusters, discussed in Section \ref{subsec:isodensity}.
	\textit{Bottom}: \tbf{Redshift histogram of objects in the dominant redshift peak, 0.525$\leq$z$\leq$0.54 (z$_{clust}$), color coded by cluster membership and normalized by area to represent surface density of each region. 
	Dotted lines are redshift centroids for the North and South clusters, defined in Section \ref{subsec:isodensity}.
	The top panel indicates a dominant redshift peak that, when examined by volume in the bottom panel, is shown to contain two high density regions of space in addition to a significantly less dense local population of galaxies. 
	Fig. \ref{fig:radecz} displays the expanse of objects in the surrounding structure which are distributed across the entire field of view.	
	Isodensity contours defining the North and South clusters, as well as the coeval population, Superfield, are seen in Fig. \ref{fig:map}. Surface densities are in Table \ref{tab:1}.}
	%Note Superfield here does not have mag or coverage cut so 171 objects
	%Surface densities and membership are in Table 1.
%	While these peaks appear dominant in the middle panel, the lower panel shows that Superfield is much less dense than  the clusters themselves.
%	one can also see objects occupying the two dominant redshift peaks in Superfield are randomly distributed on a scale as large as our field of view.Superfield peaks indicate a large population but distributed
	%(Perhaps in top hist overlay red onto hist of all photometric objects available and remove NS overlay. might look better side by side depending on final placement.)
	} 
	\label{fig:zhist}
\end{figure}

The distribution of all spectroscopic objects in our field is dominated by a peak at z$\approx$0.53, central to Fig. \ref{fig:zhist}. Two secondary peaks at z$\approx$0.5 and z$\approx$0.65 are populated by objects that are distributed across our entire
$\sim16'\times16'$ ($\sim$7 Mpc$\times$7 Mpc at z=0.53)
%$\sim$16' x 16' (6 Mpc x 6 Mpc) 
field of view, \tbf{shown by a 2D Kolmogorov-Smirnov (KS) test to have a projected spatial distribution more similar to the rest of the sample than the dominant peak.}
%and are not coherent intervening structures. \tbf{A KS test shows the spatial distribution of objects secondary redshift peaks are more similar to the rest of the field than the dominant peak.}
The dominant peak at z$\approx$0.53 spans 0.525 $\leq$ z $\leq$ 0.54, \tbf{a redshift range hereafter referred to as z$_{clust}$}. 
%, has a roughly bimodal redshift distribution. 
%%By eye, bimodal with two peaks at z=a,b
%The redshift distribution of spectroscopically confirmed objects reveals the existence of a complex structure. 
The redshift distribution \tbf{of z$_{clust}$ corresponds to a structure of galaxies encasing the clusters, in addition to the clusters themselves. The surrounding, coeval population is projected along redshift, Right Ascension (RA) and Declination in Fig. \ref{fig:radecz}, where green and magenta dots correspond to objects in the upper and lower half of the redshift range of the structure, respectively.
The clusters comprise only a subset of objects within z$_{clust}$, seen as smaller dots, which are clustered spatially in the right-hand panel. 
We identify and define the clusters using isodensity contours, described below.
%The overdensities warrant a more rigorous evaluation, as outlined in the two next Sections.
%Maybe move this to next paragraph 
%Defining cluster membership in a post-merger system is challenging due to the potential for cluster-cluster contamination and loss of spherical symmetry as a result of the collision. We mitigate this by using galaxy density contours to distinguish the merging clusters from their surroundings.
}
%spherically symmetric benchmark of virial radius 
%The presence of a local structure motivates us to systematically distinguish our clusters from such surroundings as well.

%Fig of 

\subsection{Isodensity Contours} \label{subsec:isodensity}

The sensitivity of our analysis to cluster membership requires careful delineation of overdense regions, \tbf{because we draw conclusions based on average spectral properties.
Defining cluster membership in a post-merger system is challenging, however, due to the potential for cluster-cluster contamination and loss of spherical symmetry as a result of the collision. 
In a dissociative merger like Musket Ball, for example, the gas peaks \citep[Fig. 3,][]{Dawson2012} are drastically offset from the Brightest Cluster Galaxy (BCG)/ Most Massive Cluster Galaxy (MMCG) and WL peaks (offsets listed in Table \ref{tab:4}).
We mitigate these perturbations by using galaxy density contours to distinguish the merging clusters from their surroundings.
Not only is galaxy density a fair tracer of the dark matter potential, but the presence of bright, massive, red galaxies is well-established as a tool used to identify clusters \cite{2000AJ....120.2148G}. 
%This method, described below, allows us to avoid dependening on the assumptions of spherical symmetry.
}

% especially because our merger is encased in a super structure. 
%Virial radius and luminosity-weighted number density are traditionally relied upon to define cluster boundaries, with a cluster center identified by X-ray peaks, WL centroids or the Brightest Cluster Galaxy (BCG)/ Most Massive Cluster Galaxy (MMCG). These quanitites generally coincide in relaxed systems. In a dissociative merger, however, the baryonic components are by nature disrupted. 
%Moreover, there are two BCG candidates in the Northern cluster, discussed further in Section \ref{subsec:BCG} and seen in yellow on the right of Fig. \ref{fig:map}. 
%We use galaxy number density to define the post-merging clusters and reconcile the utility of defining a center with the unique geometry. 
%When estimating number density, we employ photometric redshifts in a weighting scheme to increase our probability of analyzing galaxies within the pre-merger virial radius of each cluster. 

\tbf{We first introduce an empirical weighting scheme when estimating number density to increase our probability of analyzing galaxies within the pre-merger virial radius of each cluster. 
}
%that employs photometric redshifts 
%Isodensity contours to minimize contamination from galaxies that may not have been involved in the merger, which adapt well to the unrelaxed post-merger geometry.
Isodensity contours are created for the entire field of view using a two-dimensional histogram. Objects with spectroscopically confirmed redshifts in z$_{clust}$ are given a weight of one. Objects with photometric redshifts inside z$_{phot}$ = 0.53$\pm\sigma_{z_{phot}}$
%\cdot(1+<z_{clust}>)$ 
are weighted by 0.63, \tbf{a value calculated from the fraction of objects in our entire sampled area that were confirmed in z$_{clust}$ after being targeted based on having z$_{phot}$ = 0.53$\pm\sigma_{z_{phot}}$ (See Sections \ref{subsubsection:selection} and \ref{subsubsection:completeness}).}
%purity of results in our entire sampled area from using this range as a selection method (fraction of objects predicted to be inside z$_{phot}$ that were confirmed in z$_{clust}$).}
%based on purity of results in our entire sampled area from using this range as a selection method.
All remaining objects, including those spectroscopically confirmed to be outside the cluster range, are given a weight of zero. The addition of data from 2013 observations does not alter the overall purity. 
The resulting weighted histogram is then given a 2D spatial Gaussian smoothing of 20 pixels (32'') to eliminate practical discontinuities in the resulting contours. 
The resulting peaks and contours are superimposed on the DLS image in the left panel of Fig. \ref{fig:map}.
%, and overlayed with spectroscopically confirmed members in the right panel. 
The lowest cluster contour level in the left panel corresponds to a density of roughly 200 galaxies Mpc$^{-2}$, with increments of 50 galaxies Mpc$^{-2}$ (z$_{phot}$ = 0.53$\pm\sigma_{z_{phot}}$).
%We find that the addition of luminosity weighting yeilds roughly the same results.

\tbf{A Northern and Southern peak emerge from the isodensity contours, encased by a less dense large scale structure.} The peaks and contours are overlayed with 
%DLS image in the left panel of Fig. \ref{fig:map}, and overlayed with 
spectroscopically confirmed members in the left panel of Fig. \ref{fig:map}. 
%A number density trough between the largest contours, delineating the North and the South cluster. %We define the cluster boundaries as 
\tbf{The largest contours of the Northern and Southern peaks contain 38 and 27 spectroscopically confirmed galaxies (after magnitude and stellar mass cuts discussed in Section \ref{sec:galpop}), respectively, and divide the two clusters into separate components to prevent overlap.} Cluster membership is invariant to minor shifts in the contour separation. 
Both clusters, \tbf{hereafter referred to as the North and South}, are roughly three times the surface density of the surrounding structure (SD$_{gal}$) in Table \ref{tab:1}), hereafter referred to as Superfield and defined below. 
%The area normalized redshift distribution of all spectroscopic objects in z$_{clust}$ is shown in the bottom panel of 
%Both clusters have a dominant galaxy density peak, despite highly irregularly shaped boundaries. 
%MOVE TO RESULTS?%%%%%%%%%%%
%\textit{The contour shapes indicate that North is elongated around the dominant peak and has two minor density peaks at opposite ends of the major axis. The South has one minor density peak opposing the dominant peak, skewed in the direction of North, along the merger axis which connects the galaxy centroids.}
%%%%%%%%%%%%%%%%%%%%%%%%%%%

Two group-sized overdensities in the field meet the number density threshold, labeled as North East a and North East b in the left panel of Fig. \ref{fig:map}. We remove the objects within these contours, referred to collectively as \tbf{the} North East, from our analysis of the surrounding structure. The remaining sample is referred to as \tbf{the} Superfield, \tbf{and} provides a coeval, comparison sample of galaxies for our merger. \tbf{The Superfield boundary is shown in} the left panel of Fig. \ref{fig:map} and coincides with our DEIMOS slitmask footprint.
The Superfield is defined by all high quality galaxies in z$_{clust}$, excluding members of the North, South and NorthEast. Its boundary spans $\sim16'\times16'$ ($\sim$7 Mpc$\times$7 Mpc at z=0.53) and North, South, and NorthEast are roughly 5$\%$, 3$\%$ and 2$\%$ of its size, seen in Fig. \ref{fig:map}.
%The Superfield provides a useful comparison sample which is a non-cluster, non-group environment.
%The density of spectroscopically confirmed z$_{clust}$ objects in North and South is approximately three times greater than that of the surrounding area (Fig. \ref{fig:zhist}). A breakdown of the populations of star forming galaxies between regions is discussed in Section \ref{sec:results}. 
%%%%%%%MAYBE MOVE OR DELETE A LOT FOR THE NEXT PARAGRAPH
%The radial distribution of galaxies types in a cluster galaxies can reveal average trends in galaxy evolution when studied across redshift.
% dA common benchmark for examining radial distribution is virial radius and $r_{200}$. As with contours, the irregular shapes of our clusters are not conducive to the application of this quantity. Moreover, 
%\textit{Musket Ball galaxies have had adequate time to \tbf{traverse the length of their} host cluster in the 1.1 Gyr since core-passage.}
Fig. \ref{fig:map} (\textit{right}) shows the spatial distribution of spectroscopically confirmed objects within our cluster redshift range, scaled by observed frame R band luminosity. \tbf{The color of each dot is assigned based on membership of either the red sequence or the blue cloud, described in Section \ref{subsec:rsfit}.} 
%Kind of repeats what is written below, just describe map and confirm or affirm our decision here. Move everything else to results

%%%%%%%%%%%%%MOVED TO RESULTS%%%%%%%%%%%%%
%Maybe move because haven't defined galaxy density peaks yet
%\textit{
%We find most of the luminous red galaxies are located in the dense interior of both the North and South clusters, near the primary galaxy density peaks, which is consistent with hierarchical structure formation and cluster evolution \citep{Kauffmann1995}.}

%\textit{Cluster cores at intermediate redshifts are dominated by bright, red galaxies \citep{Kauffman2004}. 
% Moreover, as discussed in Sections \ref{fig:morph}, these objects are both early-type and massive. In the case of the North, there is a second density peak which also has a concentration of these luminous red galaxies. 
%This bimodality is suggestive of a cluster that is still forming, discussed further in Section \ref{subsec:BCG} and \ref{sec:results}. While the multicomponent structure is also suggestive of influence by the merger, its origin is difficult to determine without knowledge of the pre-merger state of the North. We investigate this further by proceeding with the spectral and dynamical analysis described in Section \ref{subsection:spectral}.}
%%%%%%%%%%%
%While the distribution of galaxies after a merger may not be spherically symmetric, galaxies are a fair tracer of the dark matter potential and provide insight into the pre-merger state of each cluster. 

Along with galaxy isodensity contours, galaxy centroids provide a method of defining a center which we use to gauge offsets and place apertures for calculation of other dynamical quantities (Section \ref{subsec:dynamics}). 
The North and South cluster galaxy centroids are estimated using an iterative procedure. An aperture is placed over a single cluster, then a centroid is calculated as it is incrementally reduced in size. The uncertainty is estimated by performing the centroid calculation on bootstrapped samples, and taking the variance of the resulting values. Uncertainties are 5$\farcs$3 and 3$\farcs$3 for North and South, respectively. 
\tbf{The WL centroids are close in proximity to the galaxy centroid locations in both clusters (Table \ref{tab:4}). A more detailed explanation of \tbf{the calculation of galaxy and WL centroids and their confidence intervals} can be found in \citet{Dawsonthesis}.}
% Relative offsets of all clusto-centric features} found in . 

\begin{figure*}
	  \begin{center}
	  \includegraphics[height=50mm]{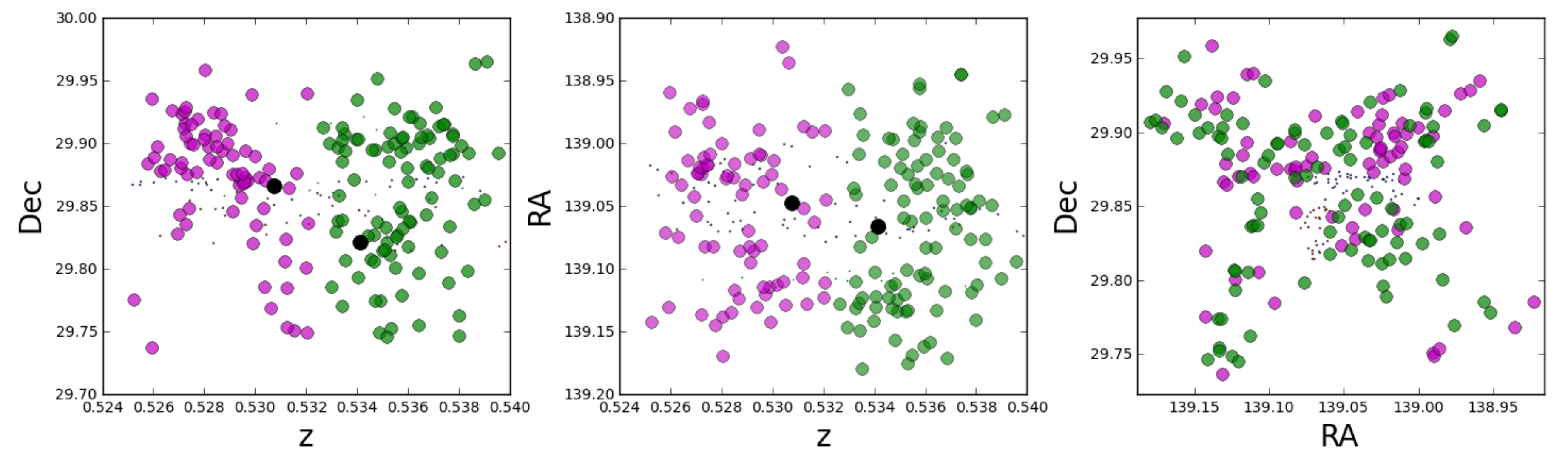}
	  \end{center}
	\caption{
	Projection of Superfield objects corresponding the yellow bars in the lower histogram of Fig. \ref{fig:zhist} along right ascension \tbf{(RA, \textit{left}), declination (DEC, \textit{center}) and in spatial dimensions (\textit{right}). Objects are divided by color into lower redshift (z $\geq$ 0.5325) as magenta dots and higher redshift (z $<$ 0.5325) as green dots.} Small dots are members of \tbf{the} North, South and North East overdensities, for reference, defined in Section \ref{subsec:isodensity}. Large black dots are redshift-galaxy centroids for \tbf{the} North and South \tbf{(lower and higher redshifts, respectively)}, described in Section \ref{subsec:dynamics}.
	Spatial scale corresponds to the field of view in Fig. \ref{fig:map}.
	% 3D plot colors are faded based on the redshift scale on the transverse axis.
	While the Superfield objects \tbf{occupy the same redshift peak, they are not spatially clustered on the scale of the North and South when projected along RA and DEC,} as discussed in Section \ref{sec:structure}. The objects corresponding to both peaks in redshift are distributed across our entire field of view (\textit{right}).
	%, which are easily distinguishable in the top two plots, are randomly distributed when projected in right ascention and declination, seen in the 3D scatter plot. 
	The galaxies within the dominant redshift peak populate a large and complex structure, motivating the use of isodensity contours to extricate the two merging clusters.
	}
	\label{fig:radecz}	
\end{figure*}

%We only 
\tbf{The systemic redshift of each cluster centroid is given as input for the TSC simulation \citep{Dawson2012,Dawson2013}}, and was calculated using the biweight statistic from \citet{Beers1990} to downweight tails of the distribution, applied to 100,000 bootstrap samples of each cluster's spectroscopic redshifts. \tbf{Results are found in Table \ref{tab:4} and seen as dotted lines in the lower redshift histogram of Fig. \ref{fig:zhist}.}
\tbf{To summarize, in this section we identified a redshift peak at 0.525 $\leq$ z$_{spec}$ $\leq$ 0.54 (Fig. \ref{fig:zhist}) containing two cluster-sized galaxy overdensities, defined as the North and South (Fig. \ref{fig:map}). The clusters are encased in a coeval population of galaxies defined as Superfield (Fig. \ref{fig:radecz}). Convention suggests that the assembly of clusters is more likely to occur in dark-matter and galaxy-rich regions of space. Coupled with the signatures of a dissociative merger, the distribution is consistent with two clusters embedded in a local structure pulled together by gravity more quickly than the surrounding structure.}

\subsection{Merger Dynamics and Substructure} \label{subsec:dynamics}

Our large sample of spectroscopic redshifts allows us to estimate \tbf{dynamical properties of the system. 
We use these properties to constrain a possible merger scenario and identify substructure that may indicate disruption of the member galaxies due to the merger.}
%, adding a dimension to our interpretation of how the merger has influenced member galaxies.

\tbf{We calculate LOS velocity dispersion to confirm that our results are consistent with those of \cite{Dawson2013}, indicating an equal mass post-merging system with a TSC of 1 Gyr.
We use spectroscopic members defined with isodensity contours (Section \ref{subsec:isodensity}), rather than circular apertures, to avoid assumptions of spherical symmetry inappropriate for a dynamically complex system.
LOS velocity dispersion is then calculated using a method described in \cite{2013ApJ...763..124R}, where the biweight statistic \citep{Beers1990} is followed by an iterative 3$\sigma$ clipping and errors are calculated with jackknife confidence intervals. Note that the clipping does not remove any galaxies from the calculation, discussed in Section \ref{subsec:redshiftdist}.
The resulting values of $\sigma_{v}$ = 801$\pm$90$\ km\ s^{-1}$ for \tbf{the} North and $\sigma_{v}$ = 755$\pm$112$\ km\ s^{-1}$ for \tbf{the} South
\tbf{are consistent within the errors of those listed in Table \ref{tab:1}}, which are the limits of the 68\% confidence \tbf{interval} from the cluster boundaries used in the dynamical simulation of \citet{Dawson2013}. 
We therefore adopt their TSC estimate which uses LOS \tbf{(Line-of-Sight)} velocity from spectroscopic redshifts, halo mass from WL mass maps and projected cluster separation from galaxy density peaks. Dynamical mass is calculated from velocity dispersion using the $M_{200}$ scaling relation from \citet{2008ApJ...672..122E}\tbf{, found in Table \ref{tab:1}. We do not use the velocity dispersion for any analysis in the paper beyond this consistency check.}
}
% and bias-corrected 68\% confidence limits . 
% $\sigma_{v}$ = 896$\pm$55$\ km\ s^{-1}$ for \tbf{the} North and $\sigma_{v}$ = 808$\pm$100$\ km\ s^{-1}$ for \tbf{the} South using 39 and 27 galaxies within a 0.5 Mpc aperture.
%With the addition of data from the follow-up observing run in March 2013 and cluster boundaries defined by isodensity contours described in Section \ref{subsec:isodensity}, our LOS velocity dispersion results are
%$\sigma_{v}$ = 801$\pm$90$\ km\ s^{-1}$ for \tbf{the} North and $\sigma_{v}$ = 755$\pm$112$\ km\ s^{-1}$ for \tbf{the} South.

%Details on confidence intervals and centroid errors in the two-body simulation are found in \citet{Dawson2013}.

\subsubsection{Dressler-Schectman test} \label{subsubsec:dstest}

The structurally and dynamically complex nature of our system increases the likelihood of subclustering.
% beyond which our number density analysis can 
We utilize our knowledge of accurate positions and redshifts to investigate subclustering using a Dressler-Schectman (DS) test \citep{DresslerShectman1988}. The DS-test indicates kinematically hot and cold areas by defining a sum of squares deviation for each galaxy, DS-$\delta$.
An array of DS-$\delta$ values is calculated using the spectroscopic redshifts of cluster members where,

\begin{equation}
\delta^{2}=\dfrac{N_{local}}{\sigma^{2}} \left[ (\overline{v}_{local} - \overline{v})^{2} + (\sigma_{local} - \sigma)^{2}  \right] ,
\end{equation}

\noindent
and $N_{local}$=$\sqrt{N_{total}}$ where $N_{total}$ is the total number of cluster galaxies. 
%to include the galaxy and its nearest neighbors. 
The set of $N_{local}$ nearest neighbors is used to calculate the local average line-of-sight velocity, $\overline{v}_{local}$, and the local velocity dispersion $\sigma_{local}$. Larger values of $\delta$ indicate a stronger chance of substructure.
%Galaxies with larger delta values are highly correlated with their neighbors and different from the parameters thus identifying local structure.

To evaluate the significance of the detection of substructure within each cluster, we calculate $\Delta$, the cumulative root-mean-square of each $\delta^{2}$. 
\tbf{To quantify statistical uncertainties, we performed a bootstrap in redshift, with 10,000 iterations,} while keeping the projected positions of the galaxies constant \citep{1992nrca.book.....P}.
%We conduct 10,000 iterations, each time randomly shuffling the redshift distribution, while keeping the galaxies' positions constant. 
% using the nearest neighbor method ($\sqrt(N)$). DS-$\delta$ is, 
%eq 1 from Dawson 2014 5.1
%Should I point out weakness due to DS test being best for redshifts with large range, thus better for front back correlations rather than mergers that are in the plae of the sky?
 
 %Maybe delete a lot of this crap talking about overal specz range
A DS plot of our entire sample of spectroscopically confirmed objects indicates there is significant substructure among objects within the dominant redshift peak.
% which coincides spatially with the North and South clusters (Fig. \ref{fig:dstest}). 
%A plot of our spectroscopically confirmed objects indicates two areas of high correlation that coincide spatially with the isodensity contours for North and South clusters (Fig. \ref{fig:dstest}). 
Objects from the subdominant redshift peaks at z=0.5 and z=0.65 show no substructure.
%The bimodal peaks within 

%DS tests of spectroscopic objects in z$_{clust}$ reveal significantly different levels of substructure between North and South clusters. 
%We perform the test using only objects that go into our spectral analysis, within the region of maximum spectroscopic coverage and on members North and South clusters. 
%Overall r2...

$\Delta$ values for the North and South are 70.9 and 27.9, indicating the presence of significant and insignificant substructure, respectively.
For the North cluster, only four of 10,000 separate realizations of the redshift distribution showed greater substructure and $\overline{\Delta}$ = 35.3.
For the South cluster, 24\% of realizations showed greater substructure, where $\overline{\Delta}$ = 24.6.
%Nick calls this a P value and mentions percent confidence level, maybe I should do this
%add something about the deviation from the mean los velocity and velocity dispersion

%There is little substructure in the South, and an area of significant substructure in the North.

Galaxies with the highest $\delta$, \tbf{indicative of substructure,} %(DESCRIPTION IN ABOVE PARAGRAPH)
are concentrated in one portion of the North cluster, indicating a clear group of objects that have a mean $v_{LOS}$ and/or $\sigma_{vdisp}$ greater than the system and cluster average. This area of highest correlation is offset from both the number density peak and BCG location in the North cluster. Offsets are in Table \ref{tab:4} and discussed in Section \ref{subsec:BCG}.
The magnitudes of these offsets are consistent with our characterization of the North cluster as unrelaxed. 

Our statistical evaluation of redshift, spatial and dynamical correlations reveals a post-merger system of two distinct clusters, North and South, encased in a local structure, Superfield. We divide our analysis according to these three regions for the remainder of the paper.

\begin{figure}
%	  \begin{left}
	  \begin{center}
	  \includegraphics[height=63mm]{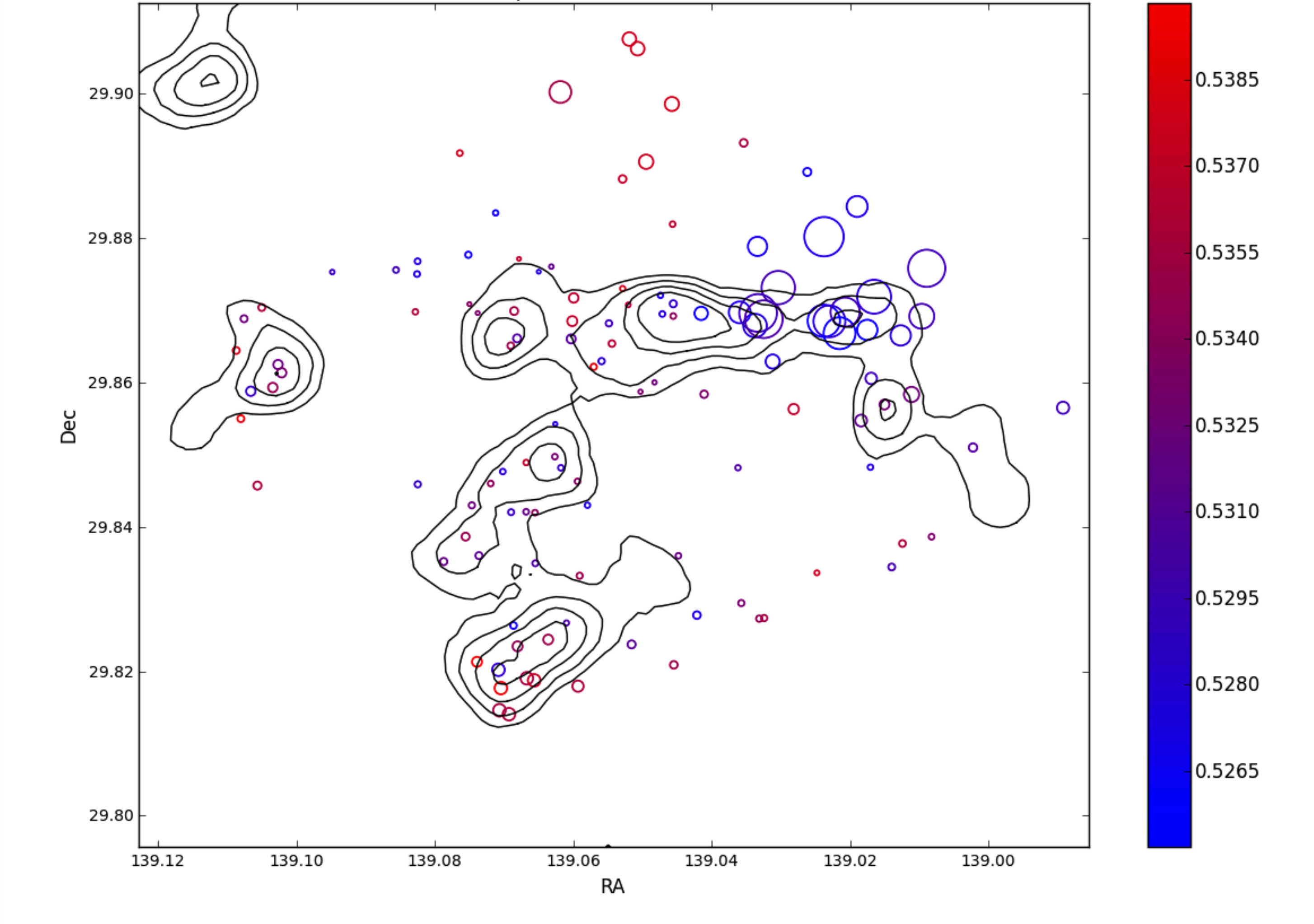}	  
	  \end{center}
	\caption{
%	\textit{Left:} DS-test of all high quality, spectroscopically confirmed galaxies in our field. 
	%\textit{Right:} 
	DS-test of spectroscopically confirmed cluster redshift members within our area of equal and maximum spectral coverage.
	Circle sizes are scaled exponentially by $\delta$ for each galaxy. Color gradient reflects redshift distribution, seen in bottom two histograms of Fig. \ref{fig:zhist}.
	%While both clusters show strong correlation relative to surrounding Superfield objects, 
	The North has significantly more substructure than the South, confirmed by our Monte Carlo calculation (Section \ref{subsec:dynamics})
	}
	\label{fig:dstest}	
\end{figure}

\begin{figure*}
%	  \begin{left}
	  \begin{center}
	  \includegraphics[height=50mm]{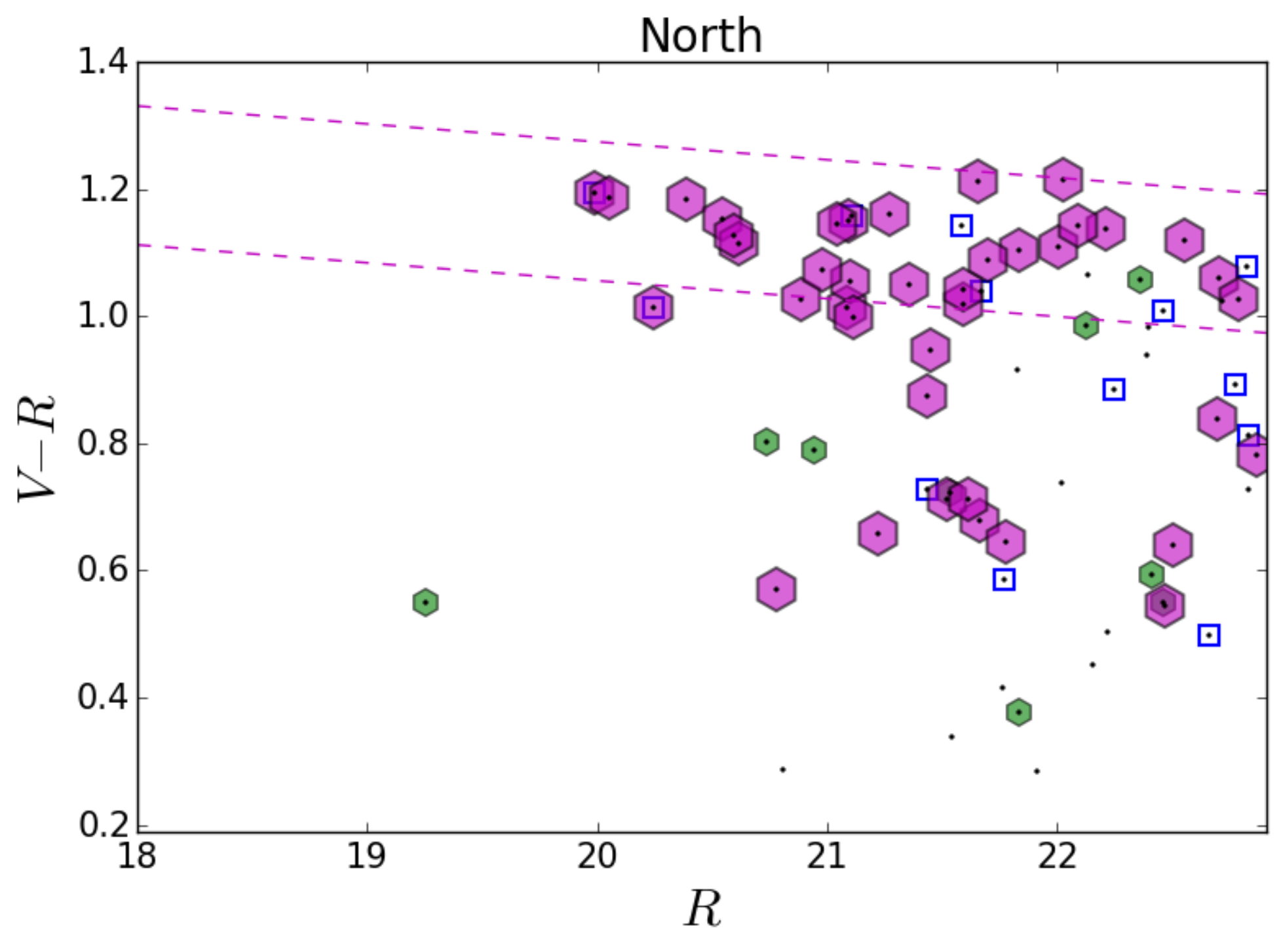}
	  \includegraphics[height=50mm]{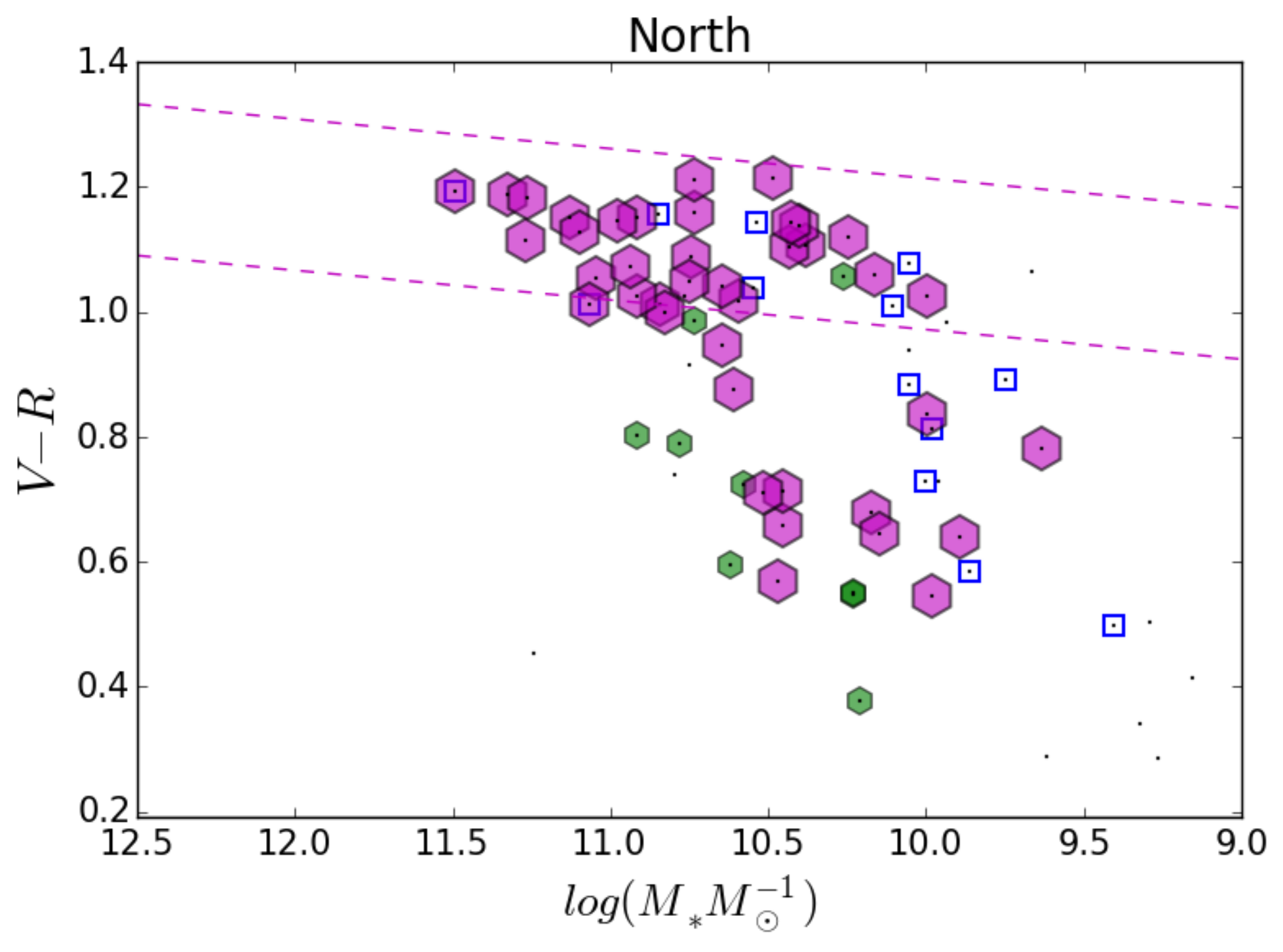}
	  \includegraphics[height=50mm]{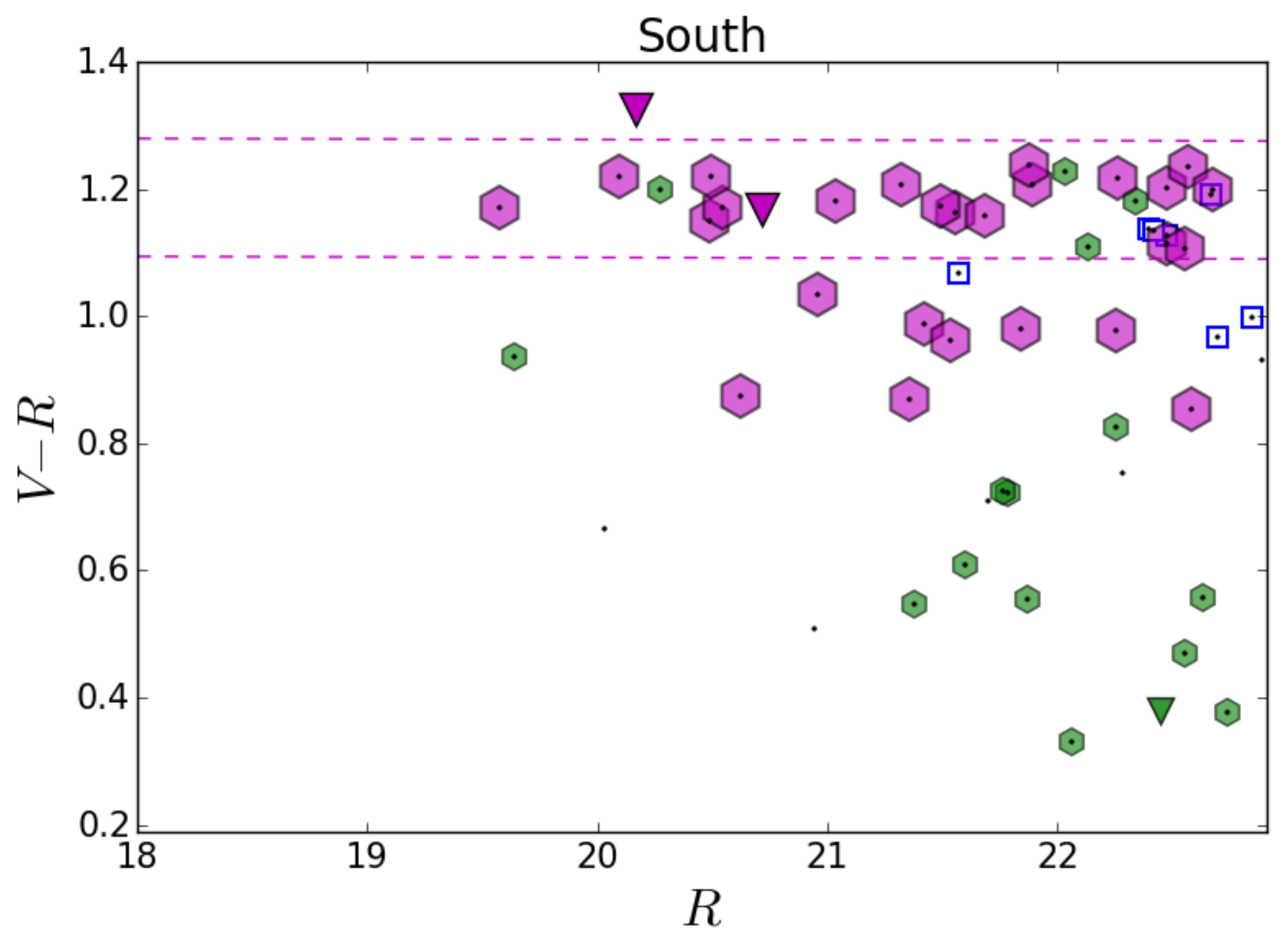}
	  \includegraphics[height=50mm]{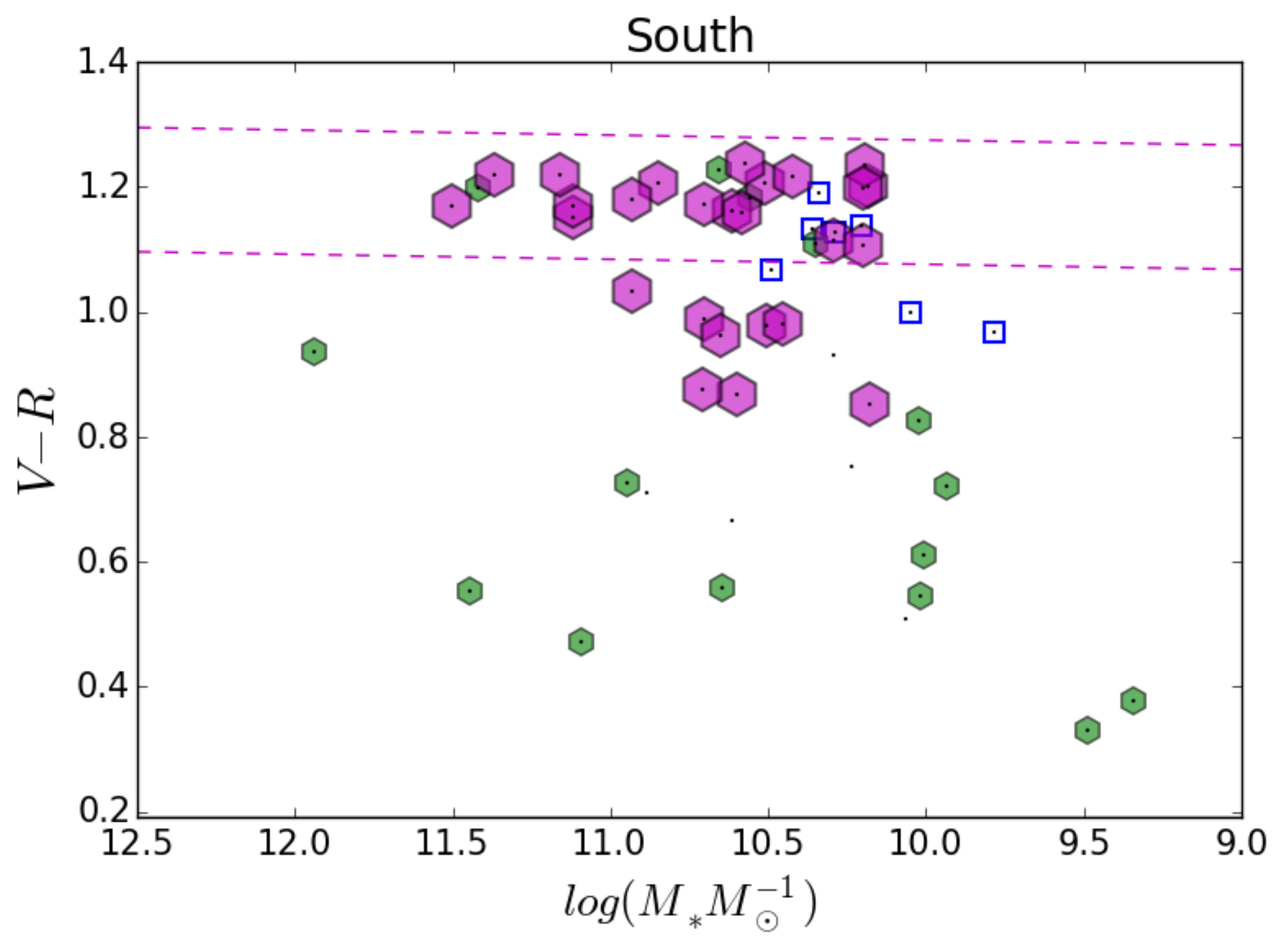}
	  \includegraphics[height=50mm]{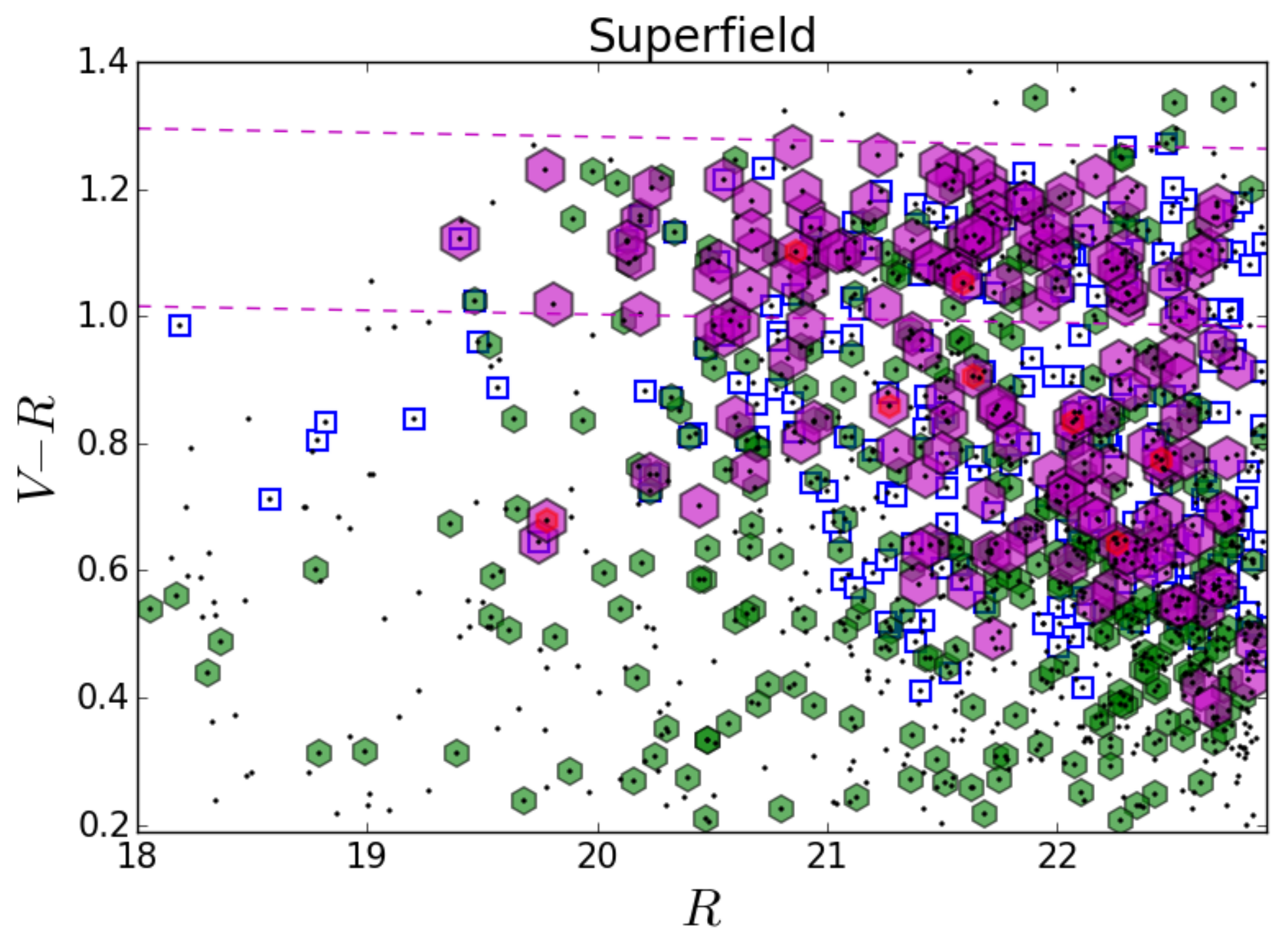}
	  \includegraphics[height=50mm]{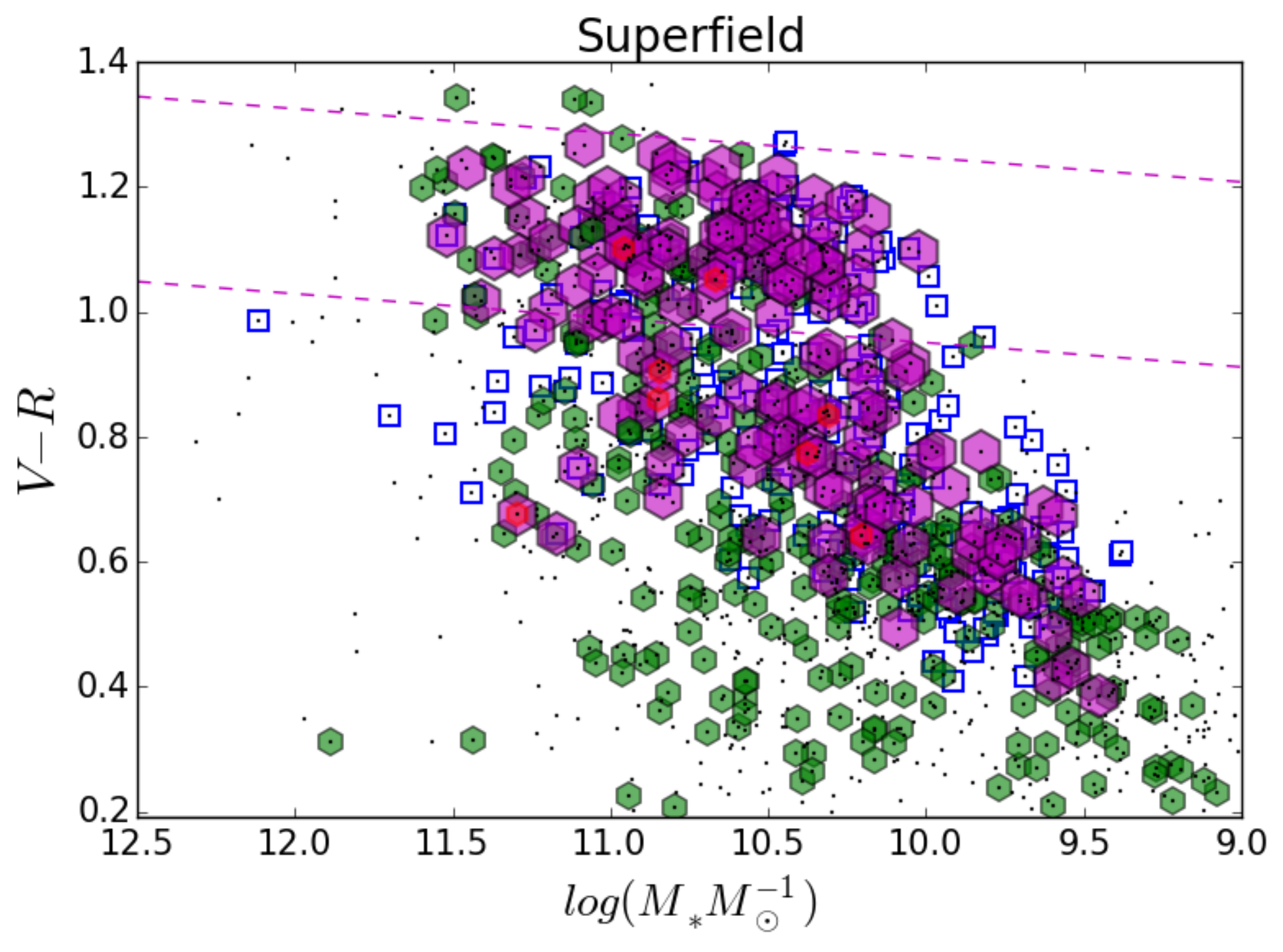}

	  \end{center}
	\caption{
		\textit{Right Four Panels:} Color-stellar mass diagrams indicating completeness and parameter space of photometric objects in our sample (black dots). Magenta and green hexagons are objects spectroscopically confirmed inside and outside our cluster redshift range, respectively. Blue squares are untargeted objects with photometric redshifts inside the photometric cluster redshift range, with a 78\% chance of being at our cluster redshift, \tbf{defined as \textit{missing potential members} (\textit{m})} in Appendix \ref{subsubsection:completeness}. Red hexagons are confirmed cluster redshift objects with photometric redshifts outside the photometric cluster redshift range. Dotted lines reflect red sequence fit boundaries using parameters in Table \ref{tab:1}, discussed in Section \ref{subsec:rsfit}. 
		\textit{Left Four Panels:} Color-magnitude diagrams corresponding to objects in the CSMD. \tbf{Individual components of the South BCG, resolved by \textit{HST} (Fig. \ref{fig:sbcg}), are represented by magenta and green triangles. The two magenta triangles correspond to the two nuclei and the green triangle represents the projected companion, discussed in Section \ref{subsec:BCG}.} 
		%Confirmed cluster redshift members span the full range of available photometric objects. 
		High completeness in the North and South clusters gives us confidence that measured differences are physical, rather than from variations due to sampling. 
	}
	\label{fig:completeness}
\end{figure*}

\section{Galaxy Population Properties} \label{sec:galpop}

In this Section we describe the procedures and motivations for isolating and comparing different galaxy populations within our structure. After first defining the North, South and Superfield in Section \ref{sec:structure}, we further divide galaxies into subpopulations to probe \tbf{star formation} histories that may differ based on color, \tbf{stellar} mass and brightness. We first divide by color by performing a red sequence fit to Color-Magnitude (CMD) and Color-Stellar Mass (CSMD) Diagrams (Section \ref{subsec:rsfit}). We then define red and blue subpopulations as high (log(${M_{*}}$) $\geq$ 10.5) or low mass and bright (R$\leq$ 21.7) or faint. Hereafter, subpopulations will be referred to using acronyms corresponding to these definitions, e. g., blue massive (BM), bright blue massive (BBM), and blue bright (BB). Finally, Equivalent widths (EWs) and \Dn are measured from composite spectra of each subpopulation, in addition to the total populations for each region (Section \ref{subsection:spectral}). A summary of these measurements, along with the subpopulation definitions, are found in Table \ref{tab:2}.

\subsection{Color-Magnitude and Color-Stellar Mass Properties} \label{subsec:CMD}

Color-Magnitude (CMD) and Color-Stellar Mass (CSMD) Diagrams allow us to identify trends and potential outliers in an entire galaxy population. They also allow us to observe trends in completeness and spectroscopic sampling, discussed in Appendix \ref{subsubsection:completeness}. Accounting for completeness and sampling effects allows us to compare galaxy populations internally, between the two cluster environments and the local Superfield
%contaminants such as dust and variability between observations. 
%view the spectral properties of a population based on empirical trends

%possibly move this to CMD Section
For CMDs we use R band magnitude and V-R colors. The central wavelength in our DEIMOS observations is optimized for R band coverage. 
%consistently has the best seeing ($<$=0.9'' vs 0.9-1.25'' in BVz), 
DLS V and R filters straddle the 4000 \AA$\ $break at our redshift and have good image quality and depth. 
R band is the deepest, and is used as the detection image for the DLS catalog \citep{Wittman2002}. 
We consider using B as the bluer filter in order to provide a longer wavelength baseline and avoid overlap with the R filter, but the depth and image quality are better in V. We therefore use V-R colors to separate the population.
%The wings of these two filter curves overlap, making B band wavelength range better for separating populations by type. However, B band photometric errors are four times that of R, whereas V errors are only twice as large. 
For CSMDs we use stellar mass determined from K-band, described in Section \ref{subsec:photoz} . 
%This method yields a reasonable estimation, despite larger errors. (check this)

We restrict our sample to eliminate poorly represented objects, particularly at the faint, blue and low mass ends. We limit magnitude, color, and stellar mass to R$\leq$ 22.9, 0.2$\leq$ V--R $\leq$ 1.4 and 9 $\leq$ log(${M_{*}}$) $\leq$ 12, respectively. By removing areas of parameter space with the worst coverage, we can compare populations between regions with consistent representation across our\tbf{magnitude} range. \tbf{For example, a comparison of blue spectroscopic members between the two clusters would be weak for R$\geq$ 22.9 because our sampling of spectroscopic objects begins to fall off drastically for galaxies fainter than this magnitude limit. The resulting uncertainties given these cuts for the blue populations are found in Table \ref{tab:2}.}

The CMD/CSMDs in Fig. \ref{fig:completeness} show the \tbf{photometric range of the spectral sample} for the North and South clusters, NorthEast and Superfield. 
All DLS photometric objects are plotted by region (black dots).
%The top four panels are CSMDs for each region, and display targets from our two Keck runs only. 
Magenta and green hexagons are objects spectroscopically confirmed inside and outside our spectroscopic cluster redshift range \tbf{(z$_{clust}$, 0.525 $\leq$ $z_{spec}$ $\leq$ 0.54, Section \ref{subsection:spectral})}, respectively. Blue squares are untargeted objects with photometric redshifts inside the photometric cluster redshift range \tbf{(z$_{phot}$ $\leq$ 0.53$\pm$0.1)}. Blue squares represent potential missing cluster members (\textit{m}, Appendix \ref{subsubsection:completeness}), which we estimate to have a \tbf{78\%} chance of being at our cluster redshift. Red hexagons are confirmed cluster redshift objects with photometric redshifts outside the photometric redshift range, \tbf{corresponding to objects that failed to be identified by our spectroscopic selection method (FTII, defined and discussed in Appendix \ref{subsubsection:completeness}). The symbols in this figure highlight the qualities of galaxies that were targeted and acquired or missed under our selection method, informing the quantitative discussion in Appendix \ref{subsubsection:completeness}.} 
%Dotted lines reflect red sequence fit boundaries, parameters in Table \ref{tab:1}, discussed in Section \ref{subsec:rsfit}. 
The plots include nine objects confirmed inside and 79 confirmed outside of the spectroscopic cluster redshift range by SHELS \citep{SHELS}, which were either untargeted or yielded low quality spectra in our DEIMOS runs. Also included are triangles indicating individually resolved components of the South BCG, identified in \textit{HST} bands as an E/E merger with a projected companion (see Section \ref{subsec:BCG}).
%Dotted lines reflect red sequence fit boundaries, where fit parameters are found in Table \ref{tab:1}. %Distinguishing between colors as red or blue
%Dividing colors allows us to tie objects in the CMD/CSMDs to physical properties, discussed in Section \ref{subsec:rsfit}.

\subsection{Red Sequence Fitting} \label{subsec:rsfit}

%why do we separate RS and BC
We separated galaxies into two populations, red sequence and blue cloud, based on the fundamental bimodal distribution observed in color-magnitude space for large galaxy populations \citep[e.g.][]{Goto2003,Kauffman2004}.
Blue cloud galaxies on average have greater \tbf{emission at bluer wavelengths, a phenomenon which is} attributable to star-forming processes. \tbf{Thus,} isolating this population effectively targets star formation activity. 
Red sequence galaxies generally have little or no star formation\tbf{, with the exception of dusty star forming galaxies where emission is absorbed at preferentially bluer optical wavelengths and re-emitted in the IR, discussed further at the end of this section}. 
The red sequence allows us to distinguish the quiescent, or passive, population, giving us insight into possible quenched galaxies. \tbf{We expect clusters to} have a well-defined red sequence by z$\sim$1 \citep{Stott09}.
The color delineation is also proxy for galaxy type, as red sequence galaxies are typically early-type, and blue cloud galaxies are typically late-type. We gain another dimension of insight we examine morphology (Section \ref{subsection:morph})\tbf{, allowing us to identify possible transition objects such as red late-type galaxies and blue early-type galaxies.} 
%As discussed in Section \ref{sec:results}, we find no late-type galaxies on the red sequence.}
% but we do see a population of early-type galaxies in the blue cloud.}
%aforementioned dusty star forming galaxies, which would introduce the likelihood that star formation indicators are a lower limit, or late-type galaxies in the blue cloud, also discussed in \ref{sec:results}}
%Stott?
%This would be a lead in for pointing out the rs is different between north and south

We performed the fitting process separately on all regions, although we compare only the two clusters in detail for this paper.
%it holds more physical meaning for comparison between the two cluster environments.
%We use R band magnitude because it has the most reliable photometric measurements of the four DLS bands for R\leq 24 at our redshift. V band allows us to separate out (not the best for straddling other bands we have, but most reliable measurements in DLS)
We fit the red sequence of both color-magnitude and color-stellar mass distributions. The scatter in the red sequence is typically less in the color-stellar mass relation. While the difference was marginal, we proceed to define our red sequences for each region using the parameters from the color-stellar mass fit (Table \ref{tab:1}).

Galaxies were split based on a linear fit to the red sequence \tbf{using a method similar to that described in \cite{Lemaux2012}. All magnitude, stellar mass and color errors were input into the fitting process, which was done using spectroscopic objects down to a magnitude limit at which photometric errors are reasonably small ($\sigma\le0.05$).}
%(R$\le22.7$ for $\sigma\le0.05$).}
First, \tbf{a least-square fit was done to objects within a color range defined by eye. The color window was consistent with that produced by \cite{2003MNRAS.344.1000B}, \cite{Maraston2005MNRAS} and \cite{Bruzal2007} models \citep[2-5 Gyr, $\tau\sim$100 Myr exponentially decaying burst, solar metallicity,][initial mass fuction (IMF)]{2003ApJ...586L.133C}.
The slope of the line with respect to color was then subtracted to create a color corrected distribution, allowing us to fit to a single Gaussian using iterative 3$\sigma$ clipping. }
%short blurb on definitions of BC v RSG
%refer to Brian's paper Appendix, of Gadders et al. 1998, and Stott et a. 2009 (read these!)
%The red sequence is bounded by $\pm$3$\sigma$, as seen by the red lines in Fig. \ref{fig:completeness}. 
A $\chi^{2}$ minimization was applied to a linear model of the form
\begin{equation}
V-R = y_{0} + m \times M_{*}
\end{equation}

\noindent
yielding fit parameters \tbf{for slope (m), intercept ($y_{0}$) and scatter ($\sigma$) in Table \ref{tab:1} \citep{Gladders98,Stott09}}.

%Talk about results a little 
Both the North and South have a well-defined red sequence \tbf{with low scatter \tbf{(0.033 versus 0.024, respectively)},} consistent with cluster evolution after z$\sim$1 \citep{Bower1992,1996MNRAS.281..985V,2004ApJ...614..679L}.
The North cluster red sequence is four times steeper than the South cluster, \tbf{however, more similar to the Superfield despite having three times the surface density (Table \ref{tab:1}). The contrast between the North and South clusters is discussed further in Section \ref{sec:results} and motivates us to investigate the star formation history of the red sequences, starting with spectral measurements the next section.}

Splitting galaxy populations by color allows us to compare the fraction of galaxies in a given environment that are star forming. We summarize blue fractions and errors in Table \ref{tab:3}, \tbf{the results of which are discussed in Section \ref{sec:results}}. 
%The North and South clusters have excellent spectral coverage and completeness for all types of galaxies, so we are confident that measured differences are physical, rather than variations due to incomplete sampling.

\tbf{As mentioned at the beginning of the section, a simple color cut may not distinguish between quiescent and dusty starburst galaxies. We attempt to resolve this ambiguity by using a V-R, R-K observed frame color-color cut which roughly mimics the U-V, V-J rest-frame color-color cut implemented by \citet{Williams2009ApJ} to separate out truly quiescent populations from star-forming populations with and without an appreciable dust content. We find that a small fraction ($\la$5\%) disperse out of the main clump formed by the quiescent population in this color-color space, lending creedence to our assumption that our red galaxies are indeed red and dead.}

\begin{figure}
	  \begin{center}
	  \includegraphics[scale=.33]{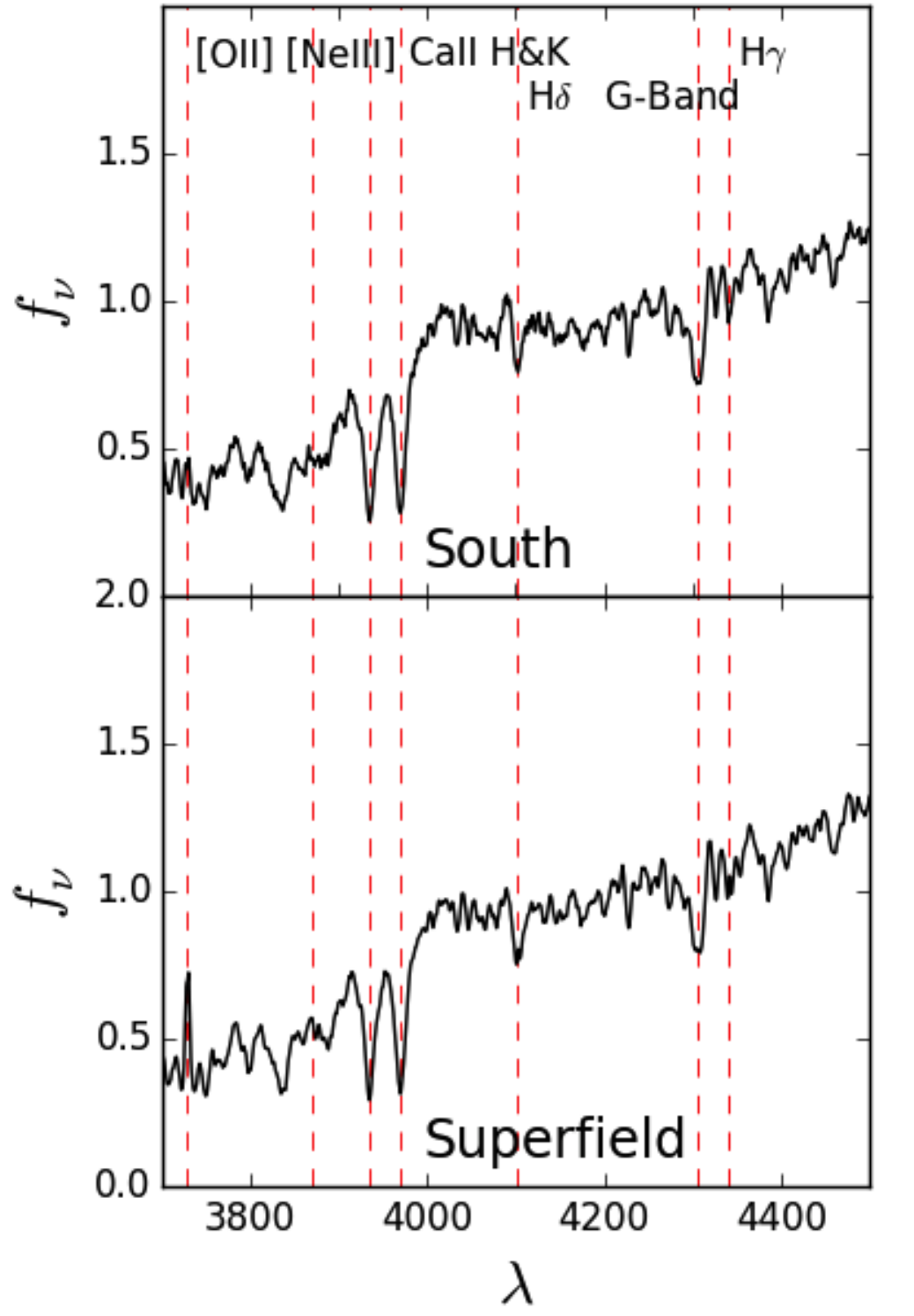}
	  \includegraphics[scale=.33]{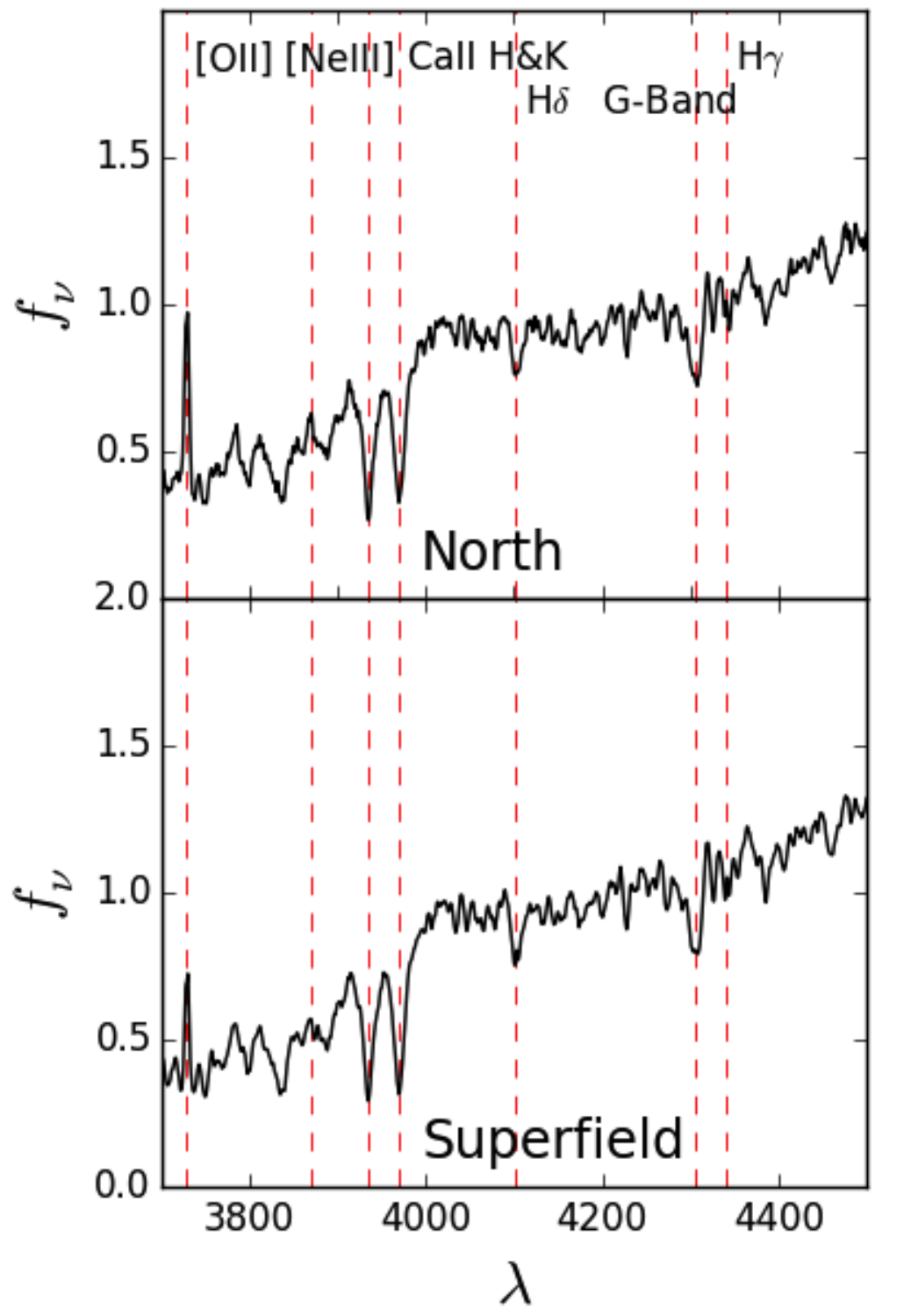}
	  \includegraphics[scale=.33]{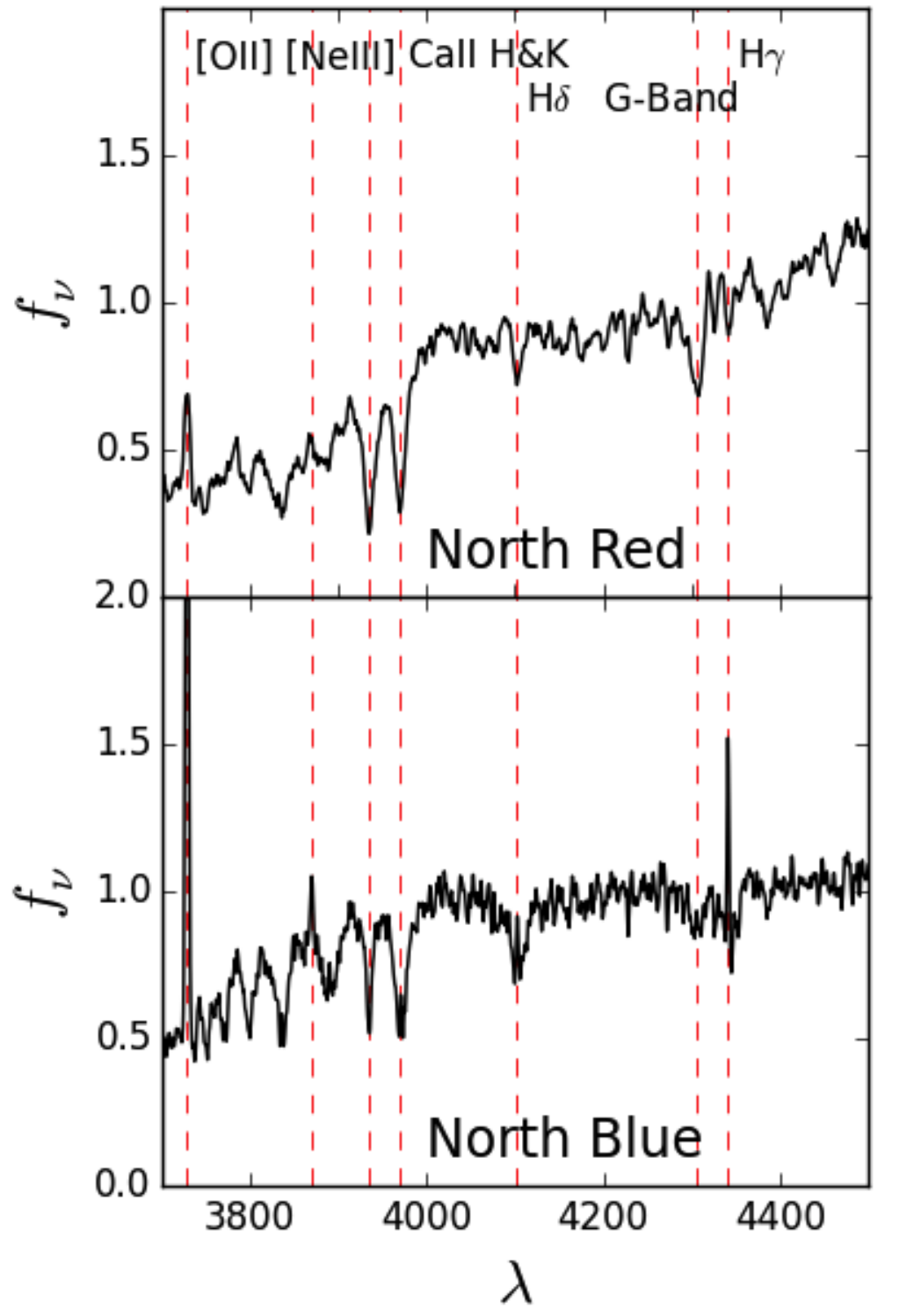}
	  \includegraphics[scale=.33]{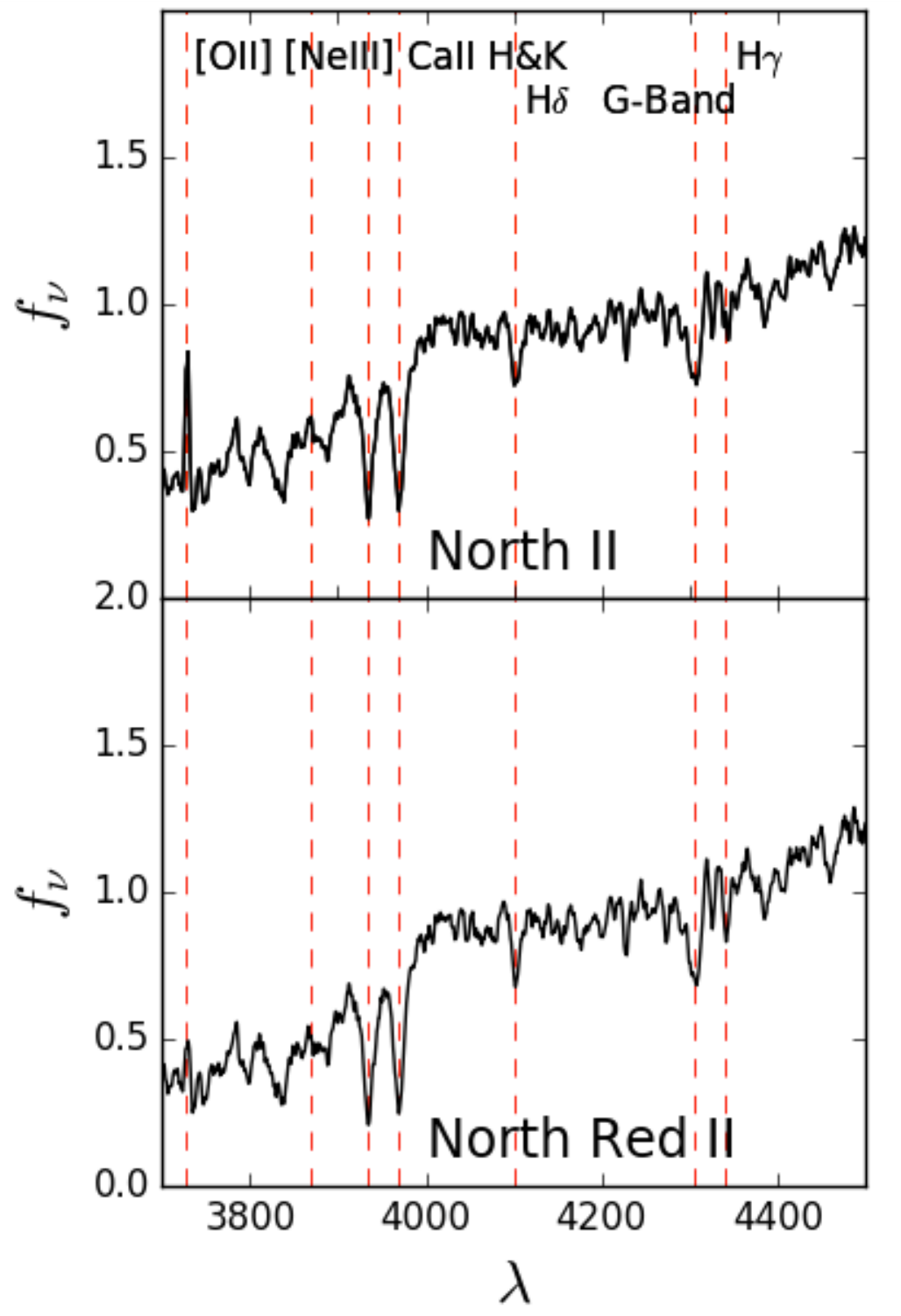} %shift de to ef and replace old d with new plot 
	  \includegraphics[scale=.33]{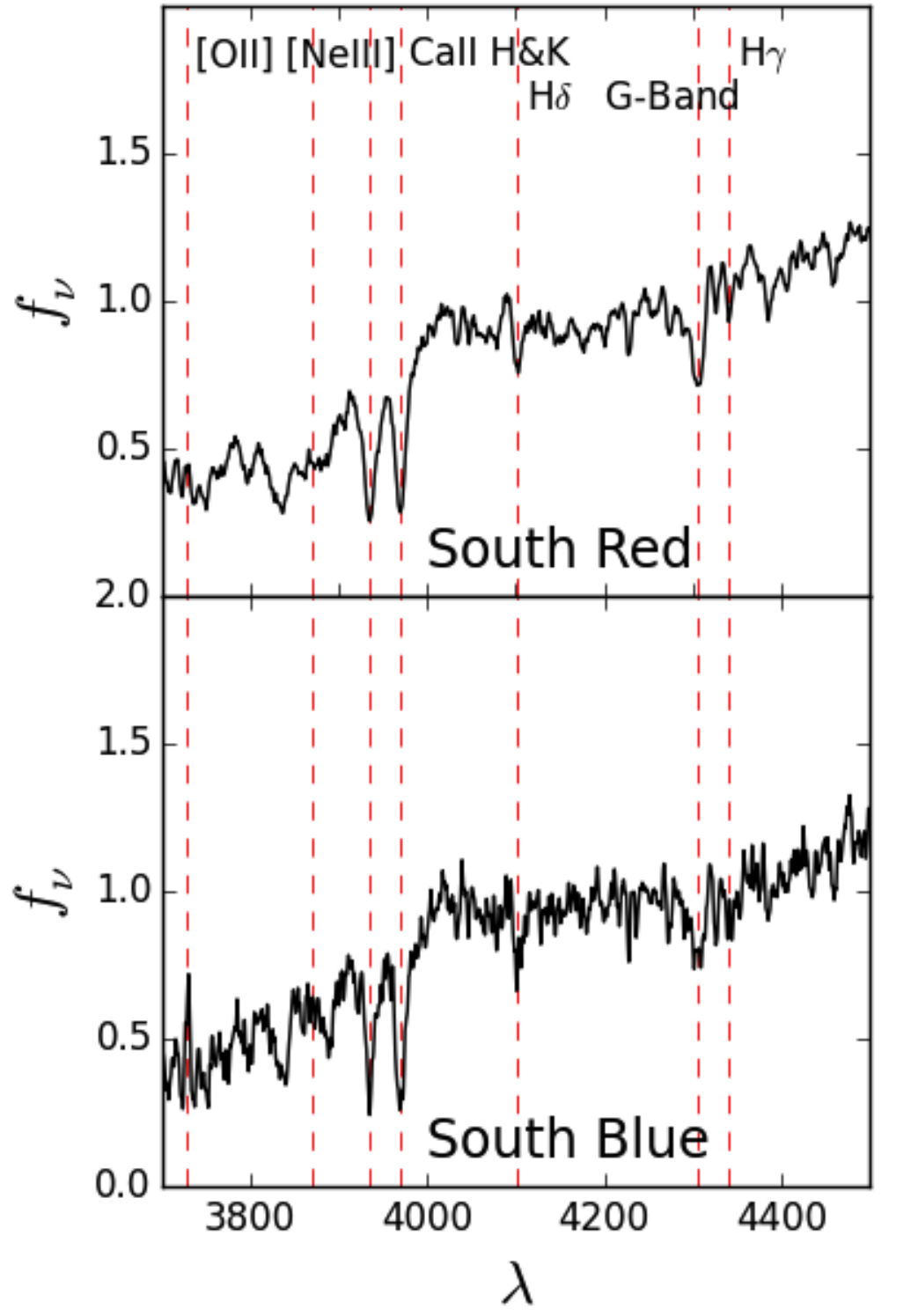}
	  \includegraphics[scale=.33]{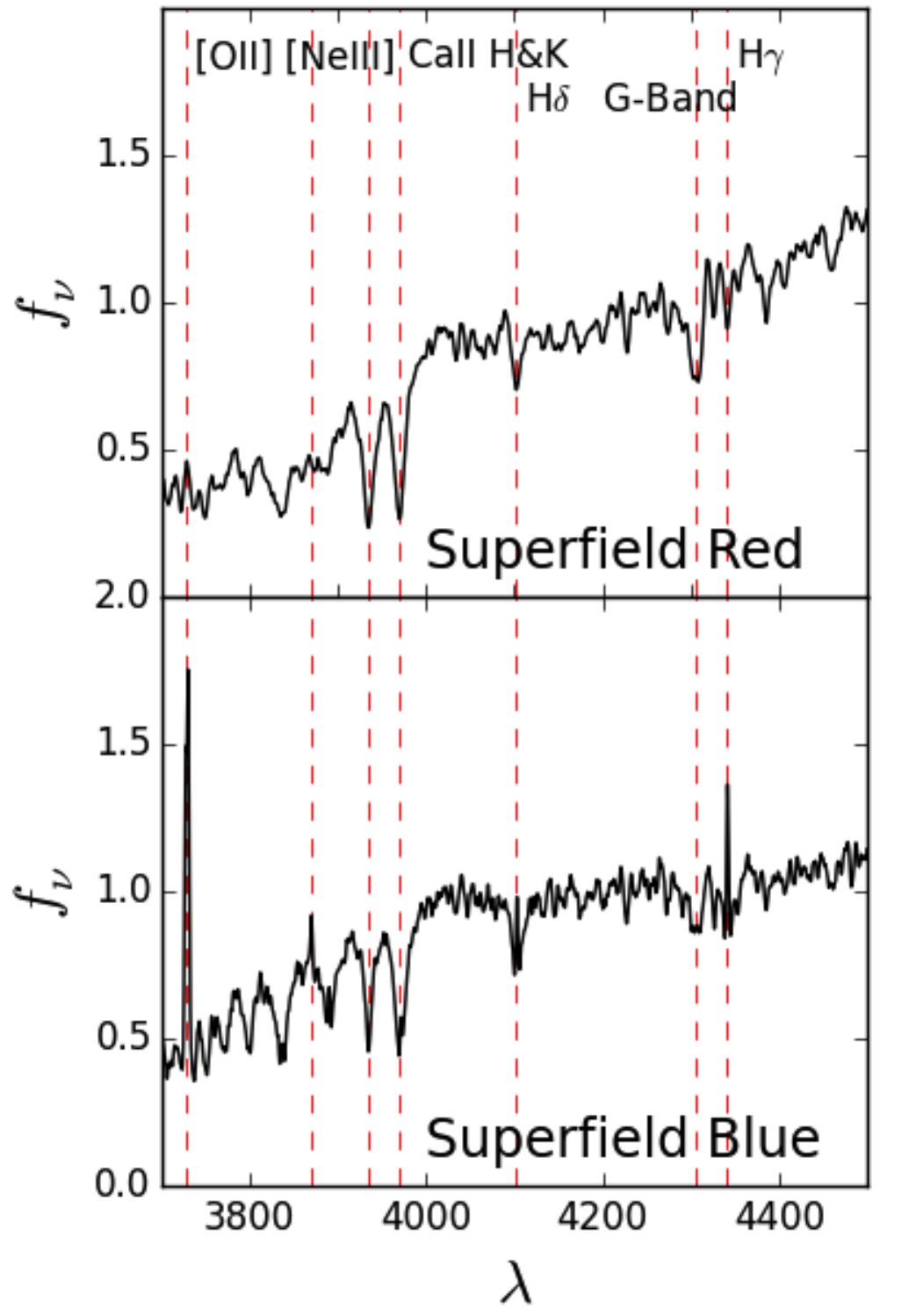}
	  \end{center}
	\caption{Composite spectra of sample done by region and color defined in Section \ref{subsection:spectral}. \tbf{North II and North Red II are subsets of North and North Red, respectively, with the two non-star forming sources removed (See Section \ref{subsubsec:AGN})}. Additional subsets by brightness and mass are not shown, but all measurements are summarized in Table \ref{tab:2}. Vertical axis is flux in arbitrary units, normalized by the median value for each composite. Spectra are smoothed by a 3 pixel Gaussian kernel. Dotted lines highlight key spectral features used to indicate star formation activity and mean age of stellar population. Overall, blue cloud composites show strong emission components, characteristic of star forming galaxies. The strong absorption components and \Dn break in red populations are characteristic of older stellar populations. The full region composites for the South and North reflect different levels of activity. The South spectrum is dominated by red features and no emission, indicating the average galaxy is no longer star forming, whereas the North is painted with emission lines, indicating the presence \tbf{of} younger, more active stellar populations. See Section \ref{subsubsection:composite} for further discussion.
	} 
	\label{fig:composite}
\end{figure}

\subsection{Spectral Measurements} \label{subsection:spectral}

In this section we discuss methods of measuring the spectral diagnostics \ewHd, \ewOII and \Dn from composite spectra and how we evaluate associated measurement errors. Sampling errors are discussed in Appendix \ref{subsubsection:completeness}.
%Note: moved sampling error stuff to appendix
% subsets of galaxy populations.
%and evaluating errors and the significance of our results. 
%talk about the post starburst qualifications physically, expand upon intro bit
We restrict our sample in accordance with the completeness limits described in the beginning of Section \ref{sec:galpop}. \tbf{We then perform our entire analysis on subsets of color, magnitude and stellar mass within the galaxy population, defined and summarized in Table \ref{tab:2}}. These restrictions and subsets allow us to confidently represent and compare properties of the average galaxy between regions, colors, brightnesses and masses.

\subsubsection{Composite Spectra} \label{subsubsection:composite}

Co-adding spectra increases the signal-to-noise ratio over individual spectra and reveals average stellar properties of the galaxies in a sample \citep[e.g.][]{Lemaux2010,Lemaux2012}. \tbf{Galaxy spectra} are co-added by region, color, stellar mass and brightness, to isolate the past and present star formation activity of different populations within our diverse system.

Each spectrum in the composite is weighted by the inverse variance of the raw data to minimize effects of poor night sky subtraction to downweight lower signal-to-noise regions and spectra, as well as the chip gap in between the red and blue CCD arrays on DEIMOS. Inverse variance is calculated for each pixel from the modified \textit{spec2d} routines used to reduce the data. 
Before co-addition, each spectrum is individually de-redshifted using its spectroscopic redshift.
%, to prevent the averaging out of key spectral properties. 
The spectra are then interpolated quadratically into a common grid with a minimum resolution of the pixel scale 0.32 \AA$\ $pixel$^{-1}$ divided by 1+z$_{min}$, so as not to artificially add resolution. Spectra are co-added using unit weighting, where each spectrum has its average flux density normalized to unity.

Composite spectra for full, red and blue, and bright massive populations by region are shown in Fig. \ref{fig:composite}, \tbf{in addition to special cases for North discussed in Section \ref{subsubsec:AGN}.} 
While the features in our blue and red composites are typical of young and old stellar populations, respectively, the full region composites indicate that the North and South galaxies on average have differing levels of star formation activity.

At first glance, in all blue cloud composites, we see \OII, [NeIII] 3869 \AA, and Balmer line (\Hd and $H\gamma$ 4340 \AA) emission coupled with a small \Dn , an indication of recent star formation. Both the North and South composites \tbf{of the full cluster populations} show evidence of G-band, Ca II H$\&$K along with Balmer line absorption, an indication of possible past star formation amongst cluster members. The North composite shows signs of emission, in addition to the presence of older stellar populations.
\tbf{The composite of the average South cluster member is dominated by features originating from late type stars (old) with no emission signatures. The composites reveal that, while both clusters have a presence of older stars with longer lifetimes as evinced by the strong continua, the North spectrum reveals the presence of younger stars with shorter lifetimes, which indicates current and recent star formation activity.}
%Note I cut off spectra in plot before OII and Hbeta to make more visually appealing, maybe change that back and add back into comments
%A priori, in all blue cloud composites, we see \OII, [OIII], [NeIII], and Balmer line (\Hd and $H\gamma$) emission coupled with a small \Dn , an indication of recent star formation. Both North and South full population composites show evidence of G-band, Ca H$\&$K and Balmer line absorption (\Hd, $H\gamma$, $H\beta$), an indication of past and rapidly truncated star formation in clusters. The North full region composite shows signs of emission, in addition to the presence of older stellar populations.
We couple these qualitative deductions with equivalent width measurements, described in Section \ref{subsubsection:EW}, for a more robust analysis.

\subsubsection{Equivalent Width Measurements} \label{subsubsection:EW}
%LaTeX won't let me make this into a subsubsub Section
%can paste from prop here...

EW measurements for all types of galaxies allow us to use composite spectra to quantitatively probe the current, recent and past modes of star formation in cluster galaxies. They largely escape the often paralyzing effects of dust contamination \citep[e.g.][]{Lemaux2010} and eliminate the need for flux calibration, as they are line strengths measured relative to the continuum. These effects are further mitigated by the scope of this paper, which compares regions and populations internally. 

\cite{D04} and \cite{P99} devise a way to differentiate between dominant modes of star formation using the relative line strengths of two features, \Hd absorption and \OII emission.  \OII emission from non-AGN/LINER sources is associated with active regions of star formation, characteristic of starburst galaxies. 
\tbf{The \OII doublet
% indicates the presence of O and B stars, which exist on short timescales ($\tau\sim$ 10 Myr). It 
 is a proxy for instantaneous star formation and recombination lines like H$\alpha$, which comes from HII regions ionized by massive early-type stars that exist on short timescales ($\tau\sim$ 10 Myr).} \Hd absorption is characteristic of B and A stars, the latter of which has a maximum lifetime of $\sim$1 Gyr. Appreciable ($>$4 \AA) absorption is observable for up to 1--2 Gyr, making \Hd an indicator of recent star formation. The absorption strength of \Hd and other Balmer lines, in the absence of associated emission features, is characteristic of a post-starburst galaxy, which has undergone a burst of star formation and/or rapid quenching. 

The EWs of these lines can be used to create a framework for the classification of the average galaxy in a composite as starburst, post-starburst, normal star-forming or quiescent. Immediately following a burst of star formation activity, \OII emission and \Hd absorption increase with starburst strength, and \Hd has yet to be infilled by new star formation. After star formation is rapidly truncated, or quenched, \OII emission \tbf{originating from star formation} ceases. As normal star formation continues or is resumed, \OII slowly increases along with \Hd infill. The EW bounds of this model are illustrated in the left-hand side of Fig. \ref{fig:EWsD4000}. Dashed lines indicate the area in this phase space which is found to contain 95\% of normal starforming galaxies observed at z$\sim$ 0.1 \citep{Oemler2009,Goto2003}. The red, light blue, green, and dark blue shaded regions correspond to quiescent, normal starforming, post-starburst, and starbursting galaxies, respectively.

EWs are measured using both bandpass and line-fitting techniques to facilitate comparisons to different models as well as previous studies. Our procedure for both is explained below. Results for line-fitting techniques are in Table \ref{tab:2}. Bandpass measurements are used to estimate incompleteness detailed in Appendix \ref{subsubsection:completeness}. 

Bandpass measurements are performed by defining three ranges around each spectral feature, adopted from \cite{Fisher1998}. The flux level of the band containing the feature is subtracted from the average of continuum bands straddling the feature. More details are found in \citep{Lemaux2010}

DEIMOS spectra have sufficient signal-to-noise for line-fitting techniques when emission and absorption lines are detected at a significance of greater than 3$\sigma$.
%This calcs distance from flux to continuum line (F-C/C)
We use a double Gaussian fit with a linear continuum for EW(\OII) and EW(\Hd). 

The \OII doublet is fit using two Gaussians at a fixed width of 2.8 \AA$\ $(rest-frame) for peaks at 3726 \AA$\ $ and 3729 \AA. Two Gaussians with the same fixed central wavelength are used for \Hd, yielding the strengths of both emission and absorption features. The Gaussian fit model contains seven free parameters: two to characterize the linear continuum, four to characterize the FWHM and amplitude of each Gaussian, and a single parameter defining the mean wavelength of the blueward Gaussian. Errors are derived using a combination of Poisson errors on the spectral feature and the covariance matrix of the linear continuum fit. 

%Addressing the different methods and theoretical corrections
%Include sentence on how using different methods affect significance of results, justification and comparison with theory, dust. Hold off on analysis of results, or keep all measurement discussion until intro plots. Maybe just refer to tables here:

The double Gaussian line-fitting technique is particularly advantageous for \ewHd because removing the emission component more accurately reflects the A star population. 
We are able to measure and remove emission infill for the majority of blue cloud composites. 
For cases where the emission component cannot be manually removed, a theoretical infill correction is calculated for red and blue galaxies, using line ratios from \cite{Schiavon2006} and \cite{Yan2006}. 
%We find that the theoretical correction that most closely agrees with the empirically measured corrections for the blue cloud is the correction developed for the red sequence, which is smaller by a factor of five. 
We find that the theoretical corrections for blue and red galaxies overestimate the measured corrections by factors of, on average, ten and five, respectively. Not only do the corrections fail to reproduce our measurements, the choice of correction does not significantly affect the age estimates, the classification of dominant star formation mode, or starburst history.
%the difference made by applying the theoretical infill correction does not translate to a significant difference in
As a result, we use the single Gaussian fit EW(\Hd) for all red composites, and all blue composites where no emission component can be removed using the double Gaussian fit method. 
%The correction is intended for comparison between populations where infill is not sufficiently pronounced to be removed manually. 
%Discussion of its utility below.
%The North cluster bright blue and bright blue massive populations were too noisy, but the theoretical infill correction from \cite{Yan2006} and \cite{Schiavon2006} matches the empirical correction measured by-eye for the populations where we were able to measure the correction by-eye. Using this proxy as an upper limit for the North EW(\Hd) emission we find that it has no significant effects on the age of the starburst.
%We do not include this correction in our final results to be consistent with using a single Gaussian fit for all composites where the infill cannot be manually removed. 
We find similar results when applying the correction to red sequence composites. Fitting a double Gaussian to potential infill, the measured correction is two orders of magnitude smaller than the theoretical correction. Moreover, the theoretical correction is on order of the errors from the fitted Gaussian method.
%The theoretical correction for Red Sequence galaxies is a closer fit to the measured Blue Cloud correction. (Do I even need to say this? Can I just quote the correction from Shaivon?) is on average four times that of the measured infill for the remaining North populations, due to the dependence on of EW(\OII), which we suspect to be inflated by AGN (Discussion in Section blah). The theoretical correction for the Red Sequence is on order of the empirically measured values, so we apply this as a lower limit (0.169317 for BBM, 0.169317 for BM table ref?).
%from the red sequence galaxies, 

%Go into detail about model here, quantify EW values associated with elevated levels, activity, explicitly how we define sb, passive, continuous and psb.
\tbf{Results for \ewHd versus \ewOII are plotted against an empirical model in Fig. \ref{fig:EWsD4000}. Coupling measures of instantaneous and recent star formation provides a metric for distinguishing passive from active galaxies, as well as relative levels of star formation activity between the average galaxy in the North, South, and Superfield. The physical motivation behind the model is described in detail in the beginning of this section. Starburst, post-starburst, normal star-forming, and quiescent are defined as \ewHd$\gtrsim$4.25\AA\  and \ewOII$<$-4\AA\ , \ewHd$\gtrsim$3.5\AA\  and \ewOII$\geq$-4\AA\ , 0\AA\ $\gtrsim$\ewHd$\lesssim$4.25\AA\  and \ewOII$<$-4\AA\ , and -0.5\AA\ $\gtrsim$\ewHd$\lesssim$3.5\AA\  and \ewOII$\geq$-4\AA\ , respectively. 
}

%These details are in the caption but warrant repeating I guess? I do not put all details on the models for the D4000 section in the text, rather in the caption, but there are too many. Here I introduce what is plotted so do not have to repeat again in the D4000 section
\tbf{EWs plotted in Fig. \ref{fig:EWsD4000}, by panel, represent full (red and blue) populations (\ref{fig:EWsD4000}a), as well as by color (red or blue)} (\ref{fig:EWsD4000}b), and for BB and BM subpopulations (\ref{fig:EWsD4000}c). BBM measurements are nearly identical to those for BM, so are not differentiated in Fig. \ref{fig:EWsD4000}c. All red subpopulations are identical, so are not differentiated in Fig. \ref{fig:EWsD4000}b. Measurements and definitions for all populations are \tbf{listed} in Table \ref{tab:2}.
%%%%%%%%%%%%%MOVED TO RESULTS
%(Insert a few conclusions but don't tie into bigger picture yet)
%\textit{

\ewOII and \ewHd measurements for the entire South and Superfield galaxy populations indicate they are on average quiescent (Fig. \ref{fig:EWsD4000}a: \textit{left}). The North measurements indicate the dominant mode of current star formation is currently normal star forming.
%}
%The motivation for listing explicit values in a table plotted against the models is to translate the quantitative to qualitative enabling us to ascertain and convey our conclusions, so we quantify and clarify the measurements that correspond to each classification and activity level but do not quote each individual value in the text of the section.
%\textit{
The average blue galaxy in the North is starbursting, indicating the North cluster is currently active. The North shows a greater level of current activity than the South and Superfield for the average galaxy in all \tbf{populations}, except for the BM Superfield galaxies (Fig. \ref{fig:EWsD4000}b,c: \textit{left}). The average North BB galaxy is significantly more active than its Superfield counterpart, which is unexpected given that galaxies in a cluster environment are not subject to the same quenching mechanisms present in a cluster environment, such as strangulation (long-acting) and ram pressure stripping (short-acting). 
%Ram pressure stripping can briefly trigger star formation, however the effect is on even shorter timescales starting from when a galaxy first encounters the ICM. \tbf{We cannot distinguish between these most common mechanisms as responsible for the star formation activity levels.}
%}
%It is unexpected that the average North cluster galaxy is more active than the average Superfield galaxy. Galaxies in less dense environments are not subject to the same quenching mechanisms that act on galaxies once they enter a cluster environment, such as ram pressure stripping and starvation.
%\textit{
While the North has a greater level of current, or instantaneous, activity than South, indicated by a higher level of \OII emission, the South \ewHd indicates an elevated level of star formation in the past for the average blue, BM and BB galaxy. The elevated \ewHd and lack of \ewOII indicate the average South blue galaxy is post-starburst, having undergone either recent, rapid quenching, or a strong level of star formation which has slowly decayed \citep{P99}. \tbf{The implications of the above results, contextualized through empirical models in the left panels of Fig. \ref{fig:EWsD4000}}, are discussed in Section \ref{sec:results}.
%}

\tbf{
The average North red galaxy \ewOII indicates normal levels of star formation. However, as discussed in Section \ref{subsubsec:AGN} we believe that this is due to non-star forming sources. While \ewOII is associated with nebular star formation activity, it can be produced by processes related to AGN/LINER activity \citep{Kauffman2003,Yan2006}. We temper the risk of contamination by using an additional indicator, \Dn, discussed in Section \ref{subsubsec:Dn}.
}

\subsubsection{\Dn Measurements} \label{subsubsec:Dn}

\Dn is a blanket absorption feature causing a depression in spectrum for wavelengths shorter than 4000 \AA$\ $indicative of the average age of a stellar population. 
We measure the strength of \Dn using the ratio of the average flux in the bandpasses of \cite{Balogh1997}. 
\tbf{Values are found in Table \ref{tab:2} and plotted on the horizontal axis in Fig. \ref{fig:EWsD4000}.} 
It has a known age-metallicity degeneracy, as well as a sensitivity to dust \tbf{which can change its value by 10\%} \citep[see discussion in][]{Lemaux2012}. 
These sensitivities are small, however, and for the purpose of internal comparisons are not restrictive if the stellar phase, metallicity, and dust properties of our sample are similar. We mitigate these caveats by combining \Dn with \Hd, \tbf{which also adds another dimension to our analysis using models seen in Fig. \ref{fig:EWsD4000} and described in this Section.}

%Talk about d4000 ressults
\Dn indicates the mean stellar age of galaxies in the South is significantly older than that of the Superfield (Table \ref{tab:2}). The North cluster galaxies, however, have an average stellar age younger than that of the Superfield galaxies. This disparity becomes even more pronounced for blue and BB subpopulations. Field galaxies are, on average, expected to be younger and more active than their cluster counterparts, so a reversal seen in the North galaxy population suggests that it is still vigorously forming.

\tbf{We combine \Dn with \ewHd (Section \ref{subsubsection:EW}) measurements to compare the evolutionary stage of the average cluster galaxy with models for galaxies undergoing a variety of different star formation histories.} The right-hand panels of Fig. \ref{fig:EWsD4000} show the average \ewHd versus \Dn of our galaxy populations against solar-metallicity synthetic spectra generated based on \cite{2003MNRAS.344.1000B} models, for \tbf{secondary bursts in star formation of varying strengths}.
\tbf{A detailed description of the models and input parameters are found in the figure caption.} 
%Do a better job connecting the models
\tbf{These models can indicate recent, elevated levels of star formation activity. If the time since the last star formation event coincides with the TSC, this suggests a possible connection with the merger event.}
\tbf{Composites representing the average galaxy} (Fig. \ref{fig:EWsD4000}a: \textit{right}) indicate the North, South and Superfield have had no recent star formation.
%Dividing the population by color (Section \ref{subsec:rsfit}) brings to light the different star formation histories of red and blue galaxies for each region. Based on well-established fundamental properties, 
Dividing by color, we examine the blue population to reveal possible elevated star formation and the red population to indicate possible quenching. 
%}
%\textit{
The average Southern blue galaxy indicates a possible star formation event within the 68\% confidence interval of the merger TSC. Evidence of the same event is also found in the Southern BB, and BM subpopulations (Fig. \ref{fig:EWsD4000}b,c: \textit{right}).
The North blue cloud galaxies show very recent elevated levels of activity, more significant than in the Superfield. This enhanced activity in the blue cloud galaxies is not within the TSC of the merger, however, so is suggestive of either latent activity due to the merger or secular cluster processes in a cluster that is young and still forming.
%}
Red composites for all regions indicate no recent quenching (Fig. \ref{fig:EWsD4000}b: \textit{right}), \textit{ruling out the possibility of a predominantly merger-extinguished galaxy population.} \tbf{These results, along with those from \ewOII in Section \ref{subsubsection:EW}, are contextualized and discussed further in Section \ref{sec:results}.}
%wide spread

%To avoid repeating, maybe just leave non repeated stuff here or say a blanket conclusion and refer to results section.

\subsubsection{Note on AGN} \label{subsubsec:AGN}

Two red sequence objects with significant \OII emission are found in the North red sequence composite, boosting the composite EW by 5 \AA\  \tbf{as seen by comparing North Red and North Red II in Fig. \ref{fig:composite}}. The resulting \ewOII is twice that of the North blue cloud population, falling outside of the pattern we expect for red populations given our other measurements and literature on clusters at this redshift. In addition, the \Hd composites have no corresponding visible emission infill, which is typical in galaxies undergoing star formation, as seen in our blue cloud population. The \ewHd and \ewOII measurements in Fig. \ref{fig:EWsD4000} indicate the average North red galaxy is continuously star forming. The sampling errors (Table \ref{tab:2}) indicate the \ewOII has a high variance, an order of magnitude greater than the variance of all other populations. In our individual spectra we are able to observe that, without the two red galaxies, our result is typical of other red sequence composites. Galaxies with high \ewOII and high \ewHd could be undergoing a starburst, but we would expect to have corresponding \Dn measurements indicating the presence a young stellar population, which is not the case as measured from the North red sequence composite.
%The composite EW values without these objects are above the average \ewOII of other regions and populations.) 
%Chung warns against assuming all high OII emitters are AGN, there are some on RS.

Only 6\% of line-emitting red sequence galaxies have \OII emission characteristic of star forming galaxies \citep{Yan2006}, a trend that extends way beyond the redshift range of the Musket Ball system \cite{Lemaux2010}. We conclude that the \OII emission in the presence of these two objects is likely from non-star forming sources, such as LINER/Seyfert galaxies. 
The \Dn and \Hd contributions for composites without these two objects \tbf{(North Red II in Fig. \ref{fig:composite})} are typical of those measured for all other red sequence populations for the North, South and Superfield (Table \ref{tab:4}). \tbf{In addition, the large [OII] emission in the North persists without the AGN candidates (seen as North II in Fig. \ref{fig:composite}), so is not likely driven by non-starforming sources.} While \tbf{we} can rule out broadline emission, we are missing the narrow lines most reliably used to distinguish between starburst and LINER/Seyfert galaxies, such as H$\alpha$ and [N II]. Regardless, including these objects makes no appreciable difference in our result for the time since starburst (Fig. \ref{fig:EWsD4000}: \textit{right}).  

\begin{figure*}
	  \begin{center}
	  \includegraphics[height=50mm]{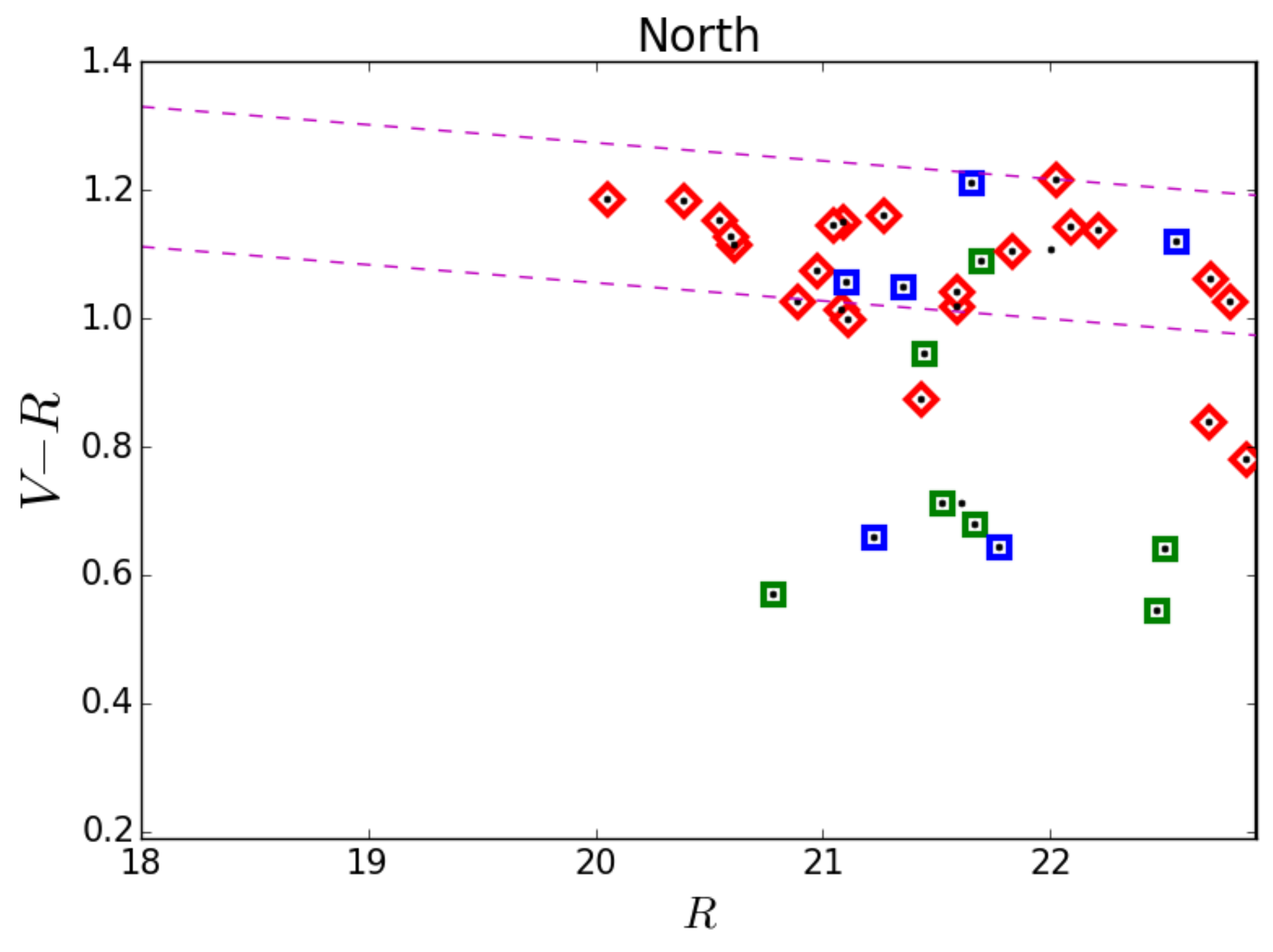}
	  \includegraphics[height=50mm]{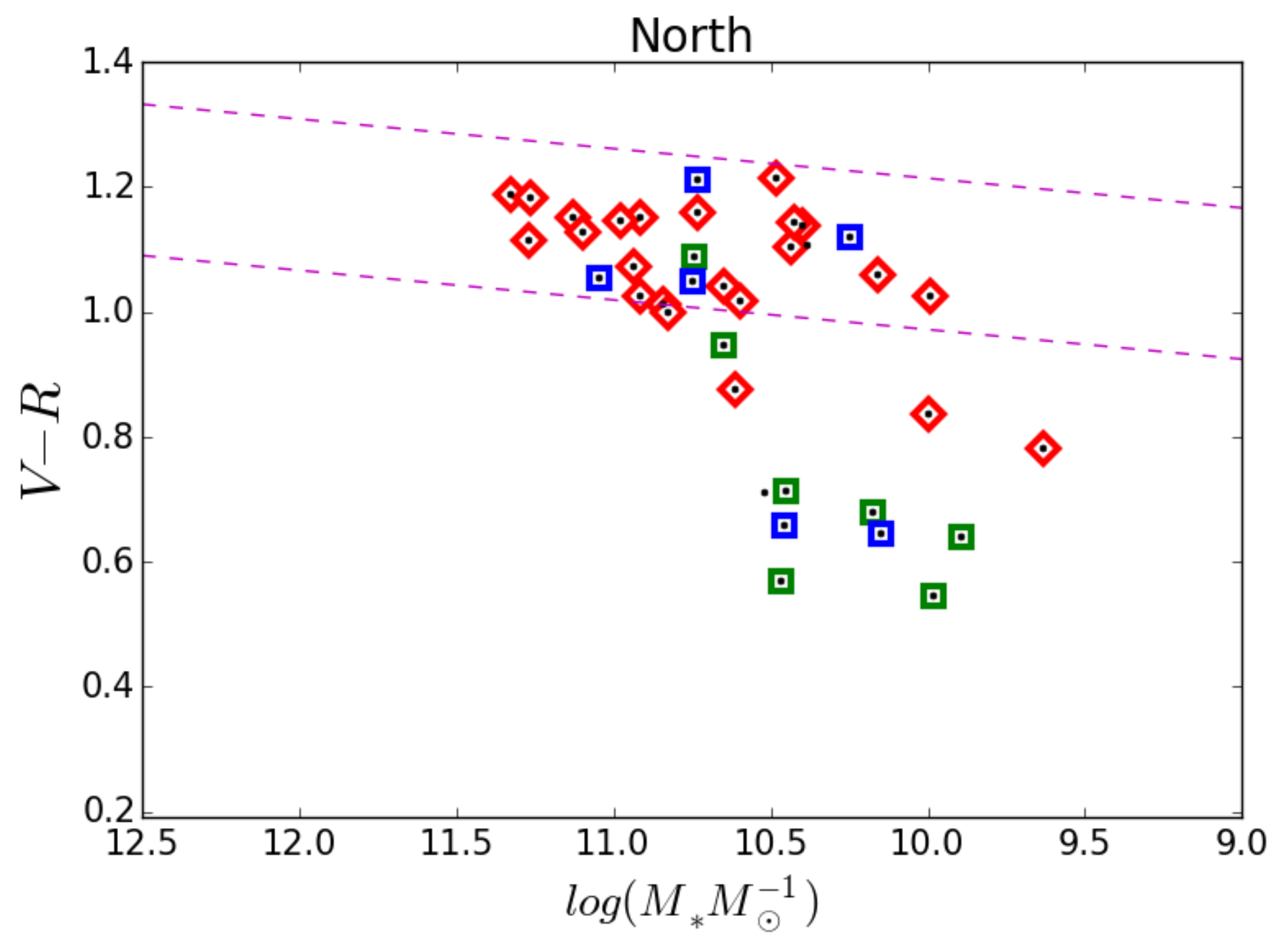}
	  \includegraphics[height=50mm]{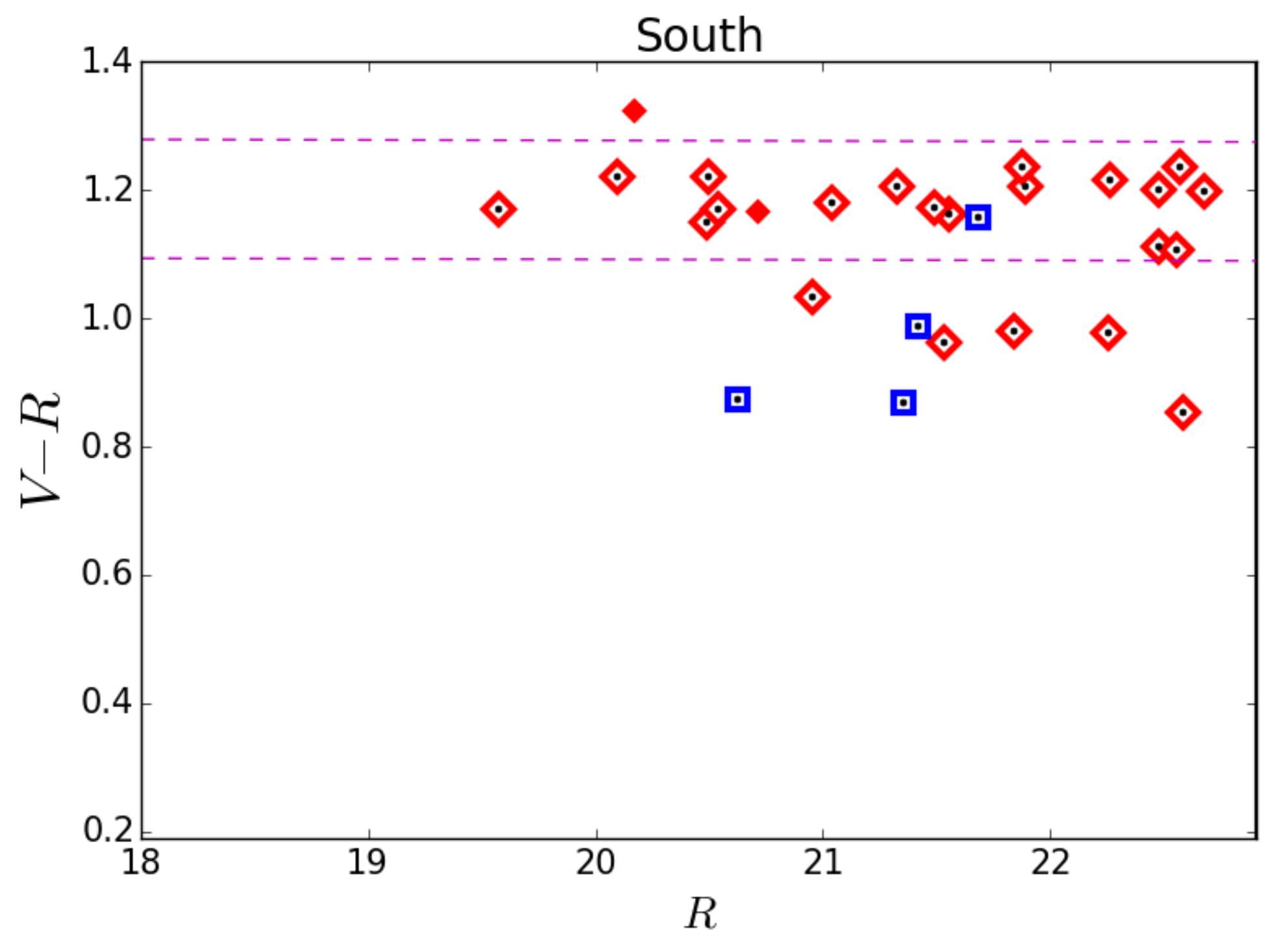}
	  \includegraphics[height=50mm]{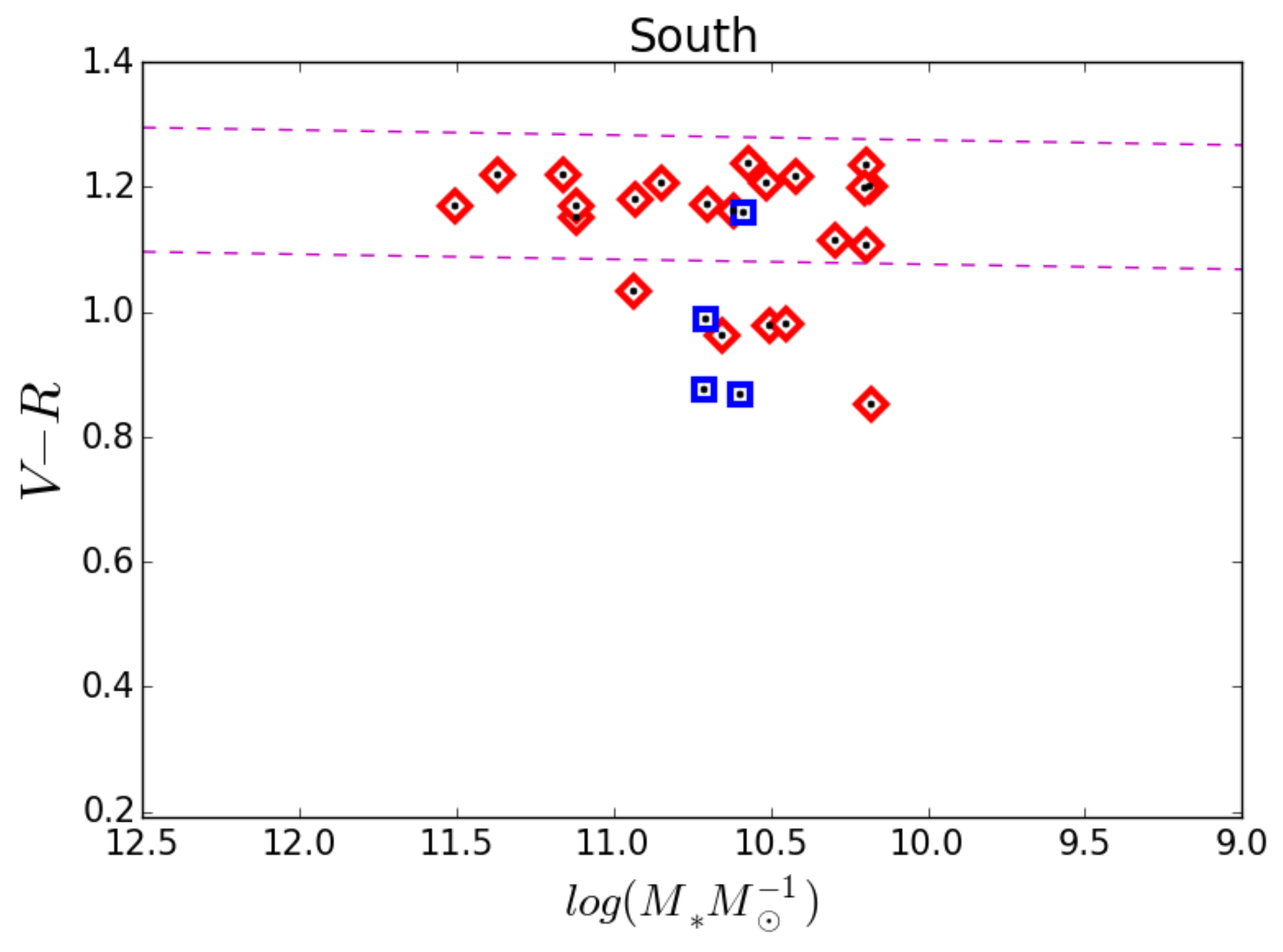}
	  \includegraphics[height=50mm]{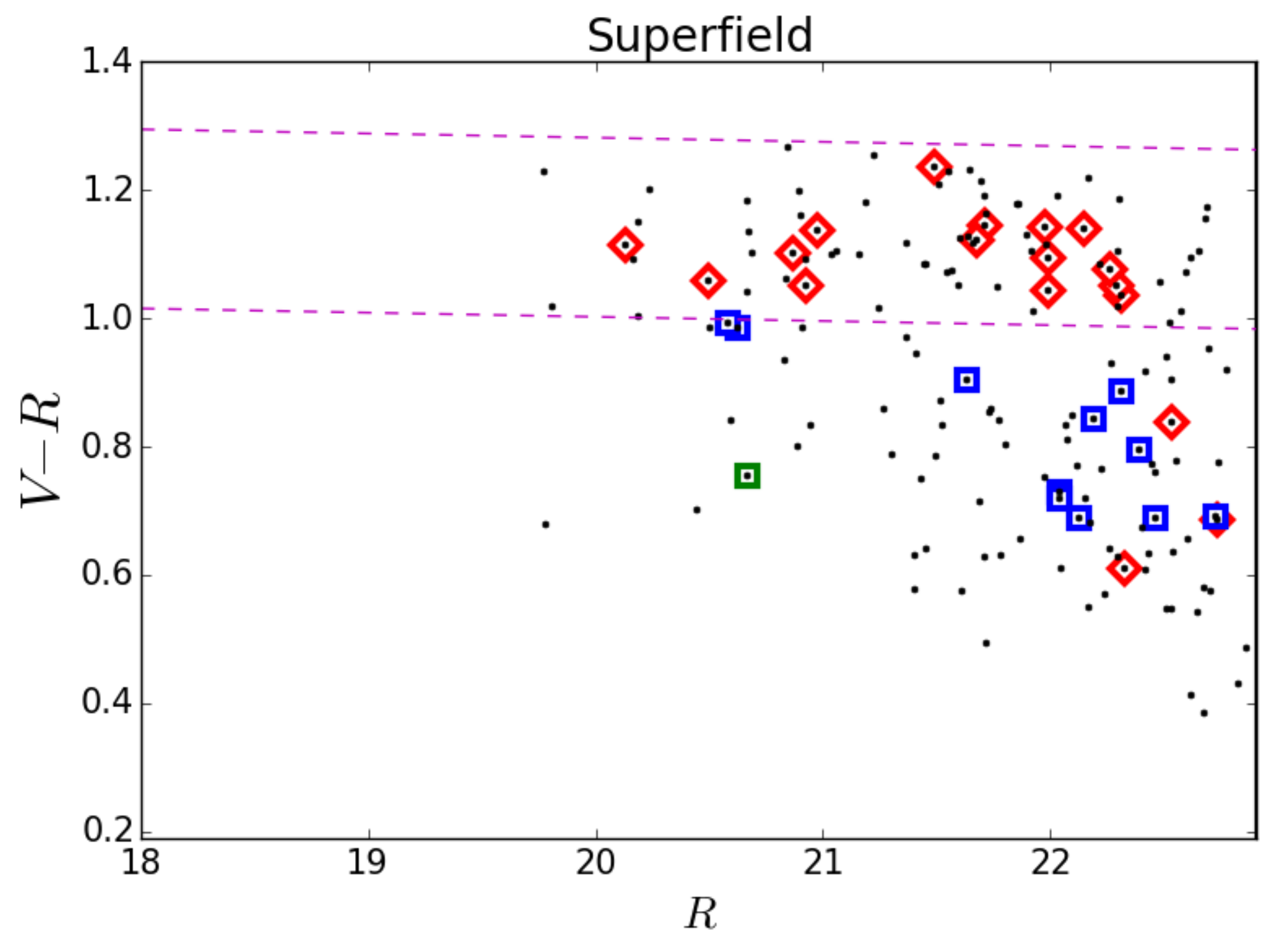}
	  \includegraphics[height=50mm]{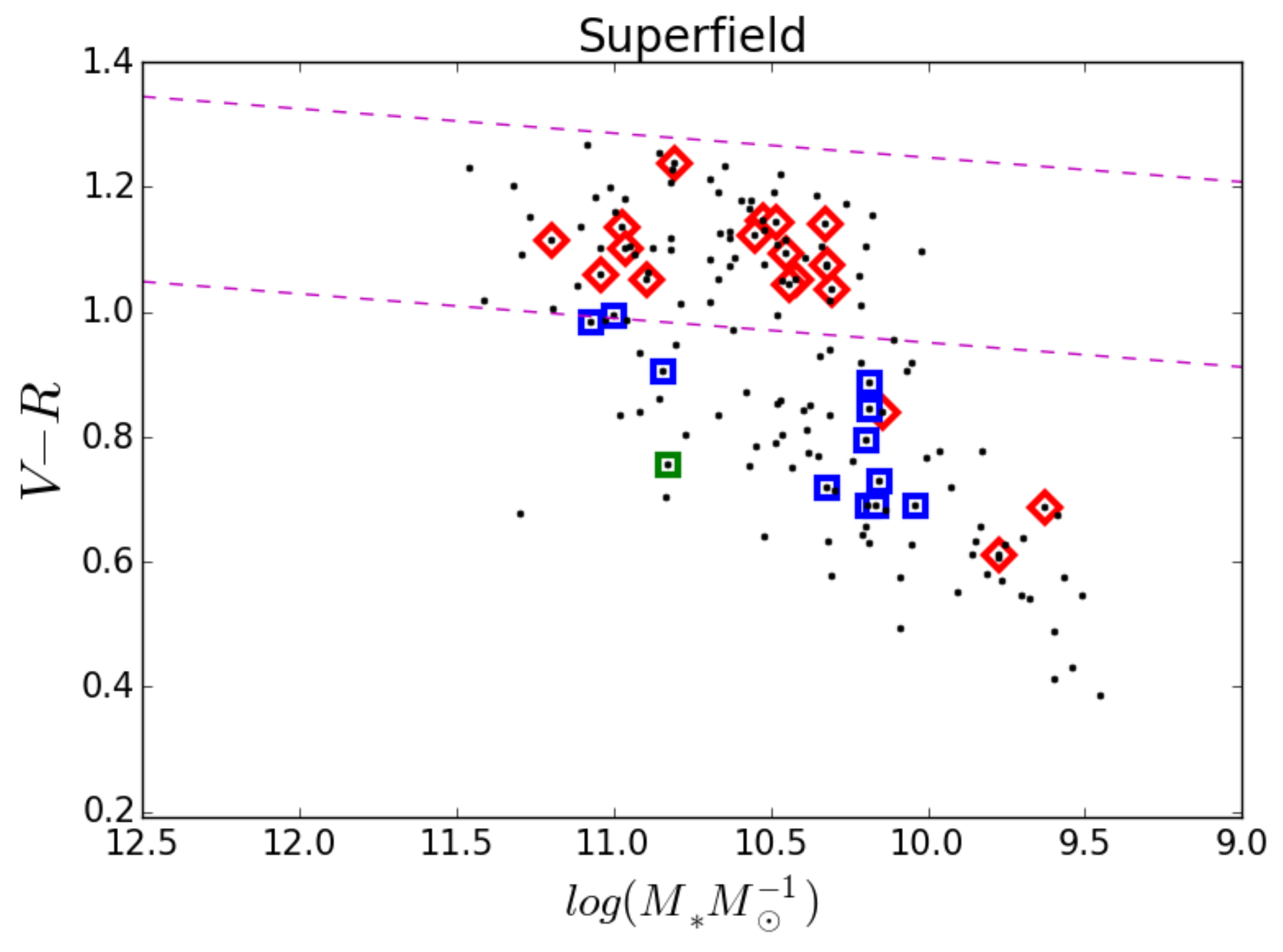}
	  \end{center}	
	  \caption{CMDs (\textit{left}) and CSMDs (\textit{right}) of all DEIMOS spectroscopically confirmed cluster redshift (black dots) circumscribed with symbols representing assigned \textit{HST} morphologies (Section \ref{subsection:morph}). Red diamonds are early-type galaxies, and blue and green squares are spiral galaxies and \tbf{disturbed} galaxies, respectively. Individually resolved components of the South BCG, which is an E/E merger, are indicated by small red diamonds (Section \ref{subsec:BCG}).
	  Superfield morphologies include only 68\% of objects due to the size of the \textit{HST} footprint. 
	  }
	\label{fig:morph}
\end{figure*}

\begin{deluxetable*}{lllcccc}
\tabletypesize{\footnotesize}
%\tablecolumns{8} 
\tablewidth{0pc} 
\tablecaption{\footnotesize{Measurement Offsets Relative to Galaxy Density Peak}}
\tablehead{ \colhead{\footnotesize{Region}}
 & \multicolumn{2}{c}{\footnotesize{Galaxy Density}}
 & \colhead{\footnotesize{BCG}}
 & \colhead{\footnotesize{SBCG}}
 & \colhead{\footnotesize{\textit{HST} WL\tablenotemark{b}}}
 & \colhead{\footnotesize{X-Ray Gas\tablenotemark{b}}} \\ 
 \colhead{\footnotesize{}}
 & \colhead{\footnotesize{RA}}
 & \colhead{\footnotesize{DEC}}
 & \colhead{\footnotesize{kpc}}
 & \colhead{\footnotesize{kpc}}
 & \colhead{\footnotesize{kpc}}
 & \colhead{\footnotesize{kpc}}
}
\startdata
North & 09\fd16\fm11\fs & 29\fd51\fm59\fs & 157\tablenotemark{a} & 494 & 145 & 554 \\
South & 09\fd16\fm16\fs & 29\fd49\fm15\fs & 369 & 117 & 168 & 757 \\
\enddata
\label{tab:4}
\tablecomments{\footnotesize{Offsets are discussed in Section \ref{subsec:BCG}.}}
%Corresponding spatial plot in Fig. \ref{fig:map} }
\tablenotetext{a}{\footnotesize{Confirmed by SHELS redshift \citep{SHELS}.}}
\tablenotetext{b}{\footnotesize{X-ray and WL maps found in \cite{Dawson2012}}}
%\tablenotetext{b}{\footnotesize{Infill corrected where removable using double Gaussian fit to emission component}}
\end{deluxetable*}
%\end{comment}

%\begin{deluxetable}{lccccccc}
\begin{deluxetable}{lccc}
\tabletypesize{\footnotesize}
%\tablecolumns{8} 
\tablewidth{0pc} 
\tablecaption{Color and Morphology Fractions}
\tablehead{ \colhead{\footnotesize{Region}}
% & \multicolumn{2}{c}{\footnotesize{\Hd}\tablenotemark{a}}
% & \multicolumn{2}{c}{\footnotesize{\OII}}
% & \multicolumn{2}{c}{\footnotesize{\Dn}}\\
% & \multicolumn{2}{c}{\footnotesize{(\AA)}}
% & \multicolumn{2}{c}{\footnotesize{(\AA)}}
% & \multicolumn{2}{c}{\footnotesize{(\AA)}}\\
%\colhead{{Population}}
%CHECK THAT THIS IS FRACTION OF TOTAL NOT RATIO
 & \colhead{\footnotesize{Blue}}
% & \colhead{\footnotesize{Blue/Blue$_S$}}
 & \colhead{\footnotesize{Early-type}\tablenotemark{a}}
 & \colhead{\footnotesize{Early-type}\tablenotemark{b}}
% & \colhead{\footnotesize{Spiral}}
% & \colhead{\footnotesize{Irregular}}
% & \colhead{\footnotesize{Mass}}
% & \colhead{\footnotesize{Brightness}}
}
\startdata
%\cutinhead{Full Region}
North & 0.34 $\pm$ 0.09 & 0.64 $\pm$ 0.13 & 0.63 $\pm$ 0.11 \\
South & 0.31 $\pm$ 0.11 & $0.85^{+0.15}_{-0.18}$ &  0.85 $\pm$ 0.16 \\
Superfield & 0.50 $\pm$ 0.10 & 0.60 $\pm$ 0.14 &  0.48 $\pm$ 0.10 \\
\begin{comment}
\cutinhead{Including z$_{phot}$=0.53$\pm$0.1}
North & \nodata & 0.63 $\pm$ 0.11 \\
South & \nodata & 0.85 $\pm$ 0.16 \\
Superfield & \nodata & 0.48 $\pm$ 0.10 \\
\cutinhead{Bright}
North & 0.47 $\pm$ 0.16 & ... \\
South & 0.5 $\pm$ 0.10 & ... \\
SuperField & 0.61 $\pm$ 0.62 & ... \\
\cutinhead{Massive}
North & 0.24 $\pm$ 0.08 & ... \\
South & 0.5 $\pm$ 0.00 & ... \\
SuperField & 0.47 $\pm$ 0.35 & ... \\
\cutinhead{Bright Massive}
North & 0.23 $\pm$ 0.08 & ... \\
South & 0.5 $\pm$ 0.10 & ...  \\
SuperField & 0.46 $\pm$ 0.39 & ... \\
\end{comment}
\enddata
\label{tab:3}
\tablecomments{\footnotesize{Early-type galaxies defined in Section \ref{fig:morph}.
Errors are Poissonian. Sampling errors discussed in Appendix \ref{subsubsection:completeness}.}}
\tablenotetext{a}{\footnotesize{\textit{HST} footprint excludes 31\% of Superfield galaxies.}}
\tablenotetext{b}{\footnotesize{\tbf{Including z$_{phot}$=0.53$\pm$0.1}}}
\end{deluxetable}

\subsection{Morphology} \label{subsection:morph}

Morphology of 95 (33/48 Superfield, 26 South, 36/38 North) cluster and Superfield members covered by our \textit{HST} footprint (Fig. \ref{fig:map}) was determined by one of the authors (L.M.L.) by visual inspection of \textit{HST} imaging in F606W and F814W. The \textit{HST} footprint is within our area of highest spectroscopic completeness.

Fig. \ref{fig:morph} shows CMD/CSMDs with assigned morphologies. 
Objects classified as spiral in at least one band were classified as spiral, seen as blue \tbf{squares}. Objects classified as chaotic or merger in at least one band were taken as \tbf{disturbed}, seen as green boxes. We define these two categories as late-type. Objects \tbf{classified as} E or S0 in one band were classified as early-type, seen as red diamonds. There was one \tbf{spectroscopically confirmed} merger with two ellipticals characterized by double nuclei and tidal tails, which was classified as early-type. This object is the BCG in the Southern cluster, discussed in Section \ref{subsec:BCG}. Early-type fractions are in Table \ref{tab:3}.
%The 5 \sigma point source magnitude limit for our ACS catalog has a mean of X.
\tbf{The visual inspection performed here has been shown to be both objective and reproducible \citep{Lemaux2012} and consistent with statistical methods \citep[G-M$_{20}$,][]{2011ApJ...736...38K} of morphological classification. Such tests were performed on a sample up to $\sim$10 times fainter than the FW814 magnitude of our faintest galaxy, using imaging twice as shallow as our own. Visual inspection is preferred over statistical measures in this work because of the relatively small size of our sample and because identifying potential merging galaxies in G-M$_{20}$ space is highly subject to galaxy detection parameters.}

\begin{figure}
	  \begin{center}
	  \includegraphics[height=35mm]{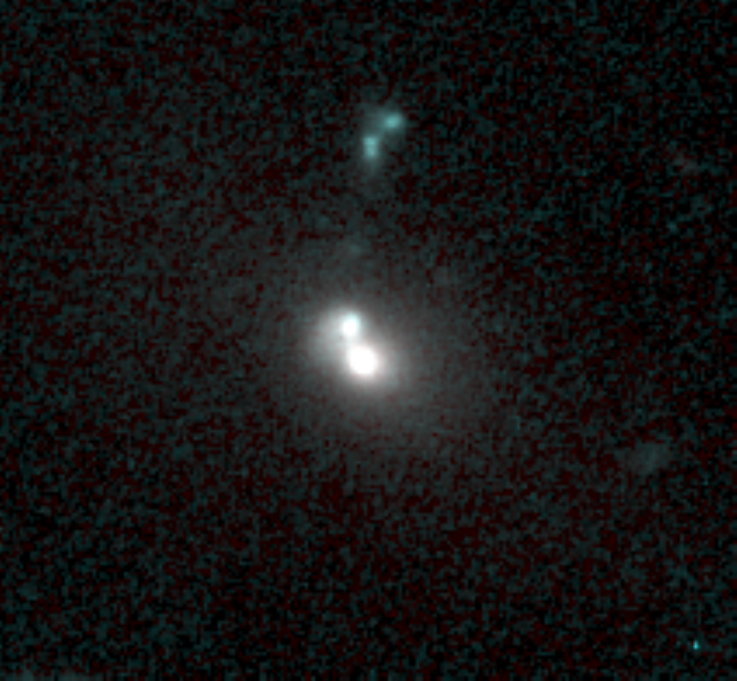}
	  \includegraphics[height=35mm]{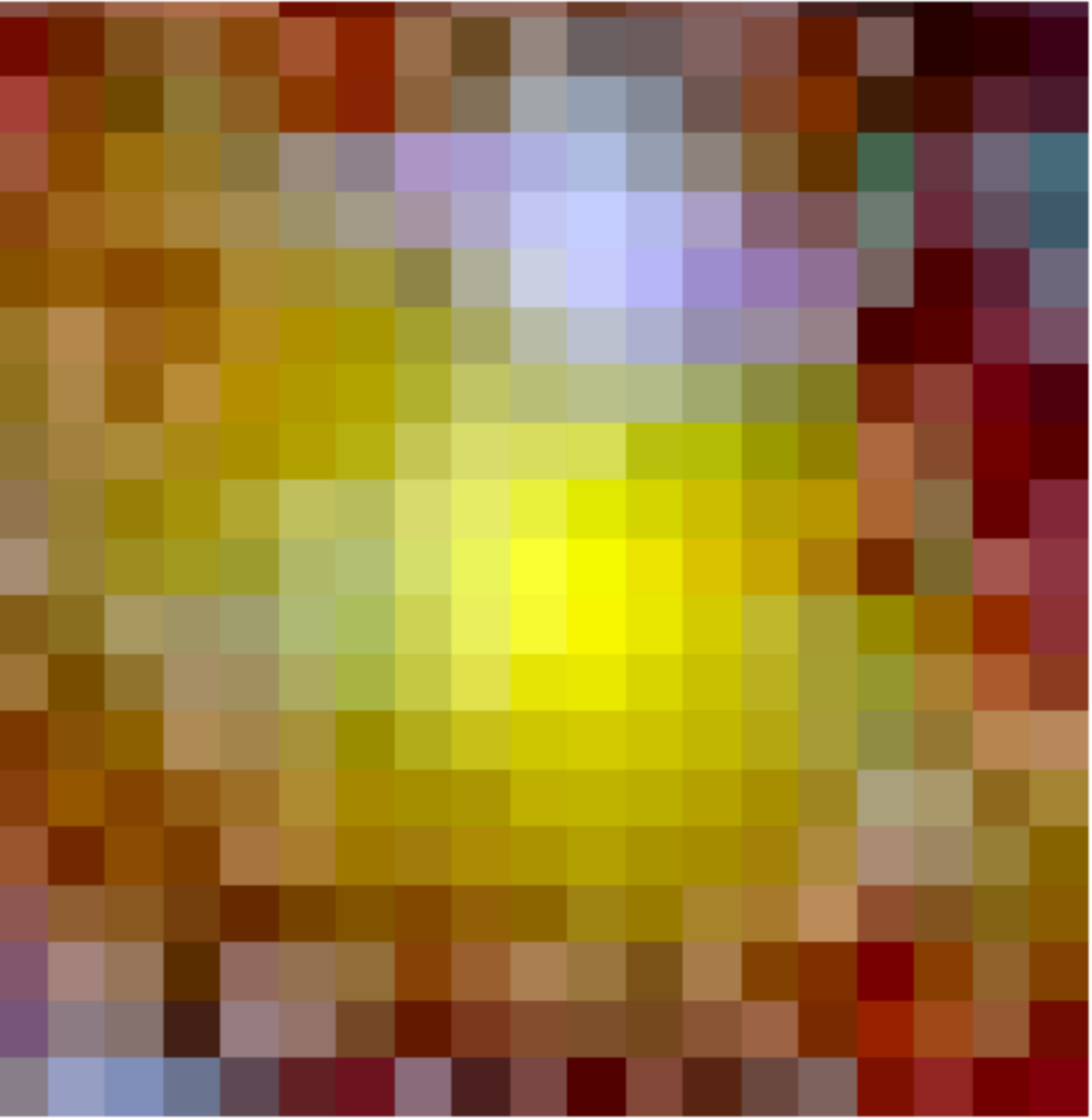}
	  \end{center}	
	  \caption{\textit{HST} (\textit{left}) and DLS (\textit{right}) $10\arcsec\times10\arcsec$ composite images of the South BCG, discussed in Section \ref{subsec:BCG}. \textit{HST} imaging resolves the object (\textit{right}) as three galaxies (\textit{left}) consisting of two merging ellipticals and a projected background galaxy (z=0.592). The decomposed magnitude of each component is shown in Fig. \ref{fig:completeness}. The individual components are all fainter than the SBCG, indicating the SBCG \tbf{was} the BCG prior to the \tbf{spectroscopically confirmed} E/E merger.
	  }
	\label{fig:sbcg}
\end{figure}

\subsection{BCG and MMCG} \label{subsec:BCG}

%I moved these definitions to an earlier section
\tbf{While the BCGs of both clusters possess signatures of the BCGs found at lower redshifts, we find several inconsistencies suggestive of changes in local environment (further discussion in Section \ref{sec:results})}.
The DEIMOS confirmed BCGs for the North and South have spectra with absorption features and a strong continuum, typical of bright, red sequence galaxies, which ultimately form into the \tbf{BCGs of regular clusters seen at lower redshifts \citep{2013hsa7.conf..433A}. For the North and South clusters, the BCG is also the MMCG, seen as the brightest and most massive object in each cluster CMD and CSMD, respectively, in Fig. \ref{fig:completeness}} and indicated with luminosity-scaled yellow dots in Fig. \ref{fig:map}. \tbf{In this section, we quantify the qualities that distinguish the North and South BCGs from those typical of relaxed clusters.}
%We find that the BCGs in each cluster have properties typical of a BCG in one way, representing another way the post-merger system is unique.}

%\tbf{However,} 
BCG candidates in relaxed clusters are typically central and 1-2 magnitudes brighter than next brightest galaxy. 
The North, \tbf{however,} has a second BCG candidate of nearly equal magnitude, for which we have \textit{HST} imaging and a redshift confirmed by SHELS. The SHELS confirmed BCG is in the densest region \tbf{of the North cluster, 157 h$^{-1}_{70}$ kpc} from the galaxy centroids, \tbf{yet} the DEIMOS confirmed BCG is isolated in a second density peak, 599 h$^{-1}_{70}$ kpc away \tbf{(Table \ref{tab:4})}. 
%The large separation of the North BCG candidates makes it difficult to establish a cluster center. 
The presence of two \tbf{close candidates for BCG} in the North suggests that it is in the process of assimilating a group or cluster, \tbf{discussed further in Section \ref{sec:results}.} 

%%%%%%MOVED TO RSEULTS
%\textit{
%This process is consistent with hierarchical structure formation and \tbf{introduces a probability that the North was already in an unrelaxed state prior to the merger \citep{Kauffmann1995}. As discussed in Section \ref{sec:results}, the large BCG offset is not entirely due to the merger. The current state increases.}
%}

The South BCG is largely offset from the galaxy density peak \tbf{(369 h$^{-1}_{70}$ kpc)}. However, morphology and \tbf{spectroscopy indicate this} object is a galaxy-galaxy merger of two ellipticals (seen individually as two small, filled red diamonds in the left-center panel of Fig. \ref{fig:morph}), so may not have been the BCG prior to the merger. Both nuclei are resolved in the \textit{HST} bands, so we perform aperture photometry to estimate the V and R magnitudes of the individual galaxies, along with a projected companion \tbf{spectroscopically confirmed} to be a background object. Fig. \ref{fig:sbcg} shows a side-by-side comparison of DLS and \textit{HST} images of the BCG. \tbf{The decomposition of each component in color-magnitude space is shown in the left-center panel of Fig. \ref{fig:completeness}. The merging galaxies are represented as a magenta triangles because they remain on the red sequence, and the deblended background galaxy is a smaller, green triangle located in the blue cloud. The magnitudes indicate that, when considered individually, each merging galaxy is fainter than the second brightest and most massive galaxy in the cluster.}
The South second BCG (SBCG) was thus the BCG prior to the E/E merger. This SBCG is in close proximity to the galaxy density peak, which is typical of a more relaxed cluster (Fig. \ref{fig:map}). \tbf{We discuss additional properties that support the picture of South as a relaxed cluster in Section \ref{sec:results}).}

%\begin{comment}
\begin{figure*}
	  \begin{center}
	  \includegraphics[height=60mm]{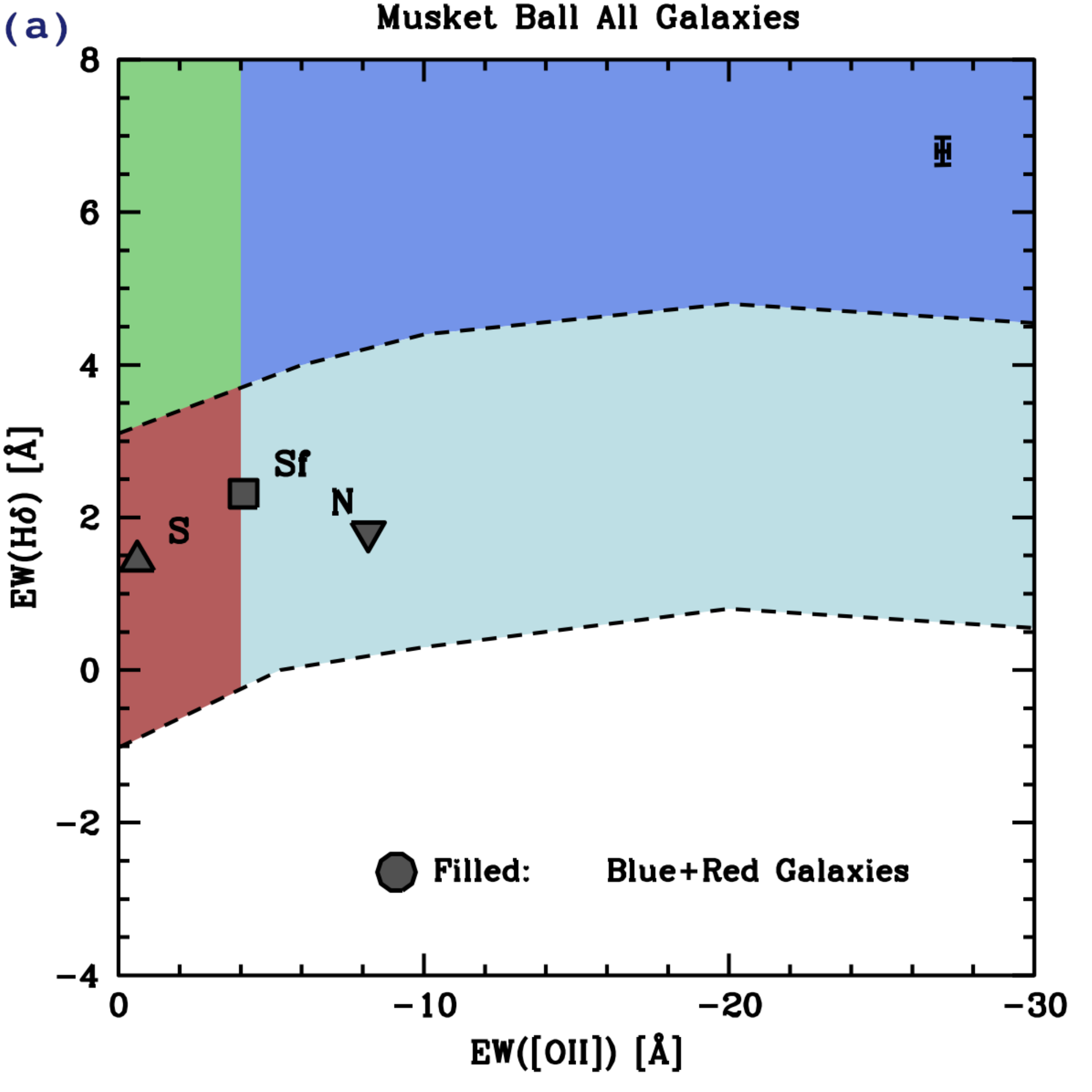}
	  \includegraphics[height=60mm]{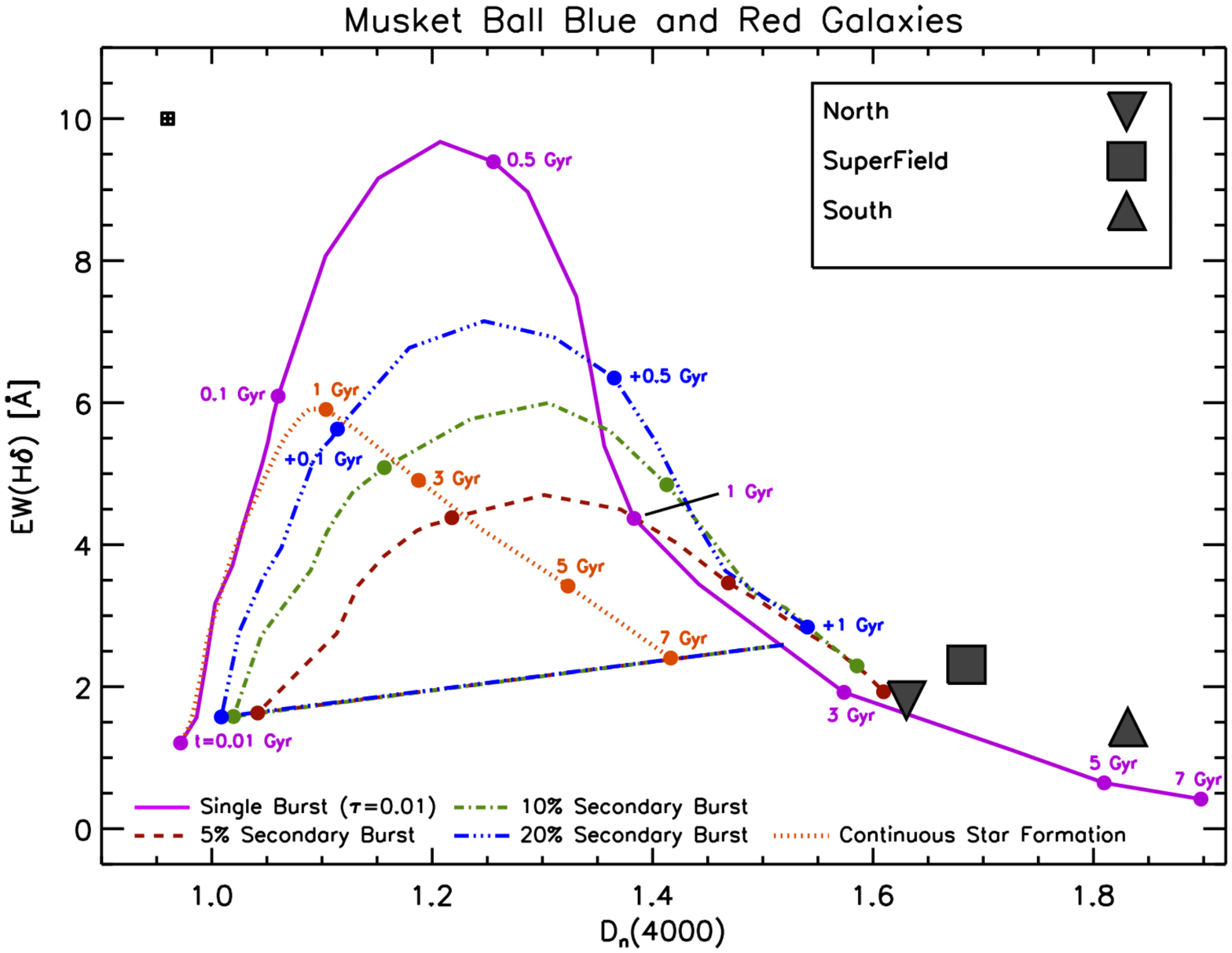}
	  \includegraphics[height=60mm]{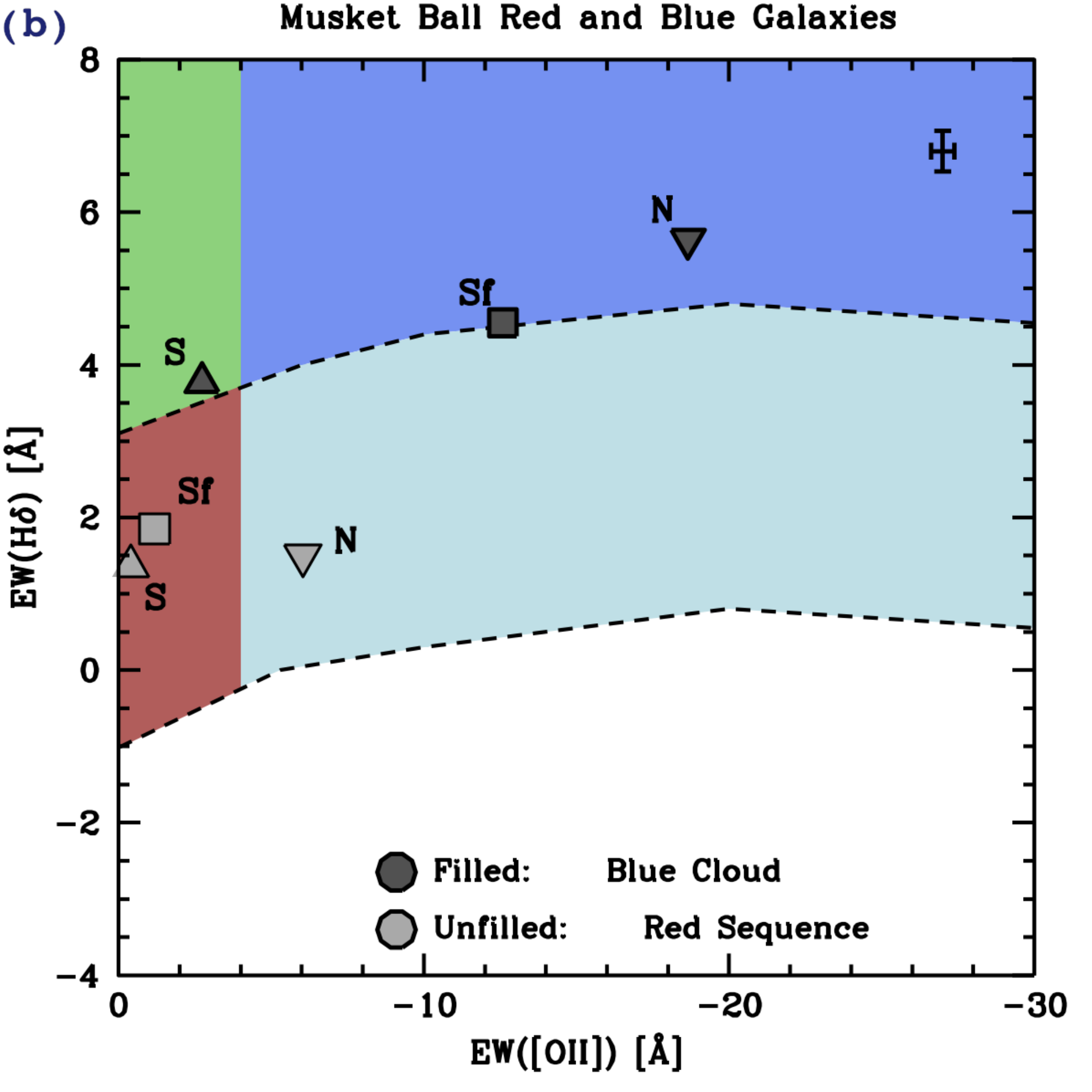}
	  \includegraphics[height=60mm]{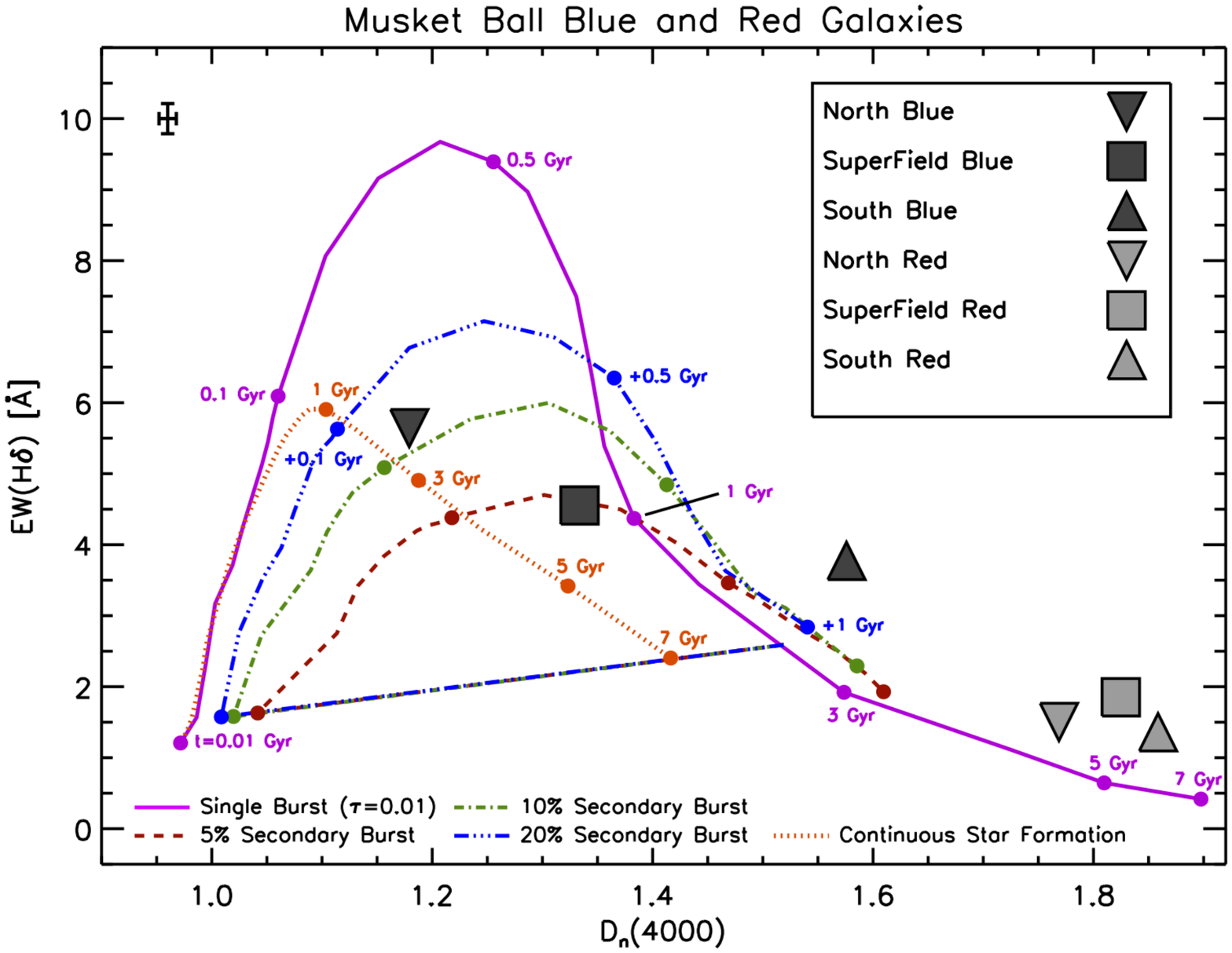}
	  \includegraphics[height=60mm]{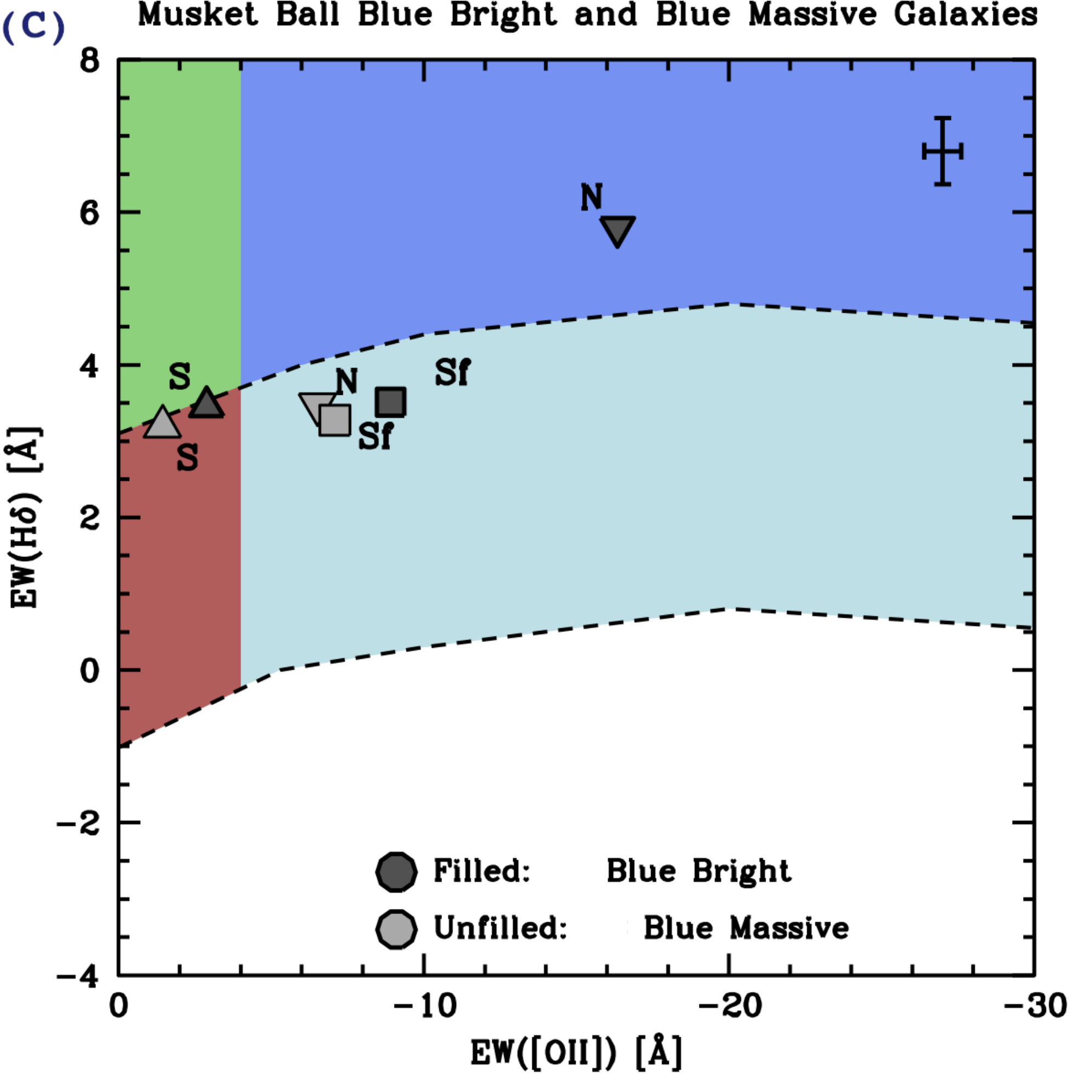}
	  \includegraphics[height=60mm]{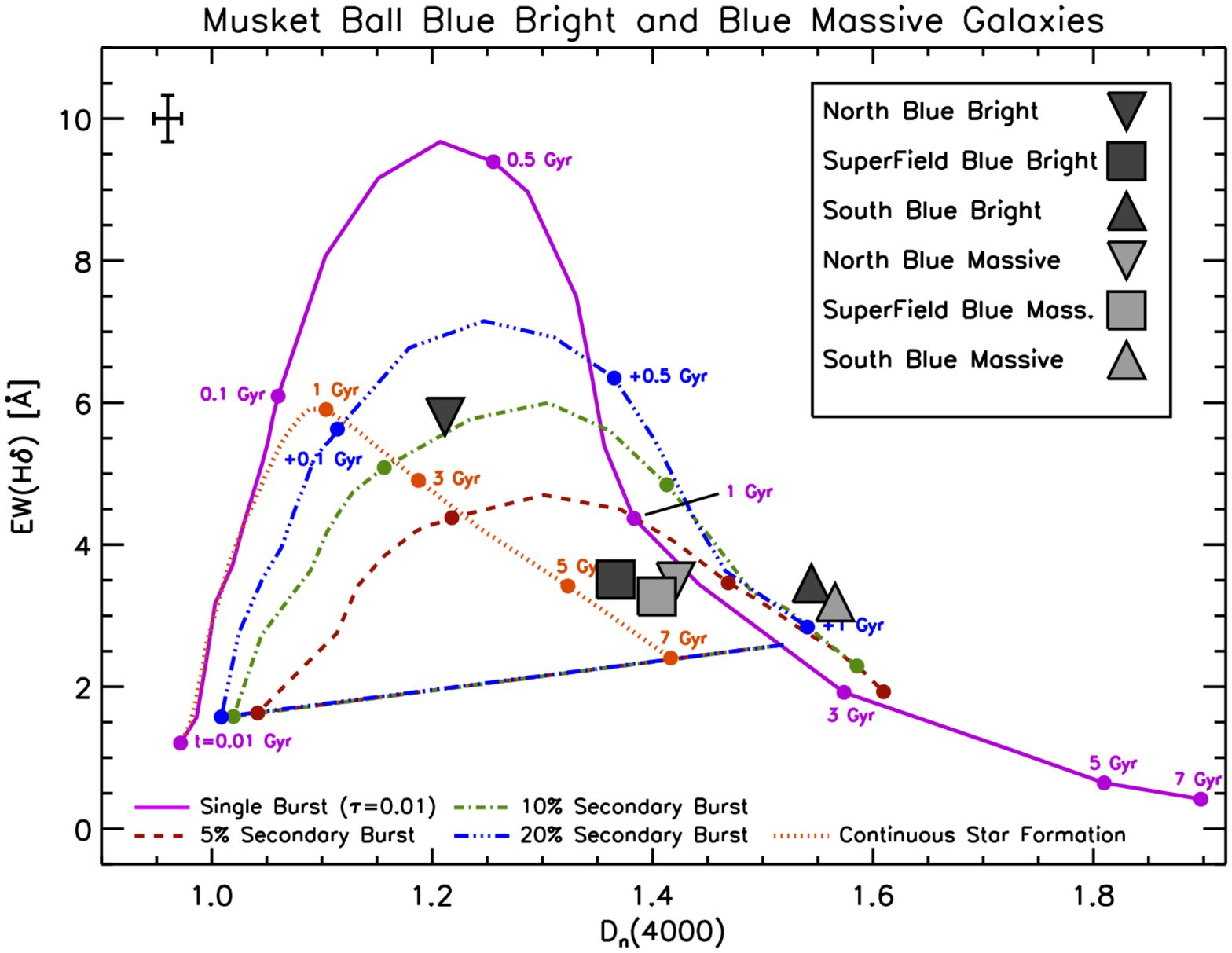}
	  \end{center}	
	  \caption{
	  \tbf{EWs measured from composite spectra of the North and South and Superfield regions contextualized with empirical and synthetic models of current and past star formation activity. EWs plotted for each region are} (a) full  (red+blue) population (b) blue and red populations (c) BB and BM. BM and BBM results are similar and not differentiated here. Red sequence subpopulations (RB, RM, BRM) are similar so only red is shown. All measurements \tbf{and definitions are in Table \ref{tab:2} and methods and models are described in Section \ref{subsubsection:EW}.}
	  \textit{Left:} Dashed lines indicate the area in this phase space found to contain 95\% of normal starforming galaxies observed at z$\sim$ 0.1 \citep{Oemler2009,Goto2003}. The red, light blue, green, and dark blue shaded regions correspond to quiescent, normal star-forming, post-starburst, and starbursting galaxies, respectively. Average measurement errors are shown in the upper right corner and do not account for completeness. 
	  \textit{Right:} 
	  \ewHd versus \Dn of various populations of the Musket Ball structure against solar-metallicity synthetic spectra generated based on \cite{2003MNRAS.344.1000B} models, for a variety of different star formation histories. Synthetic spectra are extincted using a \cite{Calzetti2000} reddening law with $A_{V}$=1. Dotted line tick marks give the ages of a secondary burst starting 2 Gyr after an initial starburst. \tbf{Dotted lines correspond to a secondary burst in star formation of strength corresponding to percentage of stellar mass found in legend.}
	  Solid line represents values for a galaxy undergoing an initial starburst. The straight, multicolored line represents values during the first 0.01 Gyr of the secondary burst, when emission from O stars dominates over \ewHd absorption.
	  Error bars located in the upper left-hand corner give average measurement error. The values for the North and South are indicative of two distinct star formation histories. 
	  While the average cluster galaxy does not show an excess of \ewHd over the Superfield, the average blue cluster galaxy indicates increased levels of current and past activity, discussed in Section \ref{sec:results}.
	  }
	\label{fig:EWsD4000}
\end{figure*}
%\end{comment}

\section{Results}\label{sec:results}

We endeavored to look for an excess or deficit of \ewHd in the galaxy populations merging Musket Ball clusters relative to the Superfield, which would indicate a change in the amount of star formation within the last 0.5-1 Gyr. This estimate, based on the lifetime of A stars, overlaps with the younger portion of the 68\% confidence interval of the estimated TSC. A collective burst of star formation in either cluster at the time of the merger would potentially indicate that the merger triggered star formation. 

\tbf{In this section, we first summarize and expand upon the results of our spectral analysis from Section \ref{subsection:spectral}. 
We then tie in results from other sections (geometry from Section \ref{subsec:isodensity}, substructure from Section \ref{subsubsec:dstest}, photometry from \ref{subsec:rsfit} and morphology \ref{subsection:morph}) to contextualize spectral results.}

\begin{itemize}
\item We find the population as a whole for each cluster does not have a greater \ewHd value relative to the Superfield.
Full population composite spectra indicate the average galaxy in the North and South has not undergone a recent star formation event, consistent with the null hypothesis that the star formation activity in these galaxies was largely unaffected by the merger. % (Fig. \ref{fig:EWsD4000}a).

\item The EWs measured from the composite spectrum of the full Southern population indicate the average Southern cluster galaxy is currently quiescent (Fig. \ref{fig:EWsD4000}a: \textit{left}) and has not undergone a major star formation event in the last 5 Gyr (Fig. \ref{fig:EWsD4000}a: \textit{right}), well outside the 2.4 Gyr upper limit of the 68\% confidence interval for our TSC. 
This constraint indicates the Southern galaxies, on average, were neither quenched nor triggered by the merger. 
Similarly, the average North galaxy has not undergone a major star formation event in the last 3 Gyr, beyond even the earliest stages of the merger indicated by our TSC (Fig. \ref{fig:EWsD4000}a: \textit{right}).
The North \Dn and \ewOII, however, indicate the average North cluster galaxy is younger and more active than not only the average South cluster galaxy, but the Superfield.

%Coupling \ewHd with the \ewOII and \Dn measurements allows us to construct a narrative that the North and South clusters are dramatically different in their current and past star forming properties. We also find that these differences carry over to additional properties of each cluster, including morphology, geometry and substructure, indicating a disturbed, active Northern cluster and Southern cluster on its way to becoming passive and relaxed. 

\item As discussed in Section \ref{subsec:rsfit}, we separate red and blue populations for the North, South and Superfield to investigate the recent star formation histories of these fundamentally different galaxy types. 
We expect the red sequence to be, on average, quiescent and for moderate levels of current star formation to be evident in the blue cloud.
%continuous or enhanced 
For populations undergoing star formation, we expect the activity level to be, on average, less in the clusters than the surrounding Superfield. 
In Fig. \ref{fig:EWsD4000}b: \textit{left} we investigate the current modes of star formation \tbf{of the average red and blue galaxy} in the North, South and Superfield. 
In Fig. \ref{fig:EWsD4000}b: \textit{right} we investigate \tbf{the recent and past of star formation of the average red and blue galaxy} in the North, South and Superfield. 
While our results for the red sequence are consistent with \tbf{the above expectations}, the North and South blue clouds show unusual levels of activity relative to Superfield within the past $\sim$1 Gyr, discussed below. 
%In Fig. \ref{fig:EWsD4000}b: \textit{right} we investigate the recent and past star formation \tbf{of the average red and blue galaxy} in the North, South and Superfield.

\item 
%In Fig. \ref{fig:EWsD4000}b: \textit{right} we investigate the current modes of star formation \tbf{of the average red and blue galaxy} in the North, South and Superfield. 
Red galaxies remain passive in all environments (Fig. \ref{fig:EWsD4000}b: \textit{left}), with the excess \ewOII in the North likely due to non-star forming processes (see Section \ref{subsubsec:AGN}). An old, quiescent red sequence is consistent with studies that show cluster red sequence galaxies have undergone little star formation since z$\sim$3 \citep{Bower1992,1996MNRAS.281..985V,1997ApJ...483..582E,1998ApJ...501..522B}. 
Our measurements indicate that red galaxies in the North and South clusters have not undergone a recent burst (Fig. \ref{fig:EWsD4000}b: \textit{right}). The age of the last star formation event exceeds the upper limit of our 68\% TSC confidence interval.
\textit{This constraint precludes the possibility of the merger affecting the star formation properties of the red sequence through the quenching of a large population of galaxies.}

\item In the South cluster, \tbf{\ewHd and \ewOII measurements indicate} the average blue galaxy is not currently forming stars \tbf{(Fig. \ref{fig:EWsD4000}b: \textit{left}), but, coupled with \Dn, \ewHd} indicates the population experienced an enhanced level of star formation in the recent past (Fig. \ref{fig:EWsD4000}b: \textit{right}). 
This event took place within the past $\sim$1 Gyr, \tbf{consistent with the timeframe of the cluster merger.} 
Fig. \ref{fig:EWsD4000}b: \textit{left} indicates this population is now post-starburst, characterized by strong \Hd absorption and little \OII emission. The values are consistent with a starburst followed by rapid quenching, or period of strong star formation that has gradually declined. The current and past activity levels are the same for all South blue subpopulations (BB, BM and BBM) as well (Fig. \ref{fig:EWsD4000}b,c).
Without further star formation these galaxies will fade to the red sequence.
\Dn indicates the Southern blue cloud galaxies are on average much older than the currently star forming blue cloud galaxies of the North and Superfield. An old mean stellar age in addition to a fading A star population suggest that the average South blue cloud galaxy is transitioning to passive. 
\end{itemize}

%originating from red sequence...
Alhough \tbf{spectral measurements suggest} the Southern blue cloud is transitioning to passive after a starburst, the population could have originated as passive. 
\citet{Abramson2013} find a starburst to be a temporary phase in lifetime of a galaxy, rather than a terminal quenching mechanism.
%originating from RS, morphology...
The morphology of the Southern blue cloud is \tbf{also} consistent with this picture.
Fig. \ref{fig:morph} indicates a blue population dominated by early-type galaxies that border the red sequence. Early-type galaxies typically dominate and define the red sequences of clusters after z$\sim$1, which lends some credence to the picture that these transition galaxies were temporarily rejuvenated from the red sequence. 

%Elliptical--no gas for SF, old morph 
A closer look at the morphology indicates the South blue cloud early-types are primarily comprised of ellipticals, which typically have old stellar populations and lack the cold gas needed to form new stars. The evolution and assembly of ellipticals is well-understood to be heterogeneous after z$\sim$0.6 \citep{1997ApJS..110..213S}.
Ellipticals that are blue and post-starburst have been found in local clusters, in a phase assumed to last 0.5 Gyr, consistent with our recent star formation event \citep{McIntosh2014}. The formation of these galaxies can be explained with minor mergers, which is consistent with our imaging indicating all of the blue ellipticals have either companions or tidal debris.

\tbf{A population of blue S0 galaxies is consistent with a post-starburst population, which presents an alternative picture to a rejuvenated red sequence.} 
%\tbf{A post-starburst population is also consistent with a population of blue S0 galaxies. It is difficult to distinguish between ellipticals and S0 galaxies, however, which presents an alternative picture to a rejuvenated red sequence.} 
In the degenerate case where the \ewHd is a result of strong star formation followed by a gradual decline, the early-type galaxies could be S0 galaxies formed from disk galaxies that have slowly faded as they have used up their gas. 
\tbf{It is difficult to distinguish between ellipticals and S0 galaxies, however.}
Without knowledge of prior morphologies, we cannot distinguish between these two possibilities. \tbf{This timeline is consistent with the findings of} \citet{P99} and \citet{Abramson2013} \tbf{in} that it is not likely a single process acting on a single timescale \tbf{that} is responsible for a post-starburst population. 

%Talk about North here now...
If an equal mass merger were the cause of the collective triggering or quenching of star formation imprinted on the South blue composites, we might expect a similar signature in the North.
Instead, we find the North blue cloud to be dramatically different. %North spectra indicate significant current activity and none in the recent past. 

\begin{itemize}
\item While the North and South galaxies in each cluster experienced the merger at the same time, we find no enhanced activity in the North blue cloud in the past and a population that is currently highly active.
The EW(\OII) and EW(\Hd) measurements in Fig. \ref{fig:EWsD4000}b: \textit{left} indicate that the average blue galaxy in North is currently starbursting. The North shows a greater level of current activity than the South for all subpopulations, as well (Fig. \ref{fig:EWsD4000}b,c: \textit{left}). Fig. \ref{fig:EWsD4000}b: \textit{right} corroborates this measurement, indicating that the average North blue galaxy recently underwent a major star formation event $<$0.5 Gyr ago.

%Maybe rephrase cause copied from EW Section in body
\item 
%, \tbf{as indicated by Fig. \ref{fig:EWsD4000}: \textit{left}}. 
The average North cluster galaxy is more active than the average Superfield galaxy, seen in Fig. \ref{fig:EWsD4000}: \textit{left} for full \tbf{(red and blue)}, blue and BB subpopulations. 
While it is interesting that the North is more active than the South (discussed above), it is unexpected that the North is more active than the Superfield. 
Field galaxies are on average younger and more active than cluster galaxies \citep[e.g.][]{Oemler2013}. Galaxies in less dense environments are not subject to the same quenching mechanisms that act on galaxies once they enter a cluster environment, such as ram pressure stripping and starvation \citep{P99}. 
\tbf{The exception to this case would be if} the North is still a forming \tbf{as a} cluster.
In young, still forming clusters, the process of cluster formation can induce star formation as galaxies fall into the potential.
The physical mechanisms responsible for triggering activity in the very recent past could be due to either secular cluster processes or latent merger activity \tbf{\citep{Dressler1980,2009ApJ...693..112P,2013ApJ...763..124R,Dressler2013}.} %Don't go into specifics...maybe wait til after fold in dynamics (N 2 bcgs) and young v irregular correlation refs regarding morphology (late-type, mergers, fractions)
%secular cluster processes and latent merger activity, as in clusters forming and virializing, and the galaxies infalling and evolving on timescales without a merging event, and secondary or nonlinear effects of merger forces, galaxies moving through areas of transient gas then hitting denser gas causing and short or longer term starburst

\item Coupled with \Dn, the mean stellar age of the average blue galaxy in the North appears to be significantly younger than the average blue galaxy in Superfield and South. 
As our observations \tbf{reveal} that the North is not only active, but irregular \tbf{(discussed below)}, we must \tbf{continue to} consider that its current state may be correlated with cluster age, rather than coupled to the merger. 
Unlike the South, the spectral properties of the North are far departed from relaxed clusters we see at the similar and lower redshifts \citep{Beers1990,DGunn1985}. Moreover, the presence of a young cluster at intermediate redshifts is consistent with hierarchical structure formation, as cluster formation is still an active process \citep{Kauffmann1995}.
\end{itemize}

%North and South differences: CMD, morph..tie in later to spectral or before:
The photometric and morphological properties of the North and South clusters reflect populations that correspond to their distinct spectral histories: young and active versus old and passive.
While the North and South have a well-defined red sequence dominated by early-type galaxies
%, consistent with cluster evolution after z$\sim$1 
(Fig. \ref{fig:morph}), the color distribution and morphology in the North cluster share more properties with the surrounding Superfield than the South cluster. 
The North contains a population of blue, faint, low mass galaxies, also seen in the Superfield. The South blue cloud does not, and instead curtails at V-R$\sim$0.8, giving it half the span and a redder mean color than North. The similarity of the North to the Superfield is consistent with spectral results for the BM populations (Fig. \ref{fig:EWsD4000}c).

The morphology CMD/CSMDs reveal that the North blue cloud is dominated by \tbf{disturbed galaxies} (Fig. \ref{fig:morph}). These morphologies are consistent with our EWs, indicative of an active, dynamic population. 
%insert age bit
%Maybe insert disturbed galaxies instead of paired
While a large number of \tbf{disturbed} galaxies can indicate a starburst in the wake of a merger, they are also an indication of a younger system that is still in the process of forming \citep[e.g.][]{Mei2006,2009ApJ...697L.137P}. 
The South, on the other hand, has a blue cloud dominated by early-type galaxies of both high and low masses and bright and faint magnitudes. A high early-type fraction is consistent with our spectral results indicating the average blue galaxy is no longer active. While it is unusual for massive, early-type galaxies to be in the blue cloud, a red sequence \textit{and} blue cloud dominated by early-type galaxies is also indicative of an old, relaxed cluster \citep{Couch1994,Wirth1994,Dressler1994}. 

The North and South clusters have differing geometry \tbf{(Fig. \ref{fig:map})} and substructure \tbf{(Fig. \ref{fig:dstest})}, consistent with spectral results of active and passive, respectively. 
Table \ref{tab:4} summarizes the significant offsets between key cluster components including galaxy density, WL, and X-ray peaks, and BCGs, \tbf{quantifying the extent to which each cluster is disturbed by using items} which are typically aligned in a relaxed cluster.
%Discuss active and passive here first. Maybe find stuff from other section.
Neither cluster is regular, post-merger, but North is exceptionally irregular.
The North departs from a virialized cluster in its distorted geometry and kinematics. 
%Discuss stuff from table or refer to sections
As discussed in Section \ref{fig:dstest}, a DS test reveals that the North has significant substructure and is elongated almost perpendicular to the merger axis (Fig. \ref{fig:map}). 

While irregular structure could indicate disruption due to the merger, it is also consistent with our spectral results indicating the North is young. Musket Ball is a dissociative merger during which the primary baryonic and non-baryonic components have become detached, so we largely attribute the gas offset to the cluster merger. The WL-galaxy density offset \tbf{of 145 h$^{-1}_{70}$ kpc (Table \ref{tab:4})} can also be \tbf{largely} attributed to the merger \citep{Dawson2012}. An irregular North cluster is consistent with morphology and spectral results indicating several late-type galaxy mergers and enhanced star formation activity \citep{Cohen2014, Cohen2015}. %An elevated level of star formation activity in a cluster is correlated with an irregular dynamical state.
An examination of the BCGs \tbf{also} indicates the North may be in the early stages of formation. As discussed in Section \ref{subsec:BCG}, the North has two BCG candidates which are largely offset and occupy separate galaxy density peaks. The fact that both candidates are close in magnitude and mass, yet distant, suggests that the North is in the process of assimilating two gravitationally bound components, possible a group or cluster \tbf{\cite[e.g.][]{2010ApJ...718...23S,2007MNRAS.375....2D}. This process is consistent with hierarchical structure formation and introduces a probability that the North was already in an unrelaxed state prior to the merger \citep{Kauffmann1995}}
Moreover, each density peak is populated by a concentration of luminous, red galaxies, characteristic of a cluster core \tbf{\citep{2000AJ....120.2148G}}. While the BCG offsets are suggestive of disruption due to the merger, they are also supportive of a cluster that is still forming, along with the young stellar populations as indicated by spectra. \tbf{We are unable to conclude which is responsible for the observed trends.}
%Cohen 2015 fing unable to distinguish  between a cluster being young or undergoing a merger

\tbf{The South cluster components} have minor offsets \tbf{(Table \ref{tab:4})} and \tbf{the LOS geometry} is not spherically symmetrical around the galaxy density peak \tbf{(Fig. \ref{fig:map})}. However, the DS test indicates little substructure. This regularity could indicate that it was less affected by the merger than the North cluster. \tbf{However,} combined with the old, passive stellar population and high early-type fraction, the lack of substructure is also suggestive of a more relaxed pre-merger state.

While virialization timescales can be on order of 1 Gyr, it is unlikely both clusters have evolved to be so vastly different in substructure, galaxy populations, morphology and geometry, in the time since the merger. Without a clear spectral signature connecting levels of star formation to the merger event, \tbf{we find} it is difficult to link \tbf{these} characteristics to the merger.

\section{Discussion} \label{sec:disc}

%This was redundent with the last paragraph
A plethora of physical mechanisms govern the star formation of cluster galaxies starting from when they first encounter the cluster potential and ICM \tbf{(e.g. strangulation and ram pressure stripping, discussed in detail in Section \ref{sec:intro})}. Our results (Section \ref{sec:results}) suggest these processes have been at work in both clusters.
%The question we sought to answer was, %Do our observations tell us these forces  long before the time of the merger. The question remains, c
\tbf{Can we link the conditions for these mechanisms to the merger, or were they in place long before the collision?}
%No...maybe should move suggestive stuff down here then?

The majority of studies \tbf{have sought} to do this by identifying a dearth or surfeit of activity based on location relative to a merger \tbf{\citep[e.g.][discussed below]{Chung2009,Owen2005,MillerOwen2003,Ferrari2005}}. %relative to a merger front. 
Galaxies proximal to the merger front are assumed to be those most affected. This geometrical constraint limits studies to systems in the throes of core passage, or soon after, if a TSC is available. The resulting window for observation is narrow and allows for only instantaneous effects to be observed. 
\citet{Chung2009} rule out ram pressure stripping near the Bullet Cluster shock, but insufficient time has passed to rule out longer-acting mechanisms.
In addition, clusters during or immediately after core passage are chaotic, which can cause membership contamination and bias dynamical calculations.
\citet{Owen2005}, \cite{MillerOwen2003}, and \citet{Ferrari2005} find enhanced activity close to a merger front, which increases the likelihood of correlation but is difficult to interpret. Studies that observe global star formation properties are also limited by the narrow window of time during which instantaneous effects can be observed.
\cite{Stroe2015} look for activity triggered by shocks in two cluster mergers with differing TSCs. 
They detect a boost in the luminosity function
%an excess of H$\alpha$ line emitters 
of CIZA J2242.8+5301, but not 1RJXS J0603.3+4213, and suspect this may be due to the different merger histories accompanied by a small window for observing instantaneous effects.
%and find it difficult to distinguish between a null result and merger scenarios due to the narrow window for observing triggered star formation.
Cluster mergers are an important part of hierarchical structure formation, which occurs over long timescales. In order to get a complete picture of how galaxies are affected by this process, it is necessary to extend beyond the instantaneous physical mechanisms that occur during core passage. As time progresses, the range of mechanisms that can be observed increases.

The Musket Ball system presented a promising opportunity to observe competing latent effects of the merger. tbf{As discussed in Section \ref{sec:target}, the TSC serendipitously coincides} with the lifetime of an A star, giving us a window back in time to the merger event. We were able to observe not only current star formation but a possible increased or diminished star formation due to short time-scale processes during the merger. A comprehensive set of high resolution spectra allowed us to probe the galaxy populations of both clusters after their first pass-through as well as a local comparison sample not involved in the merger.
Multi-wavelength data allowed us to investigate trends by color, mass, brightness and morphology. 
In addition, dynamical calculations added further insight into possible secular cluster processes and merger scenarios consistent with our observations. \tbf{Our strongest constraint, discussed in Section \ref{sec:results}, stemmed from the finding that no recent star formation has occured in the average North cluster red sequence galaxy,
\textit{precluding the possibility of a merger quenched, bulk migration of galaxies onto the red sequence.}}
%affecting the star formation properties of the red sequence through the quenching of a large population of galaxies.}
%that (Section \ref{sec:results})}

While our system has qualities that make it advantageous for studying the effects of the merger, such as a TSC of $\sim$1 Gyr, a low angle of inclination and a surrounding coeval sample, it is difficult to distinguish whether the current picture is an appreciable result of transformation due to the merger. 
As discussed in Section \ref{sec:results}, it is unexpected that the average North cluster galaxy is younger and more active than the average Superfield galaxy, a difference further amplified in blue and BB populations.
%Galaxies in less dense environments are not subject to the same quenching mechanisms that act on galaxies once they enter a cluster environment, such as ram pressure stripping and starvation.
A burst significantly younger than the 0.7 Gyr lower bound of our 68\% confidence interval could be due to longer acting processes such as torquing of galaxy pairs or groups, or merger induced shock-waves. \citet{Stroe2015} find that shocks preserve the ionized material in galaxies that fuels starbursts, potentially compressing it and triggering the collapse of star forming clouds. Shocks may act on longer timescales in a merger due to low cluster collision velocity, resulting in a low Mach number and delayed interaction with areas of the cluster trailing the merger. Simulations by \cite{Roettiger1996} suggest that during a merger a bow shock can temporarily protect galaxies in an infalling cluster, delaying the short and long-term effects of ram pressure stripping due to the ICM. 
%As time passes, the shock wave continues on to have a global effect, reaching all cluster galaxies as it travels through the cluster before dissipating. 
However, we cannot rule out the possibility of elevated activity due to infalling, low-mass galaxies, or of a population of gas-rich galaxies un-stripped by the merger and simply undergoing secular processes seen in young clusters like North.
% given that North has a young stellar population, and may still be forming, 

Recall that the South blue cloud show evidence of a star formation event coinciding with the merger timescale (Fig. \ref{fig:EWsD4000}b).
Our results (Section \ref{sec:results}) indicate the South \tbf{had} a blue cloud population dominated by early-type galaxies of both high and low masses and bright and faint magnitudes, but on average redder than counterparts in the North and Superfield, which is consistent with a transition population. 
By way of secular cluster processes, blue galaxies of different masses and brightnesses are shown to transition on different timescales across different epochs \citep[e.g.][]{Balogh2016,Muzzin2014,Ilbert2015}. 
A variety of short and long-term mechanisms could result in a starburst in the past $\sim$1 Gyr followed by rapid quenching, to produce the current post-starburst population.
The formation of a blue elliptical is consistent with a merger of a gas poor-elliptical with a gas-rich galaxy. Morphology reveals all of the blue ellipticals have either companions or tidal debris. An ensemble of minor mergers corresponding to a major star forming event is uncommon, especially given the high relative velocity dispersions in clusters. The coincidence of triggering and cessation with the merger is suggestive, but without pre-merger morphologies we cannot draw a prior connection.
%Combined with dynamics, our results are suggestive that the Southern cluster is inactive and becoming virialized, but also old, and the Northern cluster active and disturbed, but also young and still forming. 
%However, we cannot determine whether or not these were initial conditions, 

In the absence of a strong, definitive signal, without knowledge of cluster properties prior to the merger, we can only offer suggestive lines of evidence on how the merger may have transformed the Musket Ball galaxies.
%We can only speculate on the corresponding physical mechanisms.
The large spread of possibilities given competing effects of short and long term quenching and triggering mechanisms demands a more tightly constrained TSC, a coeval sample of non-merging clusters, and a large sample of similar merging systems.
%While our TSC provides a window back to the time of core-passage, a tighter constraint is still necessary to rule out vastly different scenarios.
The 2.4 Gyr upper limit of our 68\% confidence interval exceeds both the lifetime of A stars and the $\sim2$ Gyr span for observable merger effects. A simulation constraining the merger towards the 0.7 Gyr lower limit would increase the likelihood that current and recent activity levels are tied to the physical conditions produced by the merger, and still allow us to detect A stars.

%Insert bit on general studies which say we need a large sample size
While our Superfield galaxies provide a coeval, non-merger sample, a true control sample would include clusters of similar mass and redshift and a true field population not embedded in a larger structure. 
The population of A stars in Superfield provides a barometer for activity in North and South at the time of the merger, but galaxies that exist outside of groups or clusters, are not subject to the same interactions that arise in overdense environments.
A control sample of clusters would provide much tighter constraints on the pre and post-merger properties of Musket Ball. An excess or dearth of A stars in North or South compared to clusters at z$\sim$0.5 would imply that a fast-acting mechanism triggered or quenched star formation, whereas a similar population would imply the merger had little effect on star formation \citep{Oemler2009,Poggianti2006,Treu2003,D04,Ma2010}.

%A number of studies compare systems at different merger stages, but are limited by TSC and timescale of the mechanism. 
When clusters merge, the effects of star formation on member galaxies vary with location and time, as well as initial conditions. Not knowing the exact prior state of the clusters in Musket Ball, we can only speculate on initial conditions necessary to break the post-merger degeneracy of age and cluster irregularity. 
These dependencies demand a large sample of merging systems, or hydrodynamical simulations with initial conditions.

An alternate approach could be a system where we can statistically constrain both the initial and final states of the clusters.
A transverse system in the initial stage of merging would allow us to observe short-timescale processes acting at the merger front by comparison to the galaxy properties at larger distances yet to be affected by the merger. Performing spectral analysis on composites binned by radius along the merger axis would allow us to probe the star formation history of galaxies in the sample while comparing current activity levels. We have the opportunity to do so using the merger RX J0910 (RX J0910+5419 and RX J0910+5422), comprised of two clusters at z$\sim$1.10 on a trajectory to collide \citep{2006ApJ...639...81M}. \tbf{We investigate RX J0910 in Mansheim et al. 2016 (submitted).}

\section{Acknowledgments}
%\footnotesize{
We thank the staff at W. M. Keck Observatory for their help during observing runs, James Jee for reduction of \textit{HST} data, Perry Gee for help accessing DLS data, and Adam Tomczak, Nicholas Rumbaugh and Andreas Faisst for paper feedback. We thank the referee for extensive comments on how to improve the paper.
Facilities: Keck II, \textit{HST}.
%}

%Maybe remove Red sequence values, or only leave in here if don't include in plots
\renewcommand{\arraystretch}{1.0}
%\begin{table*}
%\begin{deluxetable*}{lcccccc}
\begin{deluxetable*}{lcccc}
%\tabletypesize{\footnotesize}
%\tablecolumns{8} 
%\tablewidth{0pc} 
\tablecaption{Composite Equivalent Width and \Dn Values}
\tablehead{ \colhead{Region}
 & \colhead{\footnotesize{Galaxies}\tablenotemark{a}}
 & \colhead{\footnotesize{\Hd}\tablenotemark{b,}\tablenotemark{c}}
 & \colhead{\footnotesize{\OII}}
 & \colhead{\footnotesize{\Dn}}\\
 \colhead{{}}
 & \colhead{\footnotesize{}}
 & \colhead{\footnotesize{(\AA)}}
 & \colhead{\footnotesize{(\AA)}}
 & \colhead{\footnotesize{}}
% & \multicolumn{2}{c}{\footnotesize{\Hd}\tablenotemark{a}}
% & \multicolumn{2}{c}{\footnotesize{\OII}}
% & \multicolumn{2}{c}{\footnotesize{\Dn}}\\
% & \multicolumn{2}{c}{\footnotesize{(\AA)}}
% & \multicolumn{2}{c}{\footnotesize{(\AA)}}
% & \multicolumn{2}{c}{\footnotesize{(\AA)}}
% & \multicolumn{2}{c}{\footnotesize{(\AA)}}\\
%\colhead{{}}
% & \colhead{\footnotesize{RSG}}
% & \colhead{\footnotesize{BP}\tablenotemark{b}}
% & \colhead{\footnotesize{RSG}}
% & \colhead{\footnotesize{BP}}
% & \colhead{\footnotesize{RSG}}
% & \colhead{\footnotesize{BP}}
}
\startdata
\cutinhead{ }
North & 38	 $\pm$ 11 & 1.81 $\pm$ 0.09 $\pm$ 0.25 & -8.17 $\pm$ 0.65 $\pm$ 15.72 & 1.63 $\pm$ 0.00 \\ 
Superfield & 48	 $\pm$ 25 & 2.31 $\pm$ 0.07 $\pm$ 0.04 & -4.10 $\pm$ 0.19 $\pm$ 0.23 & 1.68 $\pm$ 0.00 \\ 
South & 26	$\pm$	6 & 1.43 $\pm$ 0.08 $\pm$ 0.09 & -0.62 $\pm$ 0.16 $\pm$ 0.13 & 1.83 $\pm$ 0.00 \\ 
\cutinhead{Color\tablenotemark{d}: Red, Blue}
\sidehead{Blue}
North & 13	$\pm$	5	& 5.64 $\pm$ 0.32 & -18.65 $\pm$ 0.54 & 1.18 $\pm$ 0.01 \\ 
Superfield & 24	$\pm$	17	& 4.55 $\pm$ 0.12 & -12.59 $\pm$ 0.22 & 1.33 $\pm$ 0.00 \\ 
South & 8	$\pm$	2	& 3.77 $\pm$ 0.61 & -2.73 $\pm$ 0.60 & 1.58 $\pm$ 0.02 \\ 

\sidehead{Red}
North & 25	$\pm$	6	& 1.50 $\pm$ 0.09 & -6.03 $\pm$ 0.63 & 1.77 $\pm$ 0.01 \\ 
Superfield & 24	$\pm$	7	& 1.85 $\pm$ 0.05 & -1.18 $\pm$ 0.17 & 1.82 $\pm$ 0.00 \\ 
South & 18	$\pm$	3	& 1.36 $\pm$ 0.08 & -0.39 $\pm$ 0.13 & 1.86 $\pm$ 0.01 \\ 

\cutinhead{Massive: log(${M_{*}}$) $\geq$ 10.5}
\sidehead{Blue \tbf{(BM)}}
North & 4	$\pm$	1	& 3.48 $\pm$ 0.41 & -6.51 $\pm$ 0.89 & 1.42 $\pm$ 0.02 \\ 
Superfield & 7	$\pm$	5	& 3.27 $\pm$ 0.16 & -7.09 $\pm$ 0.68 & 1.40 $\pm$ 0.01 \\ 
South & 6	$\pm$	0	& 3.19 $\pm$ 0.44 & -1.45 $\pm$ 0.43 & 1.57 $\pm$ 0.02 \\ 

\sidehead{Red (RM)}
North & 17	$\pm$	4	& 1.41 $\pm$ 0.10 & -6.71 $\pm$ 0.70 & 1.78 $\pm$ 0.01 \\ 
Superfield & 15	$\pm$	3	& 1.84 $\pm$ 0.06 & -1.11 $\pm$ 0.16 & 1.83 $\pm$ 0.00 \\ 
South & 12	$\pm$	0	& 1.35 $\pm$ 0.08 & -0.36 $\pm$ 0.12 & 1.86 $\pm$ 0.01 \\ 

\cutinhead{Bright: R$\leq$ 21.7}
\sidehead{Blue (BB)}
North & 8	$\pm$	2	& 5.80 $\pm$ 0.35 & -16.34 $\pm$ 0.61 & 1.21 $\pm$ 0.01 \\ 
Superfield & 8	$\pm$	8	& 3.51 $\pm$ 0.14 & -8.91 $\pm$ 0.42 & 1.37 $\pm$ 0.01 \\ 
South & 5	$\pm$	1	& 3.45 $\pm$ 0.45 & -2.89 $\pm$ 0.57 & 1.54 $\pm$ 0.02 \\ 

\sidehead{Red (RB)}
North & 17	$\pm$	4	& 1.41 $\pm$ 0.10 & -6.71 $\pm$ 0.70 & 1.78 $\pm$ 0.01 \\ 
Superfield & 13	$\pm$	2	& 1.76 $\pm$ 0.06 & -1.17 $\pm$ 0.16 & 1.84 $\pm$ 0.00 \\ 
South & 10	$\pm$ 0	 & 1.35 $\pm$ 0.08 & -0.36 $\pm$ 0.12 & 1.85 $\pm$ 0.01 \\ 

\cutinhead{Bright Massive: log(${M_{*}}$) $\geq$ 10.5, R $\leq$ 21.7}
\sidehead{Blue (BBM)}
North & 4	$\pm$	1	& 3.48 $\pm$ 0.41 & -6.51 $\pm$ 0.89 & 1.42 $\pm$ 0.02 \\ 
Superfield & 7	$\pm$	5	& 3.26 $\pm$ 0.16 & -6.93 $\pm$ 0.69 & 1.40 $\pm$ 0.01 \\ 
South & 5	$\pm$	1	& 3.45 $\pm$ 0.45 & -2.89 $\pm$ 0.57 & 1.54 $\pm$ 0.02 \\ 

\sidehead{Red (BRM)}
North & 17	$\pm$	4	& 1.41 $\pm$ 0.10 & -6.71 $\pm$ 0.70 & 1.78 $\pm$ 0.01 \\ 
Superfield & 13	$\pm$	2	& 1.76 $\pm$ 0.06 & -1.19 $\pm$ 0.09 & 1.84 $\pm$ 0.00 \\ 
South & 10	$\pm$ 0 & 1.35 $\pm$ 0.08 & -0.35 $\pm$ 0.12 & 1.85 $\pm$ 0.01 \\ 

%\cutinhead{Massive}
\enddata
\label{tab:2}
\tablenotetext{a}{\footnotesize{
Galaxies in area of equal spectroscopic coverage (Section \ref{subsection:spectral}). Errors are based on completeness discussed in Appendix \ref{subsubsection:completeness}
}
}
\tablenotetext{b}{\footnotesize{Second error value is estimated maximum variance due to spectral sampling (for details see Appendix \ref{subsubsection:completeness}). Values are not listed for additional populations as they are nearly complete.}}
\tablenotetext{c}{\footnotesize{Infill corrected values used in cases where emission component removable using double Gaussian fit (for details see Section \ref{subsubsection:EW}). }}
\tablenotetext{d}{\tbf{\footnotesize{Color (red or blue) determined by a red sequence fit discussed in Section \ref{subsec:rsfit}. } } }

%\multicolumn{7}{ | l | }{**Note this changes: Infill corrections +0.07, +0.02, and +0.06$\AA$ for North, South, and Remainder, respectively.} \\
% \multicolumn{7}{ | l | }{*.}  \\
%\multicolumn{7}{ | l | }{**Measured using bandpasses from Fisher et al. (1998). }  \\
%ADD THESE TO TABLE VALUES!!!!!!!!!!!!!!!!!!!!!! RSG Hdelta!!!!!!!!!!!!!!
\end{deluxetable*}

\appendix
%\section{} 

\section{Sampling and Completeness} \label{subsubsection:completeness}
%\label{subsubsection:completeness}
%\input{Appendix.tex}
%\subsubsection{Completeness and Significance Error Estimation} 
%While we have excellent spectral coverage, it is important to estimate uncertainties due to our selection method and incompleteness of our sample. 

\tbf{
Our spectroscopic selection method for targets relied on photometric redshifts (Section \ref{subsec:photoz}) which allowed us to predict which objects in our field of view were most likely to be at the cluster redshift. Since we acquired spectroscopic redshifts for 93\% of the objects that were targeted based on our z$_{phot}$ criterion (Section \ref{subsubsection:selection}), we were able to evaluate the effectiveness this method, and the subsequent completeness of our sample. We then propagated the associated uncertainty as an error when measuring EWs from composite spectra (Section \ref{subsubsection:EW}). 
We describe the procedures for doing so in this section. 
%153/164=0.932. Table 4 EWs is tab:2
}

\tbf{
We evaluated the success of our selection method for all cross sections of color, magnitude and stellar mass found in Table \ref{tab:2} to reflect possible differences in sampling manifested in our capability to observe galaxies with varied physical properties. 
For example, cluster cores are dominated by red sequence galaxies, but DEIMOS slitmasks are limited to have a single slit at a given position perpendicular to the 
%by slit width in the 
dispersion direction so as to not overlap spectra. Thus, the densest areas are prone to undersampling. We mitigated this by targeting the densest regions with all slitmasks, but calculate completeness for blue red and blue populations separately (in addition to together) to contain and quantify any resulting bias.
%Statistical calculations for sampling uncertainties 
Galaxies assigned completeness errors (second column in Table \ref{tab:2}) are drawn only from areas that received equal spectroscopic coverage (described in Section \ref{subsection:spectral} and central to Fig. \ref{fig:map}). This reduced area contains all of the North and South clusters and most of the Superfield and North East.
}

\tbf{We calculated the success rate of our selection method as the number of objects with photometric redshifts inside the photometric redshift range (z$_{phot}=$ 0.53$\pm$0.1, Section \ref{subsec:photoz}) that were targeted then successfully confirmed to be inside our cluster redshift range (0.525 $\leq$ z$_{spec}$ $\leq$ 0.54, Section \ref{subsec:redshiftdist}). When selecting targets these objects were given first priority because we considered them as most likely to be spectroscopically confirmed inside our cluster redshift range. Objects outside this range were considered less likely to be in the cluster, so were given second priority. We define objects that fit our selection criterion for which we do not have spectra as \textit{missing potential members} (\textit{m}), marked as blue squares in the CMD/CSMDs (Fig. \ref{fig:completeness}).} We make the assumption that missing members are drawn from the same population of the observed ones.
%Objects with z$_{phot}$ = 0.53$\pm$0.1
%We consider objects which validated our selection method as contributing to a ``success rate''.}
\tbf{Due to the larger error associated with photometric redshifts ($\sigma_{z_{phot}}>\sigma_{z_{spec}}$), any method that relies upon them to predict objects that fall inside a narrow spectroscopic redshift range will have associated failures.
As a result, objects with z$_{phot}\leq$ 0.53$\pm$0.1 confirmed outside 0.525 $\leq$ z$_{spec}$ $\leq$ 0.54 are considered Failure Type I (FTI).
Objects outside z$_{phot}$ = 0.53$\pm$0.1 which were confirmed inside 0.525 $\leq$ z$_{spec}$ $\leq$ 0.54 are Failure Type II (FTII, seen as red hexagons in Fig. \ref{fig:completeness}).}

%Check that 51 is total in max area of coverage
\tbf{
For example, of the 51 confirmed z$_{clust}$ galaxies in the bright red population (BR, Table \ref{tab:2}), 43 were targeted based on our selection criterion 
 (z$_{phot}$ $=$ 0.53$\pm$0.1).
%yeilding a 84\% \textit{success} rate. 
However, one object confirmed in z$_{clust}$ was missed by our selection method 
 (z$_{phot}$ $>$ 0.53$\pm$0.1, 
(FTII) and there were three objects for which our selection method failed
 (z$_{phot}$ $=$ 0.53$\pm$0.1, 0.525 $\leq$ z$_{spec}$ $\leq$ 0.54) 
(FTI). (The remaining four objects are outside the areas of equal coverage so we do not factor them any further into the sampling statistics.)
% We define objects that fit our selection criterion for which we do not have spectra, defined as \textit{n}) \textit{missing potential members},
%  objects which were \textit{missed} because they were either not targeted or targeted and yielded low-quality spectra), 
Thus, of the \textit{m} BR missing potential members within our area of equal coverage, we assume 2\% of them may be in z$_{clust}$ because they eluded our selection criterion, scaled by 7\% that may be outside z$_{clust}$ because they failed our criterion. 
%For bright red galaxies
% there are X \textit{missing} potential members (objects with z$_{phot}$ $\leq$ 0.53$\pm$0.1 for which we have not confirmed in or out of z$_{clust}$), or 
These rates result in an uncertainty of 3.7, 0 and 0.93 FTI galaxies and 0, 0 and 1 FTII galaxies which we add and round up to attain an upper limit on incompleteness of 4, 0 and 1 galaxies for the North, South and Superfield, respectively.}

For EW measurements, we evaluate incompleteness for each region informed by the number of potential missing members, defined above as \textit{m}. We also evaluate systematic uncertainties based on the possible variance in our results given our data set.
%Completeness correction and systematic uncertainty are 
We propagate uncertainties as EW measurement errors using a bootstrap/Monte Carlo (MC) resampling method. We estimate systematic uncertainty with 500 realizations of each composite spectrum, randomly drawn with replacement \textit{n} times, where \textit{n} is the number of objects going into each composite. A distribution is made of bandpass measurements of \ewOII and \ewHd and \Dn. The uncertainty is taken as the variance of this distribution, which is assumed to be Gaussian. We estimate incompleteness by doing the 500 realizations of each composite spectrum, resampling with replacement \textit{n+m} times.
%where \textit{n} is the number of objects going into each composite and \textit{m} is the number of missing potential members from that population. 
\tbf{When making composites for EW measurements, we include spectra from the entire field, not just the area of maximum spectroscopic coverage, so \textit{m} = 13, 7 and 218 for the North, South and Superfield
%corresponding to the magenta dots shown in CMD/CSMDs in Fig. \ref{fig:completeness}). }
, corresponding to the blue circumscribed dots shown in CMD/CSMDs in Fig. \ref{fig:completeness}. (Note that objects in \textit{m} for which we have SHELS confirmed spectroscopic redshifts are plotted as magenta hexagons.)}

The completeness correction and systematic uncertainty are both on order of the bandpass measurement error for all populations, with the exception of EW(\OII) for North red sequence composites. We discuss the possible non-star forming origin of this feature in Section \ref{subsubsec:AGN} 

\clearpage
\bibliographystyle{apj}
\bibliography{ApJ_paper_amansheim}
\end{document}